\documentclass[a4paper,11pt]{article}
\pdfoutput=1
\usepackage{cite}
\usepackage{verbatim}
\usepackage{multirow}
\usepackage{jheppub}
\usepackage{hyperref,soul}
\usepackage{enumerate}
\usepackage{relsize}
\usepackage{mathrsfs}
\usepackage{pdflscape}
\usepackage{inputenc}
\usepackage{braket,xcolor,float}
\usepackage{bbold,amsmath,lipsum,amssymb,verbatim,cleveref}
\usepackage[T1]{fontenc}
\usepackage{graphicx,wrapfig}
\newcommand{\C}{\mathbb{C}}

\newcommand{\lb}{\left(}
\newcommand{\rb}{\right)}
\newcommand{\xr}{x^\rho}
\newcommand{\xt}{x^t}

\newcommand{\xf}{x^\phi}
\newcommand{\xtt}{x^T}
\newcommand{\xff}{x^\varphi}

\newcommand{\yf}{y^\phi}
\newcommand{\yff}{y^\varphi}

\newcommand{\A}{\mathcal{A}(z\,;\ell)}
\newcommand{\B}{\mathcal{B}(z\,;\ell)}
\newcommand{\G}{\Gamma(z;\ell)}
\renewcommand{\L}{\Lambda(z;\ell)}
\newcommand{\ra}{\rightarrow}

\newcommand{\nn}{\nonumber}

\newcommand{\be}{\begin{eqnarray}}
	\newcommand{\ee}{\end{eqnarray}}
\newcommand{\beq}{\begin{eqnarray}}
	\newcommand{\eeq}{\end{eqnarray}}

\newcommand{\beqa}{\begin{eqnarray}}
	\newcommand{\eeqa}{\end{eqnarray}}

\usepackage{mathrsfs}
\usepackage{orcidlink}

\newcommand{\h}{\mathfrak{h}}

\renewcommand{\d}{\partial}

\newcommand{\n}{\nu}

\newcommand{\m}{\mu}

\renewcommand{\t}{\tau}

\ifdefined\C
\renewcommand{\C}{\boldsymbol{C}}
\else
\newcommand{\C}{\boldsymbol{C}}
\fi

\newcommand{\Tr}{{\rm Tr}}
\allowdisplaybreaks[0]

\title{Blackhole Near Horizons through the Looking Glass}
\author[a]{Arjun Bagchi$\,$\orcidlink{0000-0002-6481-5933},}
\emailAdd{(abagchi} 
\author[a]{Arkachur Bhattacharya$\,$\orcidlink{0009-0004-3153-6470},}
\emailAdd{arkachurb25} 
\author[a]{Sharang Rajesh Iyer$\,$\orcidlink{0000-0001-5080-3117},} 
\emailAdd{siyer23)@iitk.ac.in}
\author[b]{and K. Narayan$\,$\orcidlink{0000-0003-1863-7715}.}
\emailAdd{\ narayan@cmi.ac.in}\author{\\}
\affiliation[a]{Indian Institute of Technology Kanpur, Kanpur 208016, India.\\}
\affiliation[b]{Chennai Mathematical Institute,
H1 SIPCOT IT Park, Siruseri 603103, India.\\}

\abstract{We show that the near horizon of a generic non-extremal black hole (BH) can be understood in terms of a Carrollian geometry with two null directions, also called a String-Carroll (SC) geometry. The base space of this fibre-bundle structure is a sphere (or a plane for a black brane) and the fibre is the two-dimensional Rindler spacetime. We launch a detailed study of probes in this geometry. We study particle geodesics and scalar fields. The first part of the paper constructs geodesics and probe scalar fields directly in the SC geometry. We then look at a wide class of examples, including the Schwarzschild BH in asymptotically flat spacetimes, the black brane in AdS spacetimes, and Lifshitz black holes and construct the explicit maps to the SC geometry to obtain results specific to each case. These results are reproduced by considering the probe particles and fields in the original BH background and taking the near-horizon limit of the solutions. Our encyclopedia of examples establishes the notion of SC geometries as near-horizon geometries of non-extremal black objects, paving the way for a detailed, intricate future analysis of the quantum aspects of this geometry.}
\begin{document}

\maketitle

\section{Introduction}
The spectacular images of M87 by the Event Horizon Telescope has ushered a new era in black hole physics. Now, even more than ever, there is a need for understanding what happens when one approaches the event horizon of a black hole. This is the question we are interested in addressing in this paper, and we do this through the new lens of Carrollian symmetry. 

\medskip

Near horizon limits and near horizon geometries are best understood in the context of extremal black holes. Here, e.g. in the context of four dimensional black holes, it is known that the geometry becomes $AdS_2 \times S^2$ \cite{Kunduri:2013gce}, where it is understood that in the general case, the sphere is non-trivially fibred on the $AdS_2$ . This is a spacetime in its own right as it solves the Einstein's equations. The dynamics of the near horizon region decouples from the bulk spacetime and physics here can be understood without referring to the original black hole spacetime. Near horizon geometries of extremal black holes have been the playground of a whole host of interesting avenues of research. We refer the reader to the comprehensive review \cite{Kunduri:2013gce} for a flavour of this wide field of research. 

\medskip

Non-extremal black holes, surprisingly, are more difficult to handle from the point of view of near horizon limits. It has long been appreciated that in the near horizon region, there is a two-dimensional (2d) Rindler spacetime that appears and various properties of this 2d Rindler plays a crucial role in understanding the black hole \cite{PhysRevD.7.2850,PCWDavies_1975,PhysRevD.14.870,PhysRevD.11.1404,10.1143/PTP.88.1} . But, unlike the extremal case, the near horizon geometry does not solve Einstein's equations and the dynamics of the near horizon does not separate from the parent black hole. 

\medskip

We revisit this problem of understanding the near horizon limit of non-extremal black holes, and in general non-extremal black objects, armed now with new intuition of how to tackle singular limits and deal with non-Lorentzian geometries where metrics become degenerate. The most studied of these non-Lorentzian set ups is the familiar world of Galilean physics, which can be reached in the speed of light $c\to \infty$ limit of relativistic physics. Although Galilean transformations descend in this limit from Poincare transformations, the relativistic spacetime metric degenerates in the non-relativistic set up.
\begin{subequations}\label{1.1}
    \begin{align}
    \eta^{\mu\nu} &= \text{diag}(-1/c^2, 1, 1, 1) \to \h^{\mu\nu} = \text{diag}(0, 1, 1, 1). \\
    -c^{-2} \eta_{\mu\nu} &= \text{diag}(1, -1/c^2, -1/c^2, -1/c^2) \to \tau_{\mu \nu} = \text{diag}(1, 0, 0, 0) = \tau_\mu \tau_\nu. 
\end{align}
\end{subequations}
The pair $\{ h^{\mu\nu}, \tau_\rho \}$ with $h^{\mu\nu} \tau_\nu = 0$, defines a Galilean spacetime. 
The diametrically opposite limit $c\to0$, called the Carroll limit, was first investigated in \cite{Leblond65, SenGupta:1966qer, Henneaux:1979vn}. Recently, this has been the cynosure of all attention because of the unexpected appearance of this symmetry in various apparently disparate fields of physics. In the Carroll world, lightcones close instead of opening up as in the Galilean case and the roles of space and time reverse from the familiar non-relativistic set up. Space is now absolute and time relative. A flat Carroll manifold is defined with 
\begin{subequations}\label{carr-met}
    \begin{align}
    \eta_{\mu\nu} &= \text{diag}(-c^2, 1, 1, 1) \to h_{\mu\nu} = \text{diag}(0, 1, 1, 1). \\
    -c^{2} \eta^{\mu\nu} &= \text{diag}(1, -c^2, -c^2, -c^2) \to \Theta^{\mu \nu} = \text{diag}(1, 0, 0, 0) = \theta^\mu \theta^\nu. 
\end{align}
\end{subequations}
The isometries of flat Carrollian manifolds form the Carroll algebra, which can also be obtained by a $c\to 0$ contraction of the relativistic Poincare algebra. 

\medskip

Carrollian symmetries manifest themselves generically on null surfaces and hence have become central to the understanding of different corners of theoretical physics starting from condensed matter physics to quantum gravity. The most important of these applications include the formulation of the holographic principle of asymptotically flat spacetimes where the dual field theories are co-dimension one Carrollian CFTs \cite{Bagchi:2010eg, Bagchi:2012xr, Barnich:2012aw, Bagchi:2014iea, Bagchi:2015wna, Bagchi:2016bcd, Donnay:2022aba, Bagchi:2022emh, Bagchi:2023fbj, Alday:2024yyj}; connections to fractons \cite{Bidussi:2021nmp}, flat bands \cite{Bagchi:2022eui} and a new class of phase transitions in condensed matter \cite{Biswas:2025dte}; connections to heavy ion collisions through ultra-relativistic hydrodynamics \cite{Bagchi:2023ysc}; and to tensionless null strings where Carrollian symmetries arise on the null worldsheet \cite{Bagchi:2013bga, Bagchi:2015nca, Bagchi:2026wcu}. There are also emerging links to cosmology and dark energy \cite{deBoer:2021jej}.  We refer the reader to the recent review \cite{Bagchi:2025vri} for an overview of the recent developments and further references to the literature, to \cite{Nguyen:2025zhg, Ruzziconi:2026bix} for reviews aimed particularly at flat holography from the Carroll perspective, and to \cite{Ciambelli:2025unn} for a nice geometric overview of Carroll symmetries. 

\medskip

The event horizons of black holes are of course one of the most important classes of null surfaces and indeed Carrollian symmetries show up here as well \cite{Penna:2018gfx, Donnay:2019jiz}. Interestingly, not only horizons, but even near horizon regions of black holes show Carrollian structures and this is what would be the focus of our work in this paper. The structures we would encounter here are generalizations of the usual Carroll structures, which now include two null directions instead of one. The previous structure was a fibre bundle $\mathcal{X}^n \times \mathbb{R}$ where $\mathcal{X}^n$ represented the base space comprising the spatial directions and $\mathbb{R}$ was the one-dimensional parent time direction which had now gone null. Near generic (non-extremal) black holes we will encounter spacetimes which are fibre-bundles again but of the form $\mathcal{X}^{n-1} \times \mathcal{Y}^2$, where now $\mathcal{X}^{n-1}$ is the base space and $\mathcal{Y}^2$ is a 2-dimensional fibre which will correspond to the two dimensional spacetime which has now become null. 

\medskip

For non-extremal black holes, we will be able to identify $\mathcal{Y}^2$ with a 2d Rindler spacetime and we will see that for field theories defined in this geometry, i.e. quantum field theories near black holes, the 2d Rindler would play a starring role, in keeping with older literature. But crucially, most of the older literature disregard the transverse space $\mathcal{X}^{n-1}$ and all the previous literature (except the references in the next line) don't take into account that the structure that emerges in a non-Lorentzian geometry. The observation of the fact that black hole near horizon expansions arrange themselves into these generalised Carrollian expansions called {\em String Carroll} (SC) expansions (borrowing terminology from non-relativistic nomenclature) was first put forward in \cite{Bagchi:2023cfp, Bagchi:2024rje} and followed up more recent in \cite{Banerjee:2025bkg} (also see \cite{Fontanella:2022gyt}). \cite{Bagchi:2023cfp, Bagchi:2024rje,Banerjee:2025bkg} considered strings in near black holes. In this paper, we ask what happens to point particles and quantum fields in these near horizon geometries and put formulation of non-extremal black hole near horizons as SC geometries on firmer footing. 

\medskip

We re-emphasise the following: 

\smallskip

{\em{Near horizon geometry of a generic non-extremal black object (black hole or black brane) is a fibre-bundle SC geometry where the 2d fibre is given by 2d Rindler spacetime and the geometry of the black object dictates the base of the fibre.}} 

\smallskip

The other parameters of the black hole determine the fibration of the bundle. In what follows, we will use probes in near horizon geometries to further strengthen the above statement. In particular, we will look at point particles and scalar fields. We will first construct the action of these probes in generic SC geometries. We will then look at particular black holes and map the near horizons to specific SC geometries to extract how these probes should behave in the near horizons of these specific from our general formulation. 

\medskip

We will complement our intrinsic analysis using SC geometries by considering probes in entire black hole geometry and then take the near horizon limit. Since these limits are inherently singular, there is no a priori guarantee that these different calculations would match. However, we will find agreement, not only at the leading order, but non-trivially to subleading order solutions as well. 
\begin{center}
\begin{figure}
\includegraphics[width=1.0\textwidth]{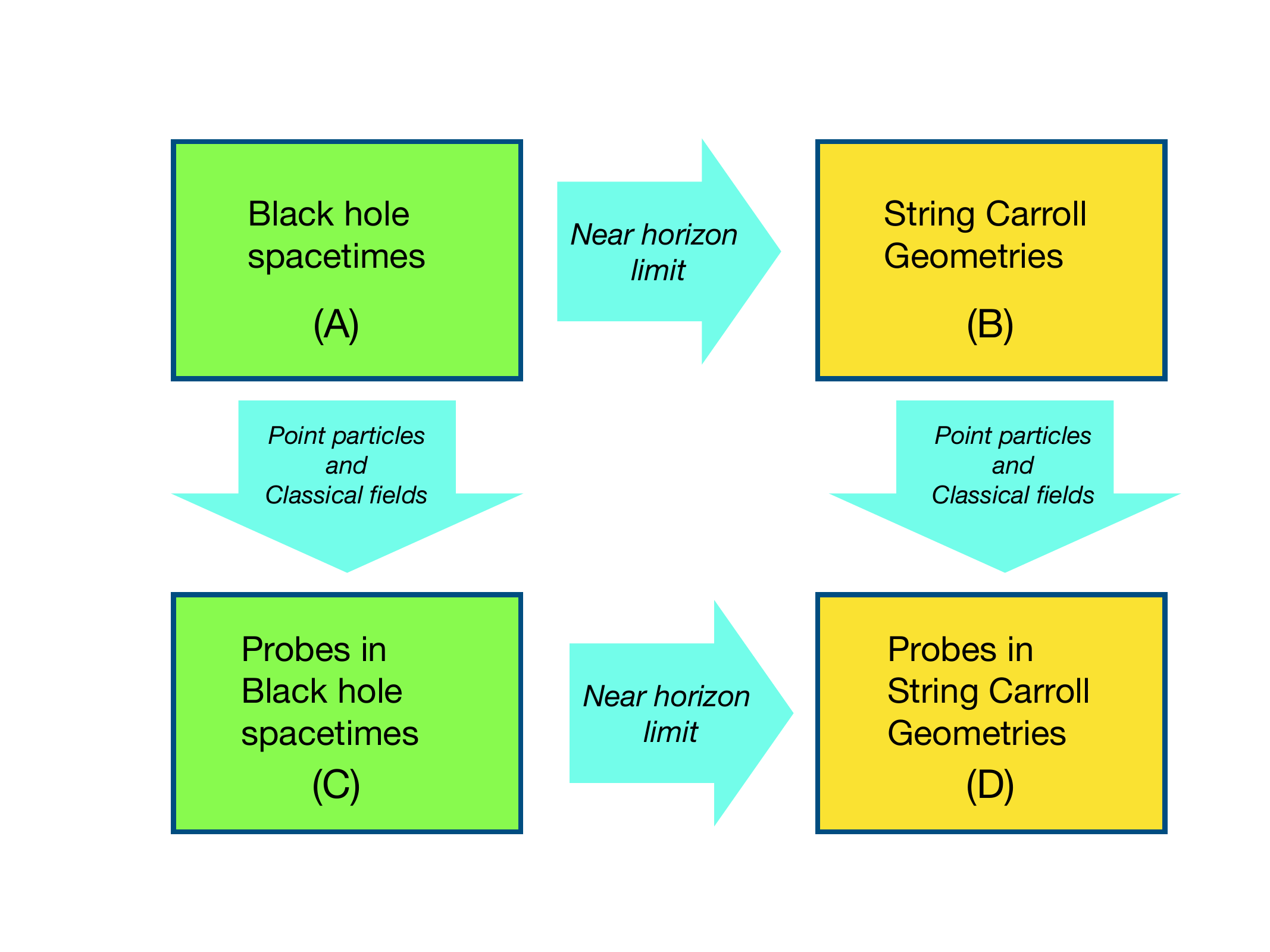}
\caption{Closing of the square -- consistency of our work is ensured by reproducing answers in two different ways of getting from Block A to D.}
\end{figure}
\end{center}
The paper is organised as follows:
\begin{itemize}
\item {\em {Sec 2: Route $A\to B$.}} -- In section \ref{sec-structure}, we demonstrate that the near-horizon geometry of a generic black object is String-Carroll (SC). We elaborate on the intrinsic structure of a SC geometry. 
\item {\em {Sec 3: Route $B\to D$.}} -- In section \ref{sec:Probes in String Carroll}, we discuss the classical behaviour of point-particles and minimally coupled scalar fields on a generic SC background. 
\item {\em {Sec 4 -- Sec 8: Route $A\to C \to D$.}} -- In \cref{sec: Checking our formulation: The algorithm,Black holes and string-Carroll 1: Schwarzschild Black hole,Black holes and string-Carroll 3: AdS Black Brane,Black holes and string-Carroll 4: Lifshitz Black hole} we examine our formulation of point particles and scalar fields on string-Carroll backgrounds across several examples. Sec~\ref{sec: Checking our formulation: The algorithm} gives the algorithm of the procedure we will follow in our examples. For readers who may not want to wade through all the equations that follow, this is a useful section. Sec~\ref{Black holes and string-Carroll 1: Schwarzschild Black hole} details the Schwarzschild example, Sec~\ref{Black holes and string-Carroll 3: AdS Black Brane} considers the AdS black brane and Sec~\ref{Black holes and string-Carroll 4: Lifshitz Black hole} focuses on the Lifshitz black hole.    
\item {\em {Sec 8: }} Here we discuss reaching the horizon from the near-horizon regime. 

\item {\em {Sec 9: }} We conclude with a summary of our paper, a discussion of universal symmetry aspects of our analysis and a list of immediate open questions. 

\item{\em Appendices:} Our paper contains a number of appendices for non-experts and people interested in explicit computations. These can of course be omitted in a first reading. 
\item[] Appendix~\ref{Carroll-sym} gives a brief account of Carrollian symmetries for the Carroll non-experts. Appendix~\ref{app: string-Carroll expansion} is an extension of \cref{sec-structure}, where we provide further details regarding the transformation of the metric data for the case of SC geometries. Ap.~\ref{app:Particle Carroll Expansion} provides details on the particle-Carroll expansion. In Ap.~\ref{App:scalar field on Carroll background}, we give a short summary of scalar fields on the Carroll background following \cite{deBoer:2021jej,carrollstories}. Ap.~\ref{D} provides details of the rotating BTZ black hole. In Ap.~\ref{String Carroll Maps}, we provide the SC metric data for several other black holes. Finally, in Ap.~\ref{App:Determinant of the Metric}, we worked out the expansion of the metric determinant in detail.   
\end{itemize}

\section{Carroll near the horizon}\label{sec-structure}

We have earlier emphasised that Carrollian structures show up on generic null surfaces and hence also on the horizon of generic black holes. In this section, we first collect the geometric data on and near the horizon of non-extremal black holes and we will then see how these fit into the fibre bundle structures related to Carrollian symmetry. 

\subsection{Structures near the horizon}

The near-horizon geometry of a $d$-dimensional black hole, which is a solution of the vacuum field equations, can be expressed in terms of the null-Gaussian coordinates $(v,\rho,x^A)$. Here, $v$ is the advanced time coordinate and the horizon $\mathcal{H}$ is situated at $\rho=0$. The angular coordinates are denoted by $x^A$. The near-horizon metric in the null-Gaussian coordinates takes the form \cite{Moncrief:1983xua,Chrusciel:2020fql}
\begin{eqnarray}\label{eq:null gaussian metric}
    ds^2=-2\kappa\rho ~dv^2+2dv~d\rho+2\omega_A\rho~dv~dx^A+(\Omega_{AB}+\rho\lambda_{AB})~dx^Adx^B+\mathcal{O}(\rho^2)\,,
\end{eqnarray}
where $\kappa$, $\omega_A$, $\Omega_{AB}$ and $\lambda_{AB}$ are metric functions which, in general, can depend on $(v,x^A)$. The induced metric on $\mathcal{H}$ is obtained by setting $\rho=0$ in \eqref{eq:null gaussian metric}. It is degenerate and is given as 
\begin{eqnarray}\label{eq:degenerate metric}
    ds^2\Big|_{\mathcal{H}}=0~dv^2+0~dv~dx^A+\Omega_{AB}~dx^Adx^B\,.
\end{eqnarray}
The horizon is a null surface and is thus associated with Carrollian symmetries \cite{Penna:2018gfx, Donnay:2019jiz}. 

\medskip

We now look near and not on the horizon. We will achieve this by dialling $\rho$ to zero in lieu of setting it to 0. To accomplish this, we will attach a dimensionless parameter $\epsilon$ to the $\rho$ coordinate and arrange the metric in orders of $\epsilon$ and tend $\epsilon\to 0$ to get,
\begin{eqnarray}\label{eq:epsilon expansion of null gaussian metric}
    ds^2=\Omega_{AB}dx^Adx^B+\epsilon~(-2\kappa\rho ~dv^2+2dv~d\rho+2\theta_A\rho dvdx^A+\rho\lambda_{AB}~dx^Adx^B)+\mathcal{O}(\epsilon^2)\,.
\end{eqnarray}
We now make the following redefinitions which would be very important in what follows. We will rewrite the metric as
\begin{align}
ds^2 = h_{\mu\nu}dx^\mu dx^\nu + \epsilon\left( \tau_{\mu\nu}dx^\mu dx^\nu +  \Phi_{\mu\nu}dx^\mu dx^\nu \right) + \mathcal{O}(\epsilon^2)\,. 
\end{align}
where we have made the following identifications: 
\begin{subequations}\label{eq:null gaussian metric mapped with String-Carroll}
    \begin{align}
        h_{\mu\nu}dx^\mu dx^\nu&=\Omega_{AB}dx^Adx^B\,,\\
        \tau_{\mu\nu}dx^\mu dx^\nu&=-2\kappa\rho ~dv^2+2dv~d\rho\,,\\
        \Phi_{\mu\nu}dx^\mu dx^\nu&=2\theta_A\rho dvdx^A+\rho\lambda_{AB}~dx^Adx^B \,.        
    \end{align}
\end{subequations}
We will now do formal Carrollian expansions for generic metrics. 

\subsection{Carroll symmetries and multiple null directions}
We begin with the definition of a Carroll manifold. As mentioned in the introduction, starting with Minkowski spacetimes and send $c\to0$, the metric and its inverse degenerate in the limit and give us \eqref{carr-met}. It is clear from here that $h_{\mu\nu} \theta^\nu = 0$. A flat Carroll manifold is thus defined with both $(h_{\mu\nu}, \theta^\nu)$. We now define a general Carroll structure based on these ideas \cite{Henneaux:1979vn, Duval:2014uoa}. 

\medskip

\paragraph{Carroll manifolds.} A $d$-dimensional Carrollian manifold is a smooth manifold with a degenerate metric $h_{\mu\nu}$ and a nowhere-vanishing vector field $\theta^\mu$ that generates the kernel of the metric \cite{Duval:2014uoa, Duval:2014lpa}, i.e.
\begin{align}\label{htht}
h_{\mu\nu} \theta^\nu = 0.
\end{align}
For flat Carroll manifolds, the isometries of this doublet 
\begin{align}\label{Carr-Kill}
\pounds_\xi h_{\mu\nu} = 0, \quad \pounds_\xi \theta^\nu = 0,
\end{align}
generates the Carroll algebra:
\begin{align}\label{xi}
\xi^i = \omega^i_{\, j} x^j + a^i, \quad \xi^0 = f(x) + a^0. 
\end{align}
Notice that this is generically infinite dimensional for all spacetime dimensions. We recover the finite Carrollian algebra when we restrict the function $f(x)$ to be linear in $x$. For the reader uninitiated to Carrollian symmetries, we have collected the relevant information in Appendix~\ref{Carroll-sym}.  

\medskip

\paragraph{String-Carroll manifolds.} A natural geometric question is what happens when we have multiple null directions. The geometry is obvious: for a $(n+m)$ dimensional manifold with $n$-nonnull and $m$-null directions, we will have a fibre bundle with a $n$-dimensional base made of the non-null directions and $m$-dimensional fibres made of the null directions. In our application to the theory of black holes, as we have stated in the introduction, we will be interested in two null directions and hence will focus on this\footnote{For the general formulation for arbitrary null directions, see \cite{Bergshoeff:2026cxt}.}. We will call the null directions longitudinal and the non-null directions transverse. 

\medskip

We first define string-Carroll manifolds generalising our discussion of Carroll manifolds above. A string-Carroll structure is defined as the triplet $(\mathcal{M},h_{\mu\nu},v^{\mu\nu})$, where  $\mathcal{M}$ is a smooth $d$-dimensional manifold, $h_{\mu\nu}$ is a symmetric, degenerate tensor that defines the metric on the transverse space, while $v^{\mu\nu}$ is a symmetric, degenerate tensor that defines the metric on the longitudinal space. $v$ generates the kernel of $h$: 
\begin{eqnarray}
    h_{\mu\nu}v^{\mu\rho}=0\,.
\end{eqnarray}
Here, In the above expansion, $h_{\mu\nu}$ has the signature $(0,0,+,\cdots,+)$, while the signature of $\tau_{\mu\nu}$ and $v^{\mu\nu}$ is $(-,+,0\cdots,0)$. 
The generalization to arbitrary number of null directions is obvious. 

\medskip

\paragraph{Carroll expansions.} We will now briefly describe the string-Carroll formalism to methodically expand a $(d+2)$ - dimensional Lorentzian metric. Some other details of the transformations are collected in Appendix \ref{app: string-Carroll expansion} and the generalisations thereof can be found in \cite{Bagchi:2024rje}. Any Lorentzian metric can be decomposed in terms of the vielbeine as follows 
\begin{eqnarray}
g_{\mu\nu}=\eta_{A'B'}\mathcal{E}^{A'}_\mu\mathcal{E}^{B'}_\nu\,,\quad g^{\mu\nu}=\eta^{A'B'}\mathcal{E}_{A'}^\mu\mathcal{E}_{B'}^\nu\,,
\end{eqnarray}
where $A',B'\in \{0,\cdots,d+1\}$ are the tangent space indices of the spacetime manifold and $\mu,\nu\in\{0,\cdots,d+1\}$  are the spacetime indices. The vielbeine, $\mathcal{E}^{A'}_\mu$ and $\mathcal{E}_{A'}^\mu$ can be split into longitudinal and transverse sectors in the following way 
\begin{equation}
    \mathcal{E}^{A'}_\mu=\left(c \,T_\mu^{A},E_\mu^{\bar A}\right)\,,\quad \mathcal{E}_{A'}^\mu=\left(c^{-1}V^\mu_{A},E^\mu_{\bar{A}}\right)\,,
\end{equation}
where $A\in \{0,1\}$ are the tangent space indices of the longitudinal submanifold and $\bar{A}\in\{2,\cdots,d+1\}$  are the tangent space indices of the transverse submanifold. The longitudinal submanifold admits a tangent space metric $\eta_{AB}$ while the transverse submanifold admits a tangent space metric $\delta_{\bar{A}\bar{B}}$. The decomposed vielbeine satisfy the following relations
\begin{eqnarray}
    V^\mu_{A} E_\mu^{\bar{A}}=T_\mu^AE^\mu_{\bar{A}}=0\,,~~T^A_\mu V^\mu_B=\delta^A_B\,,~~E^\mu_{\bar{A}} E^{\bar{B}}_\mu=\delta^{\bar{B}}_{\bar{A}}\,,~~\delta^\mu_\nu=T^A_\nu V^\mu_A+E^\mu_{\bar{A}}E^{\bar{A}}_\nu\,.
\end{eqnarray}
The Lorentzian metric in terms of the decomposed vielbeine takes the form 
\begin{eqnarray}\label{eq:SC metric decomposition}
g_{\mu\nu}=c^2\,\eta_{AB}T^{A}_\mu T^{B}_\nu+\delta_{\bar{A}\bar{B}}E^{\bar{A}}_\mu E^{\bar{B}}_\nu\,,\quad g^{\mu\nu}=c^{-2}\,\eta^{AB}V_{A}^\mu V_{B}^\nu+\delta^{\bar{A}\bar{B}}E_{\bar{A}}^\mu E_{\bar{B}}^\nu\,.
\end{eqnarray}
The expansion of the Lorentzian metric is realised when the vielbeine are expanded in powers of $c^2$ as follows 
\begin{eqnarray}\label{eq:precarrollian variables expansion stringCarroll geometry}
\begin{aligned}
&V^\mu_A=v^\mu_A+c^2m^\mu_A+\mathcal{O}(c^4)\,,\quad T_\mu^A=\tau_\mu^A+c^2m_\mu^A+\mathcal{O}(c^4)\\
&E^\mu_{\bar{A}}=e^\mu_{\bar{A}}+c^2\pi^\mu_{\bar{A}}+\mathcal{O}(c^4)\,,\quad E_\mu^{\bar{A}}=e_\mu^{\bar{A}}+c^2\pi_\mu^{\bar{A}}+\mathcal{O}(c^4)\,.
\end{aligned}
\end{eqnarray}
The metric expansion in terms of the expanded vielbeine, takes the form
\begin{eqnarray}\label{SCmetric}
g_{\m\n}=h_{\m\n}+c^2\tau_{\m\n}+c^2\Phi_{\m\n}+\mathcal{O}(c^4)\,,~~~g^{\m\n}=\frac{1}{c^2}v^{\m\n}+\bar{h}^{\m\n}+\mathcal{O}(c^2)\,.
\end{eqnarray}
Here, the various terms occurring in the SC expansion are
\begin{subequations}\label{data}
    \begin{align}
    &h_{\m\n}=\delta_{\bar{A}\bar{B}}e^{\bar{A}}_{\m}e^{\bar{B}}_{\n}\,,~~\tau_{\mu\nu}=\eta_{AB}\tau_\m^A\tau_\n^B\,,~~\Phi_{\m\n}=2\delta_{\bar{A}\bar{B}}e^{\bar{A}}_{(\mu}\,\pi^{\bar{B}}_{\nu)}\\
    &v^{\m\n}=\eta^{AB}v^\mu_Av^\nu_B\,,~~\bar{h}^{\m\n}=\delta^{\bar{A}\bar{B}}e^\mu_{\bar{A}}e^\nu_{\bar{B}}+2\eta^{AB}v^{(\mu}_Am^{\nu)}_B\,.
    \end{align}
\end{subequations}
The orthogonality relations of the decomposed vielbeine imply
\begin{eqnarray}
    v^\mu_Ae_\mu^{\bar{A}}=\tau_\mu^Ae^\mu_{\bar{A}}=0\,,\quad \tau_\mu^Av^\mu_B=\delta^A_B\,.
\end{eqnarray}

\paragraph{Identifying the expansions.}

It is clear from the above that the near horizon expansion that we had performed in the previous subsection is naturally identified with the generic Carroll expansions we have performed here by identifying 
\begin{align}
\epsilon = c^2, 
\end{align}
i.e. the distance to the horizon can be treated as an effective speed of light \cite{Donnay:2019jiz}. In particular for the near horizon expansion \eqref{eq:epsilon expansion of null gaussian metric} is to be identified with \eqref{SCmetric} and \eqref{data}.

\bigskip \bigskip

\section{Probes in String-Carroll}\label{sec:Probes in String Carroll}
In the previous section, the near-horizon geometry of a black hole was shown to be string-Carroll. Now we are interested in studying the effects of the SC geometry, when various probes are analysed on the SC background. These probes serve as useful tools for extracting features of theories near black hole backgrounds. We will study point-particle geodesics and scalar fields on the SC background. Throughout the analysis, we will assume the background to be fixed. 

\medskip

An important underlying assumption in what follows is that all fields can be Taylor expanded in terms of the speed of light (which is also near horizon parameter in the black hole context). While this assumption is something that should be scrutinized in more detail (e.g. under what circumstances terms like $e^{-1/c^2}$ or $c^2\ln{c}$ can be neglected) , this also has been the working assumption in all works considering Carrollian expansions in the literature and we will continue to follow this in our discussions below. 

\subsection{Point particles}\label{sec:Point particles}
The point particle action in general relativity can be expressed in a simple form as follows 
\begin{equation}\label{quadratic particle action}
    S = -\frac{m}{2}\int g_{\m\n}(X)\,\dot{X}^\mu\dot{X}^\nu \, d\t\, , 
\end{equation}
where $X^\mu$ is the spacetime coordinate and $\dot{X}^\mu = \frac{dX^\m}{d\t}$. This action gives rise to the following geodesic equation
\begin{eqnarray}
\ddot{X}^\mu+\Gamma^\mu_{\nu\rho}\dot{X}^\nu\dot{X}^\rho=0\,.
\end{eqnarray}
In what follows, we will expand the quadratic action in the SC background and compare the leading-order and subleading-order results.

\medskip

Consider the quadratic particle action on a Lorentzian manifold $(M,g)$ that admits a SC expansion  \cite{Bagchi:2023cfp,Bagchi:2024rje}. The worldline embedding coordinates appearing in the action \eqref{quadratic particle action} are expanded as follows 
\begin{eqnarray}\label{eq:coordinate expansion for QPA}
    X^\mu(\tau)=x^\mu(\tau)+c^2y^{\mu}(\tau)+\mathcal{O}(c^4)\,,
\end{eqnarray}
where $\tau$ is a parameter along the worldline of the particle, $x^\mu$ is the leading order coordinate and $y^\mu$ is the next-to-leading order coordinate. The SC expansion of the metric and its derivative take the form
\begin{subequations}\label{eq:metric expansion for QPA}
\begin{align}
    g_{\mu\nu}(X) &= h_{\mu\nu}(x) + c^2 \left( \t_{\mu\nu}(x) + \Phi_{\mu\nu}(x) + y^\rho\partial_\rho h_{\mu\nu}(x) \right) + \mathcal{O}(c^4),\\
    \partial_{\rho}g_{\mu\nu}(X) &=\partial_{\rho}h_{\mu\nu}(x)+c^2\partial^{(x)}_{\rho}\Lambda_{\mu\nu}(x,y),
    \end{align}
\end{subequations}
where, $\partial^{(x)}_{\rho}$ indicates a derivative with respect to the $x^\rho$ coordinate and
\begin{eqnarray}
\Lambda_{\mu\nu}(x,y)= \t_{\mu\nu}(x) + \Phi_{\mu\nu}(x) + y^\rho\partial_\rho h_{\mu\nu}(x)\,.    
\end{eqnarray}
Substituting the expansion for the metric \eqref{eq:metric expansion for QPA} and the coordinates \eqref{eq:coordinate expansion for QPA}, the action \eqref{quadratic particle action} expands as
\begin{equation}
    S = S_{LO} + c^2 S_{NLO} + \mathcal{O}(c^4),
\end{equation}
where,
\begin{subequations}\label{eq:string-Carroll point particle action}
    \begin{align}
        S_{LO} &= -\frac{m}{2}\int h_{\m\n}(x) \dot{x}^\m\dot{x}^\n \, d\t\,, \label{lo quadratic action}\\
        S_{NLO} & = -\frac{m}{2}\int \left[ \left(  \tilde{\Phi}_{\m\n}(x) + y^\rho\partial_\rho h_{\m\n}(x) \right)\dot{x}^\m\dot{x}^\n + 2h_{\m\n}(x) \dot{x}^{(\m} \dot{y}^{\n)}  \right]\, d\t. \label{nlo quadratic action}
    \end{align}
\end{subequations}
Here we defined 
\begin{equation}
    \tilde{\Phi}_{\m\n}(x) := \t_{\m\n}(x) + \Phi_{\m\n}(x).
\end{equation}
Our next job is to solve the leading order (LO) and next-to-leading-order (NLO) theories and obtain the equations of motion.

\medskip

\underline{\em{Leading Order:}} The following geodesic equation is obtained from \eqref{lo quadratic action} 
\begin{equation}\label{lo quadratic eom}
    h_{\m\rho}(x) \ddot{x}^\rho + \Gamma^{(h)}_{\m\n\rho}(x)\dot{x}^\n \dot{x}^\rho = 0\,.
\end{equation}

Here we define the Christoffel symbol of 1st kind from the degenerate metric $h_{\m\n}$ as{\footnote{In all our explorations, we will use the Christoffel symbols of the first kind, which will ensure that we do not end up mixing the metric and its inverse because they degenerate at the leading order when expanded in the vicinity of the event horizon.}}
\begin{equation}
    \Gamma^{(h)}_{\m\n\rho }(x):=\frac{1}{2} \left( h_{\rho\m,\n}(x) + h_{\m\n,\rho}(x) - h_{\n\rho,\m}(x) \right).
\end{equation}
So, in our LO analysis, we notice the emergence of an equivalent form of the geodesic equation.

\medskip

\underline{\em{Sub-leading Order:}} In the NLO action \eqref{nlo quadratic action}, we have two coordinates $x^\m(\t)$ and $y^\m(\t)$. 
Varying $x^\m$ we get,
    \begin{align}\label{eq: PP NLO EOM1}
    \Lambda_{\m\rho}(x,y) \ddot x^\rho + 2h_{\m\rho,\n}(x)\dot x^{(\n}\dot y^{\rho)} 
    + &h_{\m\rho}(x)\ddot y^\rho+ \partial_\n^{(x)}\Lambda_{\m\rho}(x,y) \dot x^\n\dot x^\rho \nn \\&= h_{\n\rho,\m}(x)\dot x^{(\n}\dot y^{\rho)} + \frac{1}{2}\partial_\m^{(x)}\Lambda_{\n\rho}(x,y) \dot x^\n\dot x^\rho \,,
\end{align}
where 
\begin{equation}
\Lambda_{\m\n}(x,y) = \tilde{\Phi}_{\m\n}(x) + y^\rho\partial_\rho h_{\m\n}(x).
\end{equation}
Thus, the NLO equation of motion \eqref{eq: PP NLO EOM1} becomes
\begin{equation}\label{nlo quadratic eom}
    \Lambda_{\m\n}(x,y)\ddot x^\n + \Gamma^{(\Lambda)}_{\m\n\rho }(x,y)\dot x^\n \dot x^\rho + h_{\m\n}(x)\, \ddot y^\n + 2 \Gamma^{(h)}_{\m\n\rho}(x)\, \dot x^{\n}\, \dot y^{\rho} = 0\,.
\end{equation}
In the above equation, $\Gamma^{(\Lambda)}_{\rho\mu\nu}$ is the Christoffel symbol of the first kind defined as
\begin{equation}
    \Gamma^{(\Lambda)}_{\m\n\rho }(x,y):=\frac{1}{2} \left( \Lambda_{\rho\m,\n}(x,y) + \Lambda_{\m\n,\rho}(x,y) - \Lambda_{\n\rho,\m}(x,y) \right)
\end{equation}
and the derivatives are with respect to the $x$ coordinate. On varying the NLO quadratic action with respect to $y^\m$, we get
\begin{align}
  h_{\m\rho}(x)\, \ddot{x}^\rho + \Gamma^{(h)}_{\m\n\rho}(x)\,\dot{x}^\n \dot{x}^\rho = 0\,,
\end{align}

which is again the LO equation of motion \eqref{lo quadratic eom}.
We thus derived the geodesic equations from the quadratic action for a point particle in a SC expanded metric.

\subsection{Scalar field}\label{sec:Scalar Field}
We will now consider classical field theories and analyse a minimally coupled free scalar field on a general SC background. The action of a minimally coupled free scalar field on a generic curved background is given by
\begin{equation}\label{eq:scalar field action}
		S =  -\frac{1}{2}\int d^4 x\sqrt{-g}\left(g^{\mu\nu}\partial_\mu \Phi\partial_\nu\Phi + m^2\Phi^2\right)\,.
\end{equation} 
Here, $\Phi$ is the scalar field, $\mu, \nu$ are spacetime indices in four dimensions and the metric $g_{\m\n}$ carries a mostly plus signature.      

\subsubsection*{Near-horizon expansion of the action}

Following the works of \cite{Bagchi:2023cfp,Bagchi:2024rje}, we expand \eqref{eq:scalar field action} in analytical powers of $c^2$ for a general Lorentzian metric which has a small $c^2$ expansion of the form \eqref{SCmetric}. We postulate that $\Phi$ has the following analytical expansion in powers of $c^2$
		\begin{eqnarray}
		\Phi = \phi^{(0)} + c^2 \phi^{(1)} + c^4 \phi^{(2)}+\mathcal{O}(c^6)\,.
        \end{eqnarray}
		 The scalar field Lagrangian in \eqref{eq:scalar field action} arranges in the following $c^2$ expansion
	
	\begin{equation*}
	    \mathcal{L} = \frac{1}{c^2}\mathcal{L}_{LO} + \mathcal{L}_{NLO} + c^2 \mathcal{L}_{NNLO} + \mathcal{O}(c^4).
	\end{equation*} 
	where the leading order $\mathcal{L}_{LO}$, next-to-leading order $\mathcal{L}_{NLO}$ and next-to-next-leading order $\mathcal{L}_{NNLO}$ Lagrangian are
    \begin{subequations}
        \begin{align}
           \mathcal{L}_{LO}&= \frac{1}{2}e\:v^{\mu\nu}\partial_\mu \phi^{(0)} \partial_\nu \phi^{(0)} \label{leading}\,,\\
            \mathcal{L}_{NLO}&= \frac{1}{2}e\:\Big(2v^{\mu\nu}\partial_{\mu} \phi^{(0)} \partial_{\nu} \phi^{(1)}  +\bar{h}^{\mu\nu}\partial_\mu \phi^{(0)}\partial_\nu \phi^{(0)}+m^2\left(\phi^{(0)}\right)^2\nonumber\\&~~~~~+\frac{1}{2}\left(v^{\rho\sigma}+h^{\rho\sigma}\right)\left(m_{\rho\sigma}+\Phi_{\rho\sigma}\right)v^{\mu\nu}\partial_\mu\phi^{(0)}\partial_\nu\phi^{(0)} \Big)\label{nextleading}\,.
    \end{align}
    \end{subequations}
Here $e$ is the leading-order determinant of the expanded metric (see \cref{App:Determinant of the Metric}) and $m_{\rho\sigma}=\eta_{AB}\tau^A_{(\rho}m_{\sigma)}^B$ is the sub-leading order metric contributing at $\mathcal{O}(c^4)$. The SC expansion of the inverse metric \eqref{SCmetric} focuses along the longitudinal directions at the leading order. Therefore, the leading order Lagrangian \eqref{leading} only contains the kinetic term along the longitudinal directions. To give the reader some context on the difference between ``vanilla''-Carroll expansions and SC expansions from the point of view of classical field theories, we invite the reader to compare the leading and next-to-leading order scalar field actions on the SC background to their Carrollian counterparts, which we have discussed in appendix \ref{App:scalar field on Carroll background}.

\quad

\underline{{\em{Leading Order Action:}}} The leading-order action is 
\begin{equation}\label{eq:LO string-Carroll action}
S_{LO}=-\frac{1}{2}\int d^4x\:e\:v_A^\mu v_B^\nu \eta^{AB}\partial_\mu \phi^{(0)} \partial_\nu \phi^{(0)}.
\end{equation}
The equation of motion of the leading-order action is given as 
\begin{equation}
\partial_\mu(e\,v^{\mu\nu}\partial_\nu\phi^{(0)})=0
\end{equation}
Using the expansion of the diffeomorphism generating vector field \eqref{eq:expansions of the transformations}, it can be shown that the leading-order action is invariant under arbitrary leading-order diffeomorphisms.

\quad

\underline{\em{Next to Leading Order Action:}} The NLO action is given as 
\begin{multline}\label{eq:NLO string-Carroll action}
S_{NLO}=-\frac{1}{2}\int d^4x\:e\:\Big(2v^{\mu\nu}\partial_{\mu} \phi^{(0)} \partial_{\nu} \phi^{(1)}  +\bar{h}^{\mu\nu}\partial_\mu \phi^{(0)}\partial_\nu \phi^{(0)}+m^2\left(\phi^{(0)}\right)^2\\+\frac{1}{2}\left(v^{\rho\sigma}+h^{\rho\sigma}\right)\left(m_{\rho\sigma}+\Phi_{\rho\sigma}\right)v^{\mu\nu}\partial_\mu\phi^{(0)}\partial_\nu\phi^{(0)} \Big).
\end{multline}
The equation of motion of the next-to-leading order action is given as 
\begin{subequations}
\begin{align}
\phi^{(1)}:&\;\partial_\mu(e\,v^{\mu\nu}\partial_\nu\phi^{(0)})=0\,,\\
\phi^{(0)}:&\;\frac{1}{e}\partial_\mu\Big(\frac{e}{2}\left(v^{\rho\sigma}+h^{\rho\sigma}\right)\left(m_{\rho\sigma}+\Phi_{\rho\sigma}\right)v^{\mu\nu}\partial_\nu\phi^{(0)}\nn\\&\quad\quad\quad\quad\quad\quad\quad\quad\quad+ev^{\mu\nu}\partial_{\nu} \phi^{(1)}  +e\bar{h}^{\mu\nu}\partial_\nu \phi^{(0)}\Big)-m^2\phi^{(0)}=0\,.
\end{align}
\end{subequations}

\quad

\underline{\em{Diffeomorphisms, SC Boosts:}} Up to this point, we have started from a minimally coupled scalar field action on a curved background and systematically expanded each constituent in powers of $c^2$. The original action \eqref{eq:scalar field action} is invariant under diffeomorphisms and local Lorentz transformations. The scalar field $\Phi$, is inert under local Lorentz transformations but transforms under diffeomorphisms generated by $\zeta$ as $\delta_\zeta\Phi=\zeta^\mu\partial_\mu\Phi$, while the vielbeine $\mathcal{E}^{A'}_\mu$, transform under both diffeomorphisms and local Lorentz transformations.

\medskip

After expanding the action in powers of $c^2$, it is not immediately obvious that the resulting LO and NLO actions remain invariant under diffeomorphisms. Moreover, expanding the background alters the local structure of the spacetime (see appendix \ref{app: string-Carroll expansion}). Therefore, one must ensure that the LO and NLO actions are invariant under the new local symmetries. Among the new local transformations are Carroll boosts ($\lambda^A_{(0)\bar{B}}$), SC boosts ($L_{(0)}\epsilon^A_{~~B}$) and their subleading contributions. Finally, expansion of the scalar field does not necessarily imply that the subleading fields $\phi^{(i>0)}$ continue to transform as scalars under diffeomorphisms.  

\medskip

To check for diffeomorphism invariance of the action, we will expand the diffeomorphism generating vector $\zeta$ as done in \cref{app: string-Carroll expansion}. The action of diffeomorphisms on the metric fields are given in \eqref{eq:transformations}. While the scalar fields transform under diffeomorphisms as follows
\begin{equation}\label{eq:Diffs}
\delta_{\zeta}\phi^{(0)}=\zeta_{(0)}^\mu\partial_{\mu}\phi^{(0)}\,,~~\delta_{\zeta}\phi^{(1)}=\zeta_{(0)}^\mu\partial_{\mu}\phi^{(1)}+\zeta_{(2)}^\mu\partial_{\mu}\phi^{(0)}\,.
\end{equation}
It can be shown that the LO and NLO actions \cref{eq:LO string-Carroll action,eq:NLO string-Carroll action} are invariant under the diffeomorphisms generated by the $\zeta_{(0)}$ and the $\zeta_{(2)}$ vectors. However, it should be noted that the field $\phi^{(1)}$ is not a scalar under diffeomorphisms generated by the $\zeta_{(2)}$ vector. The LO and NLO SC boosts and Carroll boosts act on the metric fields as shown in \eqref{eq:transformations}. It can be shown that the LO and NLO actions \cref{eq:LO string-Carroll action,eq:NLO string-Carroll action} are invariant under these transformations.

\section{Checking our formulation: The algorithm}\label{sec: Checking our formulation: The algorithm}

Our explorations have so far focused on generalising Carrollian structures to more general null structures to multiple null directions. We showed in \cref{sec-structure} that Carroll expansions with two null directions were directly related to the near horizon of generic non-extremal black holes. We have then focussed on understanding particle trajectories and scalar fields in these SC geometries. 

\medskip

We now move on from this abstract formulation to definite examples to show the validity of our analysis. In the main body of our paper we will address the following cases: Schwarzschild black holes in asymptotically flat 4d spacetimes, AdS black branes and Lifshitz black holes. Appendix \ref{D} gives details of the rotating BTZ black holes while we list the SC data of some other black holes including the very general Plebanski-Demianski class in Appendix \ref{String Carroll Maps}. 

\medskip

For each example in the coming few sections (Sec~\ref{Black holes and string-Carroll 1: Schwarzschild Black hole} -- \ref{Black holes and string-Carroll 4: Lifshitz Black hole}), we will discuss the explicit map of the near horizon geometry to the SC data. We will then consider point-particle geodesics and free scalar field theory in the near-horizon background. 

\begin{itemize}

\item {\em The map from NH to SC:} For each case, we will specify the String Carroll data $\{h_{\mu\nu}, \Phi_{\mu\nu}, \tau_{\mu\nu}, v^{\mu\nu}, {\bar{h}}^{\mu\nu} \}$. This will construct the explicit map between a particular black hole or black brane and a particular SC geometry. 

\item {\em Geodesics:} Given a particular SC map, we will use the analysis of LO and NLO geodesics in Sec.~\ref{sec:Probes in String Carroll} to write out equations explicitly. This is the analysis that we call ``intrinsic''. 
We will then consider the {\em entire} black hole geometry and write down geodesics there. By considering the near-horizon limit of these geodesics of the full geometry (we call this the ``limiting analysis''), we will reproduce the answers obtained from the SC analysis. Near horizon limits are singular limits and it is far from obvious that geodesics in the limiting geometry given by the SC data would reproduce the NH limit of geodesics of the full geometry. Indeed, we will see that although the leading order matching is straight forward, the NLO matching requires us to use the LO data. 

\item{\em Field theory:} We will follow the same procedure for the free scalar field theory. Again the leading order matching is straightforward, and the NLO is not. At NLO, the sub-leading equations of motion match by adding the leading-order equation of motion to the limiting analysis. 

\end{itemize}

We now dive headlong into examples. 

\section{Schwarzschild Black hole}\label{Black holes and string-Carroll 1: Schwarzschild Black hole}
We begin with the simplest example, our favourite Schwarzschild black hole in four dimensional asymptotically flat spacetimes. The metric in the Schwarzschild coordinates is given as
	\begin{equation}\label{eq: Schwarzschild Metric}
		ds^2 = -\left( 1-\frac{r_h}{r}\right)d  t^2  + \left( 1-\frac{r_h}{r}\right)^{-1}d  r^2 + r^2d\Omega^2\,,
	\end{equation}
where the Schwarzschild radius $r_h = 2GM$ is given in terms of Newton's constant $G$ and the mass of the black hole $M$. 

\subsection{Explicit map}\label{sec:Explicit map (Schwarzschild Black hole)}
The near-horizon of a Schwarzschild black hole can be reached via a change of the radial coordinate
    \begin{equation}
         r = {r_h} + \frac{1}{4r_h}\epsilon\rho^2,
    \end{equation}
    where $\epsilon$ is a dimensionless constant. The near-horizon limit amounts to rewriting the metric \eqref{eq: Schwarzschild Metric} in the new radial coordinate and dialling $\epsilon$ to zero. The Schwarzschild metric arranges in orders of $\epsilon$ as follows 	
    \begin{equation}\label{eq:schwarzschild-exp}
	ds^2=r_h^2d\Omega^2+\epsilon\left(-\frac{\rho^2}{4r_h^2}dt^2+d\rho^2+\frac{\rho^2}{2}d\Omega^2\right)+\mathcal{O}(\epsilon^2)\,.
	\end{equation}
    Comparing \eqref{eq:schwarzschild-exp} with the SC expansion \eqref{SCmetric} and setting $\epsilon=c^2$, we can deduce the following map between the near-horizon expansion of the Schwarzschild metric and the SC expansion
	\begin{equation}\label{eq:Schwarzschild Map}
			h_{\mu\nu}dx^\mu dx^\nu = r_h^2d\Omega^2\,,~~
            \Phi_{\mu\nu} dx^\mu dx^\nu = \frac{\rho^2}{2}d\Omega^2,~~\tau_{\mu\nu}dx^\mu dx^\nu=-\frac{\rho^2}{4r_h^2}dt^2 + d\rho^2 \,.
	\end{equation}
    The longitudinal space is a Rindler space of $(1+1)$ dimensions, whereas the leading order transverse space is a sphere ($S^2$) of radius $r_h$. Similarly, we can also map the inverse Schwarzschild metric with the inverse SC metric
    \begin{equation}\label{eq:Schwarzschild Map inverse}
    v^{\mu\nu}\partial_\mu\partial_\nu=-\frac{4r_h^2}{\rho^2}\partial_t^2+\partial_\rho^2\,,~~~\bar{h}^{\mu\nu}\partial_\mu\partial_\nu=r_h^{-2}\partial^2_{\Omega}-\partial_t^2-\frac{\rho^2}{4r_h^2}\partial_\rho^2.
    \end{equation}
The inverse map will be of importance when we study scalar fields in the vicinity of the Schwarzschild black hole.


\subsection{Geodesics using the map}\label{sec:Geodesics using the map (Schwarzschild Black hole)}
In this section, we study the geodesics of point particles near the event horizon of a Schwarzschild black hole using the SC formalism devised in section \ref{sec:Point particles}.
The leading (LO) and next-to-leading (NLO) point particle actions in the vicinity of the Schwarzschild black hole are found to be
\begin{subequations}\label{eq:point-particle actions in the Schwarzschild background}
\begin{align}
    S_{LO} &= -\frac{mr_h^2}{2} \int \left( \left( \dot x^\theta \right)^2 + \sin^2x^\theta \left( \dot x^\phi \right)^2 \right)\, d\lambda\,, \label{eq:LO point-particle action in the Schwarzschild background}\\
    S_{NLO}&= -\frac{m}{2} \int \bigg( -\left( \frac{ x^\rho}{2 r_h} \right)^2 \left( \dot x^t \right)^2 + \left( \dot x^\rho \right)^2 + \frac{ \left( x^\rho \right) ^2 }{2} \left( \left( \dot x^\theta \right)^2 + \sin^2 x^\theta \left( \dot x^\phi \right)^2 \right)\nonumber\\
    &\quad + r_h^2 y^\theta \sin\left( 2 x^\theta \right) \left( \dot x^\phi \right)^2 + 2r_h^2 \dot x^\theta\dot y^\theta + 2r_h^2\sin^2 \left(  x^\theta \right)\dot x^\phi\dot y^\phi \bigg) \, d\lambda \label{eq:NLO point-particle action in the Schwarzschild background}\,,
\end{align}
\end{subequations}
where $\lambda$ is an affine parameter and $\dot x^\mu$ denotes the derivative with respect to $\lambda$. Varying $x^\theta$ and $x^\phi$ in \eqref{eq:LO point-particle action in the Schwarzschild background} gives the geodesic equations
\begin{equation}\label{eq:LO point-particle geodesics near Schwarzschild BH}
    \ddot x^\theta - \sin x^\theta\cos x^\theta\left( \dot x^\phi \right)^2 = 0\, ,\quad
    \dot x^\phi \,\sin^2 x^\theta  = \ell_\phi^{(0)}\,, 
\end{equation}
where $\ell_\phi^{(0)}$ is an integration constant. Observe that \eqref{eq:LO point-particle geodesics near Schwarzschild BH} are the usual geodesics on a spatial sphere, which describe great circles on the surface of a sphere. We get spatial geodesics on the sphere, since the leading-order near-horizon Schwarzschild metric \eqref{eq:schwarzschild-exp} is only sensitive to the sphere. We cannot comment on the dynamics of the point particle, as seen by an asymptotic observer, from these geodesics. Therefore, we need to analyse the next-to-leading order point particle action.\\

\noindent Varying $y^\theta$ and $y^\phi$ in the next-to leading order action \eqref{eq:NLO point-particle action in the Schwarzschild background} gives the leading order geodesic equation \eqref{eq:LO point-particle geodesics near Schwarzschild BH}, respectively. The geodesic equations for $x^\theta$, $x^\phi$, $x^t$ and $x^\rho$ are
\begin{subequations}\label{eq:NLO point-particle geodesics near Schwarzschild BH}
    \begin{align}
        \frac{d}{d\lambda}\left[ \lb x^\rho\rb^2\dot x^\theta + 2r_h^2\dot y^\theta \right]-  \frac{ \lb x^\rho\rb^2 }{2} \sin\left( 2 x^\theta \right)\left( \dot x^\phi \right)^2 - 2r_h^2 \sin\left(2 x^\theta\right) \dot x^\phi\dot y^\phi  &\nonumber\\
        - 2r_h^2  y^\theta \cos \left( 2 x^\theta \right) \left( \dot x^\phi \right)^2  &= 0 \, , \label{eq:NLO point-particle geodesic 1 near Schwarzschild BH}\\
        \lb\xr\rb^2\sin^2 x^\theta \dot x^\phi + 2r_h^2 y^\theta\sin \left( 2  x^\theta \right) \dot x^\phi + 2 r_h^2\sin^2 \left(  x^\theta \right) \dot y^\phi  &=\ell^{(1)}_\phi \,,\label{eq:NLO point-particle geodesic 2 near Schwarzschild BH}\\
        \lb x^\rho\rb^2 \dot x^t & = E_{(1)}\,,\label{eq:NLO point-particle geodesic 3 near Schwarzschild BH}\\
        \frac{d}{d\lambda}\left( 2 \dot\xr  \right) + \frac{ \xr }{2r_h^2}\left( \dot x^t  \right)^2 - \xr\left( \left( \dot x^\theta \right)^2 + \sin^2 x^\theta \left( \dot x^\phi \right)^2 \right) &= 0\,,\label{eq:NLO point-particle geodesic 4 near Schwarzschild BH}
    \end{align}
\end{subequations}
where $\ell_\phi^{(1)}$ and $E_{(1)}$ are integration constants. To comment on the observations made by an asymptotic observer, we need to evaluate the derivatives of the coordinates with respect to the coordinate time. Using \eqref{eq:LO point-particle geodesics near Schwarzschild BH} and \eqref{eq:NLO point-particle geodesic 3 near Schwarzschild BH}, the following relations can be inferred
\begin{equation}\label{eq:Freezing of point particles on the sphere}
    \frac{dx^\phi}{dx^t}=\frac{\left(x^\rho\right)^2\ell_\phi^{(0)}}{E_{(1)}\sin^2x^\theta}\,,\quad
    \frac{dx^\theta}{dx^t}=\pm \frac{x^\rho\sqrt{\left(\ell^{(0)}_{\theta}\right)^2\sin^2x^\theta-\left(\ell^{(0)}_\phi\right)^2}}{E_{(1)}\sin x^\theta}\,,
\end{equation}
where $\ell^{(0)}_\theta$ is an integration constant. Taking $x^\rho$ to $0$ we see that the derivatives vanish. Hence, for an asymptotic observer, the point particle freezes on the event horizon. ({\emph{Animation for these motions can be found \href{https://github.com/Arkachur/Blackhole-Near-Horizons-through-the-Looking-Glass.git}{here}}). 

\medskip

Geodesic equations \eqref{eq:NLO point-particle geodesic 3 near Schwarzschild BH} and \eqref{eq:NLO point-particle geodesic 4 near Schwarzschild BH} describe the motion of a point particle in Rindler spacetime, with additional non-trivial contributions from the spherical part arising due to the $\Phi_{\mu\nu}$ components of the near-horizon metric. Using \eqref{eq:LO point-particle geodesics near Schwarzschild BH} and \eqref{eq:NLO point-particle geodesic 3 near Schwarzschild BH}, the geodesic equation for $x^\rho$ can be simplified to
\begin{equation}\label{eq:NLO point-particle geodesic 4 near Schwarzschild BH 2}
    \ddot x^\rho + \frac{ E_{(1)}^2}{4r_h^2 \lb \xr \rb^3 } - \frac{ \lb \ell_{\theta}^{(0)}\rb^2 }{2} \xr = 0.
\end{equation}
Since $\xr(\lambda) \geq 0\,, \; \forall \lambda$, we define $H(\lambda) := \ln \xr(\lambda)$. In this redefinition, \eqref{eq:NLO point-particle geodesic 4 near Schwarzschild BH 2} can be rewritten as,
\begin{equation}\label{eq:NLO point-particle geodesic 4 near Schwarzschild BH 3}
    \lb 4\dot H^2 - 2 \lb \ell^{(0)}_\theta \rb^2 \rb e^{2H} - \frac{E_{(1)}^2}{r_h^2} e^{-2H} + k = 0\,,
\end{equation}
where $k\geq 0$ is an integration constant \footnote{Imposing positivity of the roots of the quadratic equation (in $e^{2H}$) \eqref{eq:NLO point-particle geodesic 4 near Schwarzschild BH 3}, one can obtain $k>0$.}. Solving \eqref{eq:NLO point-particle geodesic 4 near Schwarzschild BH 3} and using the solution to solve for $x^t$ we get,
\begin{subequations}\label{nlo schwarzschild sols}
\begin{align}
x^\rho(\lambda) &= \sqrt{ x^\rho_0 \pm  A\sinh f(\lambda) } \,,\label{nlo schwarzschild xr sol}\\
x^t(\lambda)&=x^t_0 \pm \frac{E_{(1)}}{\ell^{(0)}_\theta\sqrt{2\left(A^2 + \left(x^\rho_0\right)^2\right)}} \ln \left( \frac{ \left( \pm x^\rho_0 + \sqrt{A^2 + \left(x^\rho_0\right)^2} \right) e^{ f(\lambda)} - A }{ \left( \mp x^\rho_0 + \sqrt{A^2 + \left(x^\rho_0\right)^2} \right) e^{ f(\lambda)} + A } \right)\,\label{nlo schwarzschild xt sol},
\end{align}
\end{subequations}
where
\begin{equation}
    f(\lambda)=\sqrt{ 2}\ell^{(0)}_\theta(\lambda - \lambda_0)\,,~~x^\rho_0=\frac{k}{4\left(\ell^{(0)}_\theta\right)^2}\,,~~A=\sqrt{\left(\frac{E_{(1)}}{\sqrt 2r_h\ell^{(0)}_\theta}\right)^2- \lb \frac{k}{4\left(\ell^{(0)}_\theta\right)^2} \rb^2 }\,.
\end{equation}
As expected, the point particle asymptotes to the event horizon. 

\subsection{Geodesics from NH limit of Schwarzschild geodesics}\label{sec: Geodesics from near horizon limit of Schwarzschild geodesics}
We now analyse the near-horizon limit of the geodesic equations in the full Schwarzschild background. Instead of the usual geodesic equations, we will work with the following geodesic equation, which involves the Christoffel symbols of the first kind
\begin{equation}
g_{\mu\nu}\ddot{X}^\nu+\Gamma_{\mu\rho\sigma}\dot{X}^\rho\dot{X}^\sigma=0.
\end{equation}
The geodesic equations for the coordinates $X^t$, $X^r$, $X^\theta$, and $X^\phi$ in a Schwarzschild black hole are given as
\begin{subequations}\label{eq:Schwarzschild Geodesic equation}
    \begin{align}
        \left(r_h-X^r\right)\ddot{X}^t-\frac{r_h}{X^r}\dot{X}^r\dot{X}^t&=0\,,\label{eq:Schwarzschild Geodesic equation xt}\\
        \frac{X^r}{X^r-r_h}\ddot{X}^r-\frac{r_h}{2(X^r-r_h)^2}\left(\dot{X}^r\right)^2+\frac{r_h}{2\left(X^r\right)^2}\left(\dot{X}^t\right)^2-X^r\left(\dot{X}^\theta\right)^2\nonumber\\~~~~-X^r\sin^2(X^\theta)\left(\dot{X}^\phi\right)^2&=0\,,\label{eq:Schwarzschild Geodesic equation xr}\\
        \left(X^r\right)^2\ddot{X}^\theta+2X^r\dot{X}^r\dot{X}^\theta-\left({X}^r\right)^2\cos(X^\theta)\sin(X^\theta)\left(\dot{X}^\phi\right)^2&=0\,,\label{eq:Schwarzschild Geodesic equation xth}\\
        \left({X}^r\right)^2\sin^2(X^\theta)\ddot{X}^\phi+\left(X^r\right)^2\sin(2X^\theta)\dot{X}^\theta\dot{X}^\phi+2X^r\sin^2(X^\theta)\dot{X}^r\dot{X}^\phi&=0\,\label{eq:Schwarzschild Geodesic equation xf}.
    \end{align}
\end{subequations}
The Schwarzschild coordinates will be expanded as follows 
\begin{subequations}\label{eq:expansion of Schwarzschild Coordinates}
\begin{align}
&X^\theta=x^\theta+\epsilon y^\theta+\cdots, ~~X^\phi=x^\phi+\epsilon y^{\phi}+\cdots,\\&
X^t=x^t+\epsilon y^t+\cdots,~~X^r=r_h+\frac{\epsilon}{4r_h^2} \lb x^\rho \rb^2+\cdots.
\end{align}
\end{subequations}
Plugging in the expansions of the Schwarzschild coordinates \eqref{eq:expansion of Schwarzschild Coordinates} in the geodesics \eqref{eq:Schwarzschild Geodesic equation}, we get,
\begin{subequations}\label{eq:NH Limit of Sch geodesics}
    \begin{align}
        \frac{d}{d\lambda}\left( \lb x^\rho\rb^2  \dot{x}^t\right)&=0\,,\label{eq:NH Lmit of xt Sch geodesic}\\
        \frac{d}{d\lambda}\left( 2 \dot\xr  \right) + \frac{ \xr }{2r_h^2}\left( \dot x^t  \right)^2 - \xr\left( \left( \dot x^\theta \right)^2 + \sin^2 x^\theta \left( \dot x^\phi \right)^2 \right)&=0\,,\label{eq:NH Lmit of xr Sch geodesic}\\
        \ddot{x}^\theta-\cos(x^\theta)\sin(x^\theta)\left(\dot{x}^\phi\right)^2 &=0\,,\label{eq:NH Lmit of xth Sch geodesic}\\
        \frac{d}{d\lambda}\left(\sin^2(x^\theta)\dot{x}^\phi\right)&=0\,,\label{eq:NH Lmit of xf Sch geodesic}\\
        \frac{d}{d\lambda}\left[ \lb x^\rho\rb^2\dot x^\theta + 2r_h^2\dot y^\theta \right]-  \frac{ \lb x^\rho\rb^2 }{2} \sin\left( 2 x^\theta \right)\left( \dot x^\phi \right)^2 - 2r_h^2 \sin\left(2 x^\theta\right) \dot x^\phi\dot y^\phi  \nonumber\\
        - 2r_h^2  y^\theta \cos \left( 2 x^\theta \right) \left( \dot x^\phi \right)^2&=0\, , \label{eq:NH Lmit of yth Sch geodesic}\\
        \frac{d}{d\lambda}\left(\lb\xr\rb^2\sin^2 x^\theta \dot x^\phi + 2r_h^2 y^\theta\sin \left( 2  x^\theta \right) \dot x^\phi + 2 r_h^2\sin^2 \left(  x^\theta \right) \dot y^\phi  \right) &=0\,.\label{eq:NH Lmit of yf Sch geodesic}
    \end{align}
\end{subequations}
The geodesic equations after taking the near-horizon limit \eqref{eq:NH Limit of Sch geodesics} match with the ones we got before in equations \eqref{eq:LO point-particle geodesics near Schwarzschild BH} and \eqref{eq:NLO point-particle geodesics near Schwarzschild BH}.


\subsection{Scalar field using the map}\label{sec:Scalar field using the map (Schwarzschild Black hole)}
Using the map between the near-horizon Schwarzschild metric and the SC metric \eqref{eq:Schwarzschild Map inverse}, we explicitly write the action of a minimally coupled scalar field on the near-horizon region of a Schwarzschild black hole
\begin{equation}\label{eq:LO Scalar Field action near Sch BH}
    \mathcal{S}_{LO}=-\frac{1}{2}\int d^4x\:e\Big(-4\frac{r_h^2}{\rho^2}\left(\partial_t\phi^{(0)}\right)^2+\left(\partial_\rho\phi^{(0)}\right)^2\Big),
\end{equation}
where $e$ is the leading order determinant and is given as $e=\frac{1}{2}r_h\rho\sin\theta$ \ (simplifying from the leading term in (\ref{e:det-Expansion})). The equation of motion for \eqref{eq:LO Scalar Field action near Sch BH} is
\begin{equation}\label{eq:LO Scalar Field EOM near Sch BH}
    \partial_t^2\phi^{(0)}-\frac{\rho}{4r_h^2}\partial_\rho\phi^{(0)}-\frac{\rho^2}{4r_h^2}\partial_\rho^2\phi^{(0)}=0.
\end{equation}
This equation can be solved using the method of separation of variables, assuming a solution of the form $\phi^{(0)}=R(\rho)\,T(t)\,Y(\theta,\phi)$. Note that $Y(\theta,\phi)$ can be arbitrary. The equation of motion can be rewritten as follows
\begin{equation}\label{eq:Separation of variables LO EOM Sch BH}
\partial_t^2T=k^2T\,,\quad 
\rho\partial_\rho R +\rho^2\partial_\rho^2R=4r_h^2k^2R,
\end{equation}
where $k$ is an undetermined separation constant.
The solution for the scalar field after solving \eqref{eq:Separation of variables LO EOM Sch BH} is
\begin{eqnarray}
    \phi^{(0)}=\left(A\rho^\Delta+B\rho^{-\Delta}\right)\left(Ce^{kt}+De^{-kt}\right)Y(\theta,\phi)\,,
\end{eqnarray}
where, $\Delta=\pm2r_hk$. Next, we analyse the subleading action $\mathcal{S}_{NLO}$, which is given as
\begin{multline}\label{eq:NLO Scalar Field action near Sch BH}
    \mathcal{S}_{NLO}=-\frac{1}{2}\int d^4x\:e\Big(2v^{\mu\nu}\partial_\mu\phi^{(0)}\partial_\mu\phi^{(1)}+\bar{h}^{\mu\nu}\partial_\mu\phi^{(0)}\partial_\mu\phi^{(0)}+m^2\left(\phi^{(0)}\right)^2+\frac{\rho^2}{2r_h^2}v^{\mu\nu}\partial_\mu\phi^{(0)}\partial_\mu\phi^{(0)}\Big).
\end{multline}
Here, $\Omega^{ij}_{S^2}$ is the inverse sphere metric, and the indices $i,j$ only run over the coordinates of the sphere. Varying the $\phi^{(1)}$ field in the NLO action \eqref{eq:NLO Scalar Field action near Sch BH}, we obtain the LO equation of motion \eqref{eq:LO Scalar Field EOM near Sch BH}. The equation of motion of the $\phi^{(0)}$ field is 
\begin{multline}\label{eq:NLO Scalar Field EOM1 near Sch BH}
    -3\partial_t^2\phi^{(0)}+\frac{3\rho}{4r_h^2}\partial_\rho\phi^{(0)}+\frac{\rho^2}{4r_h^2}\partial^2_\rho\phi^{(0)}-\frac{1}{r_h^2}\hat{L}^2\phi^{(0)}+\partial_\rho^2\phi^{(1)}\\-\frac{4r_h^2}{\rho^2}\partial_t^2\phi^{(1)}+\frac{1}{\rho}\partial_\rho\phi^{(1)}-m^2\phi^{(0)}=0,
\end{multline}
where, $\hat{L}^2$ is the angular momentum squared operator defined as 
\begin{eqnarray}
    \hat{L}^2=-\partial_\theta^2-\cot\theta\partial_\theta-\frac{1}{\sin^2\theta}\partial_\phi^2\,.
\end{eqnarray}

\subsection{Scalar field from near horizon limit }\label{sec:Scalar field from near horizon limit (Schwarzschild Black hole)}
In this section, we will verify that, if we take near-horizon limits on the scalar field equations in the Schwarzschild background, we can retrieve the LO equations of motion \eqref{eq:LO Scalar Field action near Sch BH}. The equation of motion of the scalar field in the Schwarzschild background is
\begin{equation}\label{eq:scalar field EOM Sch}
    \frac{2}{r}\left(1-\frac{r_h}{r}\right)\partial_r\Phi-\frac{1}{r^2}\hat L^2\Phi+\frac{r_h}{r^2}\partial_r\Phi-\frac{r}{r-r_h}\partial^2_t\Phi+\left(1-\frac{r_h}{r}\right)\partial_r^2\Phi-m^2\Phi=0\,.
\end{equation}
Now, we will expand the scalar field as $\Phi\to\phi^{(0)}+\epsilon\phi^{(1)}+\cdots$. The near-horizon limit will be implemented by the following coordinate transformation of the radial coordinate $r\to r_h+\frac{\epsilon}{4r_h}\rho^2$. The equation of motion \eqref{eq:scalar field EOM Sch} can be arranged in powers of $\epsilon$ as follows
\begin{subequations}\label{eq:taking limits on scalar field EOM Sch}
    \begin{align}
        \mathcal{O}(\epsilon^{-1}):~& \partial_t^2\phi^{(0)}-\frac{\rho}{4r_h^2}\partial_\rho\phi^{(0)}-\frac{\rho^2}{4r_h^2}\partial_\rho^2\phi^{(0)}=0\,,\label{eq:taking limits on scalar field EOM Sch LO}\\
        \mathcal{O}(\epsilon^{0}):~& \partial_t^2\phi^{(0)}-\frac{\rho}{4r_h^2}\partial_\rho\phi^{(0)}+\frac{\rho^2}{4r_h^2}\partial^2_\rho\phi^{(0)}+\frac{1}{r_h^2}\hat{L}^2\phi^{(0)}-\partial_\rho^2\phi^{(1)}\nonumber\\&~+\frac{4r_h^2}{\rho^2}\partial_t^2\phi^{(1)}-\frac{1}{\rho}\partial_\rho\phi^{(1)}+m^2\phi^{(0)}=0\,\label{eq:taking limits on scalar field EOM Sch NLO}.
    \end{align}
\end{subequations}
The equations \eqref{eq:taking limits on scalar field EOM Sch LO} match \eqref{eq:LO Scalar Field EOM near Sch BH} in the leading order when we take $\epsilon\to 0$. At the subleading order, the limiting \eqref{eq:taking limits on scalar field EOM Sch NLO} and intrinsic \eqref{eq:NLO Scalar Field EOM1 near Sch BH} analysis match on adding the leading order equation of motion as follows
\begin{equation}
-\eqref{eq:taking limits on scalar field EOM Sch NLO}-2\eqref{eq:taking limits on scalar field EOM Sch LO}=\eqref{eq:NLO Scalar Field EOM1 near Sch BH}
\end{equation}

\bigskip \bigskip

\section{AdS Black brane}\label{Black holes and string-Carroll 3: AdS Black Brane}
The above section dealt with Schwarzschild black holes in asymptotically flat spacetimes in four dimensions. We treat the BTZ black hole (with rotations) in AdS$_3$ in the appendix. Moving beyond spherical black holes, we now extend the checks of our formulation to black branes and for concreteness choose the AdS black brane. The near horizon of the black brane will again be a SC geometry and now the base space will change from a sphere to a plane.

\subsection{Explicit map}\label{sec:Explicit map (AdS Black Brane)}
The metric of a 3+1 dimensional AdS black brane is 
\begin{eqnarray}\label{eq:AdS Black brane metric}
    ds^2=-\frac{r^2}{\ell^2}\left(1-\frac{r_h^3}{r^3}\right)dt^2+\frac{\ell^2}{r^2}\left(1-\frac{r_h^3}{r^3}\right)^{-1}dr^2+\frac{r^2}{\ell^2}d\vec{x}^2\,.
\end{eqnarray}
The near-horizon metric of the AdS black brane can be obtained via the following coordinate transformation
\begin{eqnarray}\label{eq:coordinate transformation}
    r\to r_h+\frac{3r_h}{4\ell^2}\epsilon\rho^2\,.
\end{eqnarray}
Plugging the coordinate transformation \eqref{eq:coordinate transformation} into \eqref{eq:AdS Black brane metric} and dialling $\epsilon$ to zero, leads to the following near-horizon metric
\begin{eqnarray}\label{eq:AdS Black brane-exp}
    ds^2=\frac{r_h^2}{\ell^2}d\vec{x}^2+\epsilon\left(-\frac{9r_h^2}{4\ell^4}~\rho^2dt^2+d\rho^2+\frac{3r_h^2}{2\ell^4}~\rho^2d\vec{x}^2\right)+\mathcal{O}(\epsilon^2)\,.
\end{eqnarray}
Comparing \eqref{eq:AdS Black brane-exp} with the SC expansion \eqref{SCmetric} and setting $\epsilon=c^2$, we can deduce the following map between the near-horizon expansion of the AdS black brane metric and the SC expansion
\begin{equation}\label{eq:AdS Black brane Map}
	h_{\mu\nu}dx^\mu dx^\nu = \frac{r_h^2}{\ell^2}d\vec{x}^2\,,~~
    \Phi_{\mu\nu} dx^\mu dx^\nu =\frac{3r_h^2}{2\ell^4}~\rho^2d\vec{x^2} \,,~~\tau_{\mu\nu}dx^\mu dx^\nu=-\frac{9r_h^2}{4\ell^4}~\rho^2dt^2+d\rho^2\,.
\end{equation}
The longitudinal space is a Rindler space of $(1+1)$ dimensions, whereas the leading-order transverse space is a two-dimensional plane ($\mathbb{R}^2$). Similarly, we can also map the inverse AdS black brane metric with the inverse SC metric as follows
\begin{equation}\label{eq:AdS black brane Map inverse}
    v^{\mu\nu}\partial_\mu\partial_\nu=-\frac{4\ell^4}{9r_h^2\rho^2}\partial_t^2+\partial_\rho^2\,,~~~\bar{h}^{\mu\nu}\partial_\mu\partial_\nu=\frac{\ell^2}{r_h^2}\partial^2_{\vec{x}}\,.
\end{equation}

\subsection{Geodesics of AdS Black Brane}
\paragraph{Using the SC map.} The LO and NLO point particle actions near the event horizon of a 3+1 dimensional AdS black brane are given below
\begin{subequations} \label{eq:point-particle action in the AdS black brane background}
\begin{align}
    S_{LO} &= -\frac{mr_h^2}{2\ell^2} \int \left[ \lb \dot x^u \rb^2 + \lb \dot x^v \rb^2 \right]d\lambda\,, \label{eq:LO point-particle action in the AdS black brane background} \\
    S_{NLO} &= -\frac{m}{2}\int \Bigg[ -\lb \frac{9r_h^2}{4\ell^4}\rb \lb x^\rho \rb^2 \lb \dot x^t \rb^2 + \lb \dot x^\rho \rb^2 + \frac{3r_h^2 \lb x^\rho \rb^2 }{2\ell^4} \lb \lb \dot x^u\rb^2 + \lb \dot x^v\rb^2  \rb\nonumber\\
    &\quad  + \frac{2r_h^2}{\ell^2}\lb \dot x^u\dot y^u + \dot x^v\dot y^v \rb  \Bigg] d\lambda \,. \label{eq:NLO point-particle action in the AdS black brane background}
\end{align}
\end{subequations}
where we have written $\vec x = (u,v)$. Varying \eqref{eq:LO point-particle action in the AdS black brane background} w.r.t $x^u$ and $x^v$, we get
\begin{align}\label{eq:LO point-particle geodesics near AdS black brane}
\ddot x^u &= 0, \quad \ddot x^v = 0 \,. 
\end{align}
They can be integrated to get
\begin{align} \label{eq:LO point-particle geodesics solutions near AdS black brane}
        x^u(\lambda) &= x^u_0 + p_u^{(0)}\lb \lambda - \lambda_0\rb , \quad
        x^v(\lambda) = x^v_0 + p_v^{(0)}\lb \lambda - \lambda_0\rb . 
    \end{align}
Here $p_u^{(0)}$ and $p_v^{(0)}$ can be interpreted as conserved momentum of the particle. To obtain the solutions for $x^t$ and $x^\rho$, we proceed to the NLO action \eqref{eq:NLO point-particle action in the AdS black brane background}. Its variation with respect to $x^u$, $x^v$, $x^t$ and $x^\rho$ gives the following equations of motion
\begin{subequations}\label{eq: NLO point-particle geodesics near AdS black brane}
    \begin{align}
        & \frac{d}{d\lambda}\lb \dot y^u + \frac{3 \lb x^\rho \rb^2 }{2\ell^2} \dot x^u \rb = 0\,, \quad 
        \frac{d}{d\lambda}\lb \dot y^v + \frac{3 \lb x^\rho \rb^2 }{2\ell^2} \dot x^v \rb = 0\,, \label{eq: NLO point-particle geodesics 2 near AdS black brane} \\
        & \lb x^\rho \rb^2 \dot x^t = E^{(1)}\,, \quad 
        \ddot x^\rho + \lb \frac{9r_h^2 x^\rho}{4\ell^4} \rb \lb \dot x^t \rb^2 - \frac{3r_h^2x^\rho}{2\ell^4} \lb \lb \dot x^u \rb^2 + \lb \dot x^v \rb^2 \rb =0 \,. \label{eq: NLO point-particle geodesics 4 near AdS black brane}
    \end{align}
\end{subequations}
The solutions to above equations are given as
\begin{subequations}\label{eq: NLO point-particle geodesics solutions near AdS black brane}
    \begin{align}
        x^\rho (\lambda) &= \sqrt{ x^\rho_0 \pm A \sinh f(\lambda) }\,, \label{eq: NLO point-particle geodesics solutions 1 near AdS black brane}\\
        x^t (\lambda)    &= x^t_0 \pm \frac{ r_h }{\sqrt 2 \ell^2} \ln \left[ \frac{ \lb \pm x^\rho_0 + \sqrt{ \frac{3 \lb E^{(1)} \rb^2 }{2 \lb p^{(0)} \rb^2 } } \rb e^{ f(\lambda)} - A }{ \lb \mp x^\rho_0 + \sqrt{ \frac{3 \lb E^{(1)} \rb^2 }{2 \lb p^{(0)} \rb^2 } } \rb e^{ f(\lambda)} + A } \right] , \label{eq: NLO point-particle geodesics solutions 2 near AdS black brane}
    \end{align}
\end{subequations}
where $f(\lambda)=\frac{\sqrt{6} r_h p^{(0)} }{\ell^2}(\lambda - \lambda_0)$, $p^{(0)} = \sqrt{ \lb p^{(0)}_u \rb^2 + \lb p^{(0)}_v \rb^2 }$ and $A = \sqrt{ \frac{3 \lb E^{(1)} \rb^2 }{2 \lb p^{(0)} \rb^2 } - \lb x^\rho_0 \rb^2 }$. Seeing the analogous structure of \eqref{eq: NLO point-particle geodesics solutions near AdS black brane} with the solutions of near-horizon Schwarzschild geodesics \eqref{nlo schwarzschild sols}, one can easily identify that the particle takes an infinite coordinate time to reach the horizon with respect to an asymptotic observer.

\paragraph{NH limit of black brane geodesics.} The geodesic equations in the full geometry of $(3+1) D$ $AdS$ black brane are 
\begin{subequations}\label{eq: geodesic equations in AdS black brane}
    \begin{align}
        &\ddot X^u+\frac{2}{X^r} \dot X^r \dot X^u = 0, \quad 
        \ddot X^v+\frac{2}{X^r} \dot X^r \dot X^v = 0,\quad 
        \ddot X^t - \frac{r_h^3 + 2 \lb X^r\rb^3 }{X^r\lb r_h^3 - \lb X^r \rb^3 \rb} \dot X^r\dot X^t = 0 \,,\\
        &\ddot X^r - \frac{r_h^3 + 2\lb X^r \rb^3}{ 2 X^r \lb \lb X^r \rb^3 - r_h^3 \rb } \lb \dot X^r \rb^2 + \frac{ r_h^6 + r_h^3 \lb X^r \rb^3 - 2 \lb X^r \rb^6 }{2 \ell^4 \lb X^r \rb^3} \lb X^t \rb^2 \nonumber \\
        & \qquad \qquad \qquad \qquad \qquad \qquad \qquad\qquad \quad\quad- \frac{ \lb X^r \rb^3 - r_h^3 }{\ell^4} \lb \lb \dot X^u \rb^2 + \lb \dot X^v \rb^2 \rb = 0.
    \end{align}
\end{subequations}

The worldline coordinates in the NH limit are expanded as
\begin{subequations}\label{eq: expansion of AdS black brane coordinates}
\begin{align}
&X^u=x^u+\epsilon y^u+\cdots, ~~X^v = x^v+\epsilon y^v + \cdots,\\&
X^t = x^t+\epsilon y^t+\cdots,~~X^r=r_h+\epsilon\frac{3r_h}{4r_h^2} \lb x^\rho \rb^2+\cdots.
\end{align}
\end{subequations}

In this expansion scheme, the geodesics equations expands upto $\mathcal O(\epsilon)$ to again give us \eqref{eq:LO point-particle geodesics near AdS black brane} and \eqref{eq: NLO point-particle geodesics near AdS black brane}. This serves as a robust cross-check of our developed formalism to instrisically consider the particle dynamics near the event horizon.

\subsection{Scalar fields in AdS black brane NH}
\paragraph{Using the SC map.}\label{sec:Scalar field using the map (AdS Black Brane)}
Using the map between the near-horizon AdS black brane metric and the SC metric \eqref{eq:AdS black brane Map inverse}, we explicitly write the action of a minimally coupled scalar field on the near-horizon region of a black brane
\begin{equation}\label{eq:LO Scalar Field action near ABB}
    \mathcal{S}_{LO}=-\frac{1}{2}\int d^4x~e\left(\left(\partial_\rho\phi^{(0)}\right)^2-\frac{4\beta^2}{\rho^2}\left(\partial_t\phi^{(0)}\right)^2\right),
\end{equation}
where $\beta=\frac{\ell^2}{3r_h}$. The equation of motion for \eqref{eq:LO Scalar Field action near ABB} is
\begin{equation}\label{eq:LO Scalar Field EOM near ABB}
    \partial_t^2\phi^{(0)}-\frac{\rho}{4\beta^2}\partial_\rho\phi^{(0)}-\frac{\rho^2}{4\beta^2}\partial_\rho^2\phi^{(0)}=0\,.
\end{equation}
Next, we analyse the subleading action $\mathcal{S}_{NLO}$, which is given as
\begin{multline}\label{eq:NLO Scalar Field action near ABB}
\mathcal{S}_{NLO}=-\frac{1}{2}\int d^4x\:e\Big(2v^{\mu\nu}\partial_\mu\phi^{(0)}\partial_\mu\phi^{(1)}+\bar{h}^{\mu\nu}\partial_\mu\phi^{(0)}\partial_\mu\phi^{(0)}\\+m^2\left(\phi^{(0)}\right)^2+\frac{3\rho^2}{2\ell^2}v^{\mu\nu}\partial_\mu\phi^{(0)}\partial_\mu\phi^{(0)}\Big).
\end{multline}
Varying the $\phi^{(1)}$ field in the NLO action \eqref{eq:NLO Scalar Field action near ABB}, we obtain the LO equation of motion \eqref{eq:LO Scalar Field EOM near ABB}. The equation of motion of the $\phi^{(0)}$ field is 
\begin{multline}\label{eq:NLO Scalar Field EOM near ABB}
    \frac{\ell^2}{r_h^2}\d_{\vec{x}}^2\phi^{(0)}-m^2\phi^{(0)}+\frac{9\rho}{2\ell^2}\partial_{\rho}\phi^{(0)}+\frac{1}{\rho}\partial_{\rho}\phi^{(1)}+\frac{3\rho^2}{2\ell^2}\partial_\rho^2\phi^{(0)}+\partial_\rho^2\phi^{(1)}\\-\frac{2\beta}{r_h}\partial_t^2\phi^{(0)}-\frac{4\beta^2}{\rho^2}\partial_t^2\phi^{(1)}=0\,.
\end{multline}

\paragraph{Scalar field from NH limit.}
We now verify that, if we take near-horizon limits on the scalar field equations in the black brane background, we can retrieve the LO equations of motion \eqref{eq:LO Scalar Field action near ABB}. EOM of the scalar field in the AdS black brane background is
\begin{equation}\label{eq:scalar field EOM ABB}
    \frac{4r^3-r_h^3}{\ell^2r^2}\d_r\Phi-\frac{\ell^2r}{r^3-r_h^3}\d^2_t\Phi+\frac{r^3-r_h^3}{\ell^2r}\d_r^2\Phi+\frac{\ell^2}{r^2}\d_{\vec{x}}^2\Phi-m^2\Phi=0\,.
\end{equation}
Now, we will expand the scalar field as $\Phi\to\phi^{(0)}+\epsilon\phi^{(1)}+\cdots$. The near-horizon limit will be implemented by the following coordinate transformation of the radial coordinate $r\to r_h + \frac{\epsilon}{4\beta}\rho^2$. The equation of motion \eqref{eq:scalar field EOM ABB} can be arranged in powers of $\epsilon$ as follows
\begin{subequations}\label{eq:taking limits on scalar field EOM ABB}
    \begin{align}
        \mathcal{O}(\epsilon^{-1}):~& \partial_t^2\phi^{(0)}-\frac{\rho}{4\beta^2}\partial_\rho\phi^{(0)}-\frac{\rho^2}{4\beta^2}\partial_\rho^2\phi^{(0)}=0\,,\label{eq:LO EOM from limiting analysis ABB}\\
        \mathcal{O}(\epsilon^{0}):~& \frac{3}{\ell^2}\rho\d_\rho\phi^{(0)}+\frac{\ell^2}{r_h^2}\d_{\vec{x}}^2\phi^{(0)}-m^2\phi^{(0)}
    -\frac{4\beta^2}{\rho^2}\d_t^2\phi^{(1)}+\frac{1}{\rho}\d_\rho\phi^{(1)}+\d_\rho^2\phi^{(1)}=0\,\label{eq:NLO EOM from limiting analysis ABB}.
    \end{align}
\end{subequations}
The leading order equations of motion \cref{eq:LO Scalar Field EOM near ABB,eq:LO EOM from limiting analysis ABB} match exactly in both analyses. Whereas the NLO equations of motion \cref{eq:NLO Scalar Field EOM near ABB,eq:NLO EOM from limiting analysis ABB} match on adding the LO equations of motion
\begin{eqnarray}
    \eqref{eq:NLO Scalar Field EOM near ABB}+\frac{2\beta}{r_h}\eqref{eq:LO Scalar Field EOM near ABB}=\eqref{eq:NLO EOM from limiting analysis ABB}\,.
\end{eqnarray}
\bigskip
\bigskip

\section{Lifshitz Black holes}\label{Black holes and string-Carroll 4: Lifshitz Black hole}

In this section, we explore a very different class of black holes, called Lifshitz black holes and encounter some interesting features about them. But, remarkably, our general formulation holds and gives further confidence about the power of our formulation. 

\subsection{Interlude: a quick review of Lifshitz spacetimes}
The last several years have seen extensive studies of theories with Lifshitz scaling and hyperscaling violation as interesting families of
nonrelativistic holography or $AdS/CMT$ with condensed matter like features, e.g.  \cite{Kachru:2008yh,Taylor:2008tg, Charmousis:2010zz,Iizuka:2011hg,Ogawa:2011bz,Huijse:2011ef,Dong:2012se}.
There is a large literature here: we refer the reader to the
reviews \cite{Taylor:2015glc,Hartnoll:2016apf}.

\medskip

Lifshitz theories exhibit the anisotropic scale invariance 
\begin{align}
t\ra \lambda^z t ,\ \ x_i\ra \lambda x_i ,\ \ r\ra \lambda r,
\end{align} 
with dynamical exponent $z$. With planar base space, the $(d+2)$-dim
dual spacetimes
\begin{align}
ds^2=-\frac{r^{2z}}{ \ell^{2z}}dt^2+\sum_{i=1}^{d}\frac{dx_i^2+dr^2}{r^2}
\end{align}
arise in gravity theories with negative cosmological constant (related
to the scale $\ell$) and massive abelian gauge fields. 

\medskip

More general hyperscaling violating Lifshitz (hvLif), or conformally Lifshitz, theories
with planar base space have an extra conformal factor
$(\frac{r_F}{r})^{2\Theta/d}$ over the metric above, with $\Theta$ the
hyperscaling violating exponent. These arise in Einstein-Maxwell-scalar theories: $r_F$ above is an extra length scale
that controls the regime of validity of these effective gravity
descriptions. 

\medskip

While these are phenomenological gravity descriptions, they do arise in various gauge/string realisations. Perhaps among the
simplest are $x^+$-reductions of highly boosted limits of $AdS$ black branes, which in the extreme limit become $AdS_{D+1}$ plane waves
\cite{Narayan:2012hk,Singh:2012un}, containing a normalizable $g_{++}$ deformation: the metric is
\begin{align} \label{AdSpw}
ds^2 = \frac{R^2}{r^2} [-2dx^+dx^- + dx_i^2 + dr^2] + R^2Qr^{D-2} (dx^+)^2
\end{align}
suppressing the transverse sphere in 10/11-dimensions. This represents a state in the dual CFT with uniform energy-momentum density 
$T_{++}\sim Q$. Under dimensional reduction on the $x^+$-direction regarded as compact, we obtain the hyperscaling violating Lifshitz
(hvLif) background
\begin{align}
ds^2 = R^2(R^2Q)^{\frac{1}{d}}r^{\frac{2\Theta}{d}} \left(-{\frac{dt^2}{ Qr^{2z}}} +\sum_{i=1}^{d}\frac{dx_i^2 + dr^2 }{r^2}\right),
\end{align}
with $z=\frac{d}{2}+2 ,\ \ \Theta=\frac{d}{2}\ ,\ \ d=D-2$. We see that the
microscopic flux $Q$ controls the effective scale $r_F$ above. These backgrounds for the $AdS_5$ plane wave case ($D=4$) 
\cite{Narayan:2012hk} in particular exhibit interesting logarithmic scaling for entanglement entropy akin to Fermi surfaces  \cite{Ogawa:2011bz,Huijse:2011ef,Dong:2012se}. Thus the gauge/string realizations above map excited CFT states to ground states of the hvLif
theories that arise in intermediate effective regimes, thereby mapping the corresponding novel entanglement scalings as well \cite{Narayan:2012ks,Narayan:2014ofl}.

\medskip

The hvLif spacetimes above pertain to zero temperature: in their deep interior they generically have singularities in curvature or tidal
forces. However, good effective field theories simulating nonrelativistic holograms do arise at finite temperature: these map to
considering black branes in these hvLif backgrounds, which in practice lead to well-defined holographic observables including entanglement, 
hydrodynamics, and so on (see \cite{Hartnoll:2016apf}). In this regard, it is also interesting to note that the hvLif black brane representing
the finite temperature hvLif theory can be obtained via $x^+$-reduction of the highly boosted $AdS$ black branes \cite{Singh:2012un} (see also \cite{Kolekar:2016pnr}), whose
extreme (zero temperature) limits are the $AdS$ plane waves (\ref{AdSpw}) above \cite{Narayan:2012hk}: the blackening factors under reduction organize themselves in
the appropriate way. The hvLif $z,\Theta$ parameters above are
constrained to satisfy relations following from the null energy conditions,\
$(z-1)(d+z-\Theta) \geq 0,\ \ (d-\Theta)(d(z-1)-\Theta) \geq 0$.
In addition, positivity of specific heat following from the
thermodynamic relation $S\sim T^{(d-\Theta)/z}$ requires $d>\Theta$ 
(we restrict to $z\geq 1$, which is required for pure Lifshitz with
$\Theta=0$).
For a wide range of $z,\Theta$ continuously connected to the $AdS$ 
case $z=1, \Theta=0$, the finite temperature near horizon structure in these
nonrelativistic cases exhibits features qualitatively similar to that in $AdS$.

\medskip

Similar features arise for black holes in these hvLif and Lifshitz
theories which correspond to the field theories on spherical base
spaces. A wide variety of such black holes with exponents $z,\Theta$
(as well as charge) were classified in \cite{Pedraza:2018eey},
building on earlier work (see eg \cite{Balasubramanian:2009rx,Iizuka:2011hg,Tarrio:2011de,Dong:2012se,Alishahiha:2012qu},
and references therein). The metric for uncharged hvLif black holes with
spherical horizon topology in $(d+2)$-dims is
\begin{eqnarray}\label{eq:hvLbh-Ped}
&&  ds^2=\left(\frac{r_F}{r}\right)^{2\Theta/d}\left(-\frac{r^{2z}}{\ell^{2z}}f(r)dt^2+\frac{\ell^2}{r^2f(r)}dr^2+r^2d\Omega_d^2\right),\nonumber \\ 
&& \qquad f(r)=1-\frac{m}{r^{d+z-\Theta}}+\frac{(d-1)^2\ell^2}{r^2(d-2+z-\Theta)^2}\,.
\end{eqnarray}
(we have suppressed other background fields required to support this 
metric.)\ 
The null energy conditions here are the same as above (for hyperbolic
topology further restrictions arise \cite{Pedraza:2018eey}). For $z=1, \Theta=0$, these
reduce to $AdS_{d+2}$ Schwarzschild black holes. The blackening factor in
(\ref{eq:hvLbh-Ped}) has a single zero, as in that case (for charged cases,
interesting extremal limits arise, with $AdS_2$ throats \cite{Kolekar:2018sba}).

\medskip

Given this, it is a reasonable expectation that the near-horizon structure of these nonrelativistic hvLif or Lifshitz black holes with
finite temperature continues to organise itself as a SC expansion with the nonrelativistic features only modifying details
thereof. This is corroborated by the fact that the Lifshitz metric can be recast via $\rho\sim r^z$ as 
\begin{align}\label{hvL-2d}
ds^2\sim \left( -\frac{dt^2}{\rho^2} + \frac{d\rho^2}{\rho^2} \right)
+ \frac{dx_i^2}{\rho^{2/z}}.
\end{align} 
These can be obtained independently as solutions in auxiliary 2-dim dilaton gravity theories \cite{Bhattacharya:2020qil}. \
In this form we see that the anisotropy has moved to the radial scaling of the transverse $d$-dim space: thus the blackening factor with $\rho_0>0$ of a finite black hole horizon (finite temperature) renders this nonsingular. The zero temperature limit $\rho_0\ra 0$ is likely to have more intricate features pertaining to the singularities mentioned earlier: 
we hope to explore this in future work.

We will focus here on 4-dimensional Lifshitz black holes at finite temperature as representative examples in the $z,\Theta$-space continuously connected to $AdS$: we will discuss these in detail towards a near horizon Carroll expansion.

\subsection{Back to Carroll: explicit map from NH to SC}\label{sec:Explicit map (Lifshitz Black hole)}
Setting $d=2, \Theta=0$ in (\ref{eq:hvLbh-Ped}) above, the metric of a 3+1 d Lifshitz black hole is \cite{Pedraza:2018eey,Tarrio:2011de}
\begin{eqnarray}\label{eq:Lifshitz metric1}
    ds^2=-\frac{r^{2z}}{\ell^{2z}}\left(1-\frac{m}{r^{2+z}}+\frac{\ell^2}{r^2z^2}\right)dt^2+\frac{\ell^2}{r^2}\left(1-\frac{m}{r^{2+z}}+\frac{\ell^2}{r^2z^2}\right)^{-1}dr^2+r^2d\Omega^2\,.
\end{eqnarray}
In what follows, all the expressions reduce to those for $AdS_4$ Schwarzschild black holes upon setting $z=1$ and $m=2M$ (see Appendix \ref{String Carroll Maps}). 
The location of the event horizon can be found by setting $g_{rr}^{-1}=0$. On doing this, we can find a relation between the mass parameter and the location of the event horizon $r_h$
\begin{eqnarray}
    m=\frac{r_h^z}{z^2}\left(\ell^2+r_h^2z^2\right)\,.
\end{eqnarray}
Using the relation between $m$ and $r_h$, the metric \eqref{eq:Lifshitz metric1} can be rewritten as
\begin{subequations}\label{eq:Lifshitz metric2}
    \begin{align}
    ds^2&=-\frac{r^{2z}}{\ell^{2z}}f(r)dt^2+\frac{\ell^2}{r^2f(r)}dr^2+r^2d\Omega^2\,,\label{eq:Lifshitz metric2a}\\
    f(r)&=1-\frac{r_h^z}{z^2r^{2+z}}\left(\ell^2+r_h^2z^2\right)+\frac{\ell^2}{r^2z^2}\label{eq:Lifshitz metric2b}\,.
    \end{align}
\end{subequations}
The near-horizon metric of the Lifshitz black hole can be obtained via the following coordinate transformation
\begin{eqnarray}\label{eq:Lifshitz coordinate transformation}
    r\to r_h+\lambda(z\,;\ell)\epsilon\rho^2\,,~~\lambda(z\,;\ell)=\frac{\ell^2+z(2+z)r_h^2}{4z\ell^2r_h}\,.
\end{eqnarray}
Plugging the coordinate transformation \eqref{eq:Lifshitz coordinate transformation} into \eqref{eq:Lifshitz metric2} and dialling $\epsilon$ to zero, leads to the following near-horizon metric
\begin{eqnarray}\label{eq:Lifshitz-exp}
    ds^2=r_h^2d\Omega^2+\epsilon\left(-\frac{4\lambda(z\,;\ell)^2}{\ell^{2(z-1)}r_h^{2-2z}}~\rho^2dt^2+d\rho^2+2r_h\lambda(z\,;\ell)~\rho^2d\Omega^2\right)+\mathcal{O}(\epsilon^2)\,.
\end{eqnarray}
Comparing \eqref{eq:Lifshitz-exp} with the SC expansion \eqref{SCmetric} and setting $\epsilon=c^2$, we can deduce the following map between the near-horizon expansion of the Lifshitz black hole metric and the SC expansion
\begin{gather}\label{eq:Lifshitz Map}
    h_{\mu\nu}dx^\mu dx^\nu = r_h^2d\Omega^2\,,~~
    \Phi_{\mu\nu} dx^\mu dx^\nu =2r_h\lambda(z\,;\ell)~\rho^2d\Omega^2\,,\nonumber\\\tau_{\mu\nu}dx^\mu dx^\nu=-\frac{4\lambda(z\,;\ell)^2}{\ell^{2(z-1)}r_h^{2-2z}}~\rho^2dt^2+d\rho^2\,.
\end{gather}
The longitudinal space is a Rindler space of $(1+1)$ dimensions, whereas the leading-order transverse space is a two-dimensional sphere ($S^2$). Similarly, we can also map the inverse Lifshitz black hole metric with the inverse SC metric as follows
\begin{equation}\label{eq:Lifshitz Map inverse}
\begin{split}
    v^{\mu\nu}\partial_\mu\partial_\nu&=-\frac{l^{2(z-1)} r_h^{2-2 z}}{4\lambda(z\,;\ell)^2\rho ^2}\partial_t^2+\partial_\rho^2\,,\\\bar{h}^{\mu\nu}\partial_\mu\partial_\nu&=\frac{1}{r_h^2}\partial_\Omega^2+\Gamma(z;\ell)\partial_t^2+\Lambda(z;\ell)\rho ^2\partial_\rho^2\,,       
\end{split}
\end{equation}
where the constants $\Gamma(z;\ell)$ and $\Lambda(z;\ell)$ are defined as follows
\begin{subequations}
    \begin{gather}
        \Gamma(z;\ell)=\frac{ \ell^{2 (z-2)} \left(\ell^2 (3 z-5)+3 r_h^2 z \left(z^2+z-2\right)\right)}{32r_h^{2z}z\lambda(z\,;\ell)^2},\\\Lambda(z;\ell)=\frac{1}{8}  \left(-\frac{z^2+z-2}{\ell^2}-\frac{z+1}{r_h^2 z}\right)\,.
    \end{gather}
\end{subequations}
\subsection{Geodesics of the Lifshitz BH}

\paragraph{Using the SC map.} The LO and NLO particle actions near the event horizon of a 3+1 dimensional Lifshitz black hole become
\begin{subequations}\label{eq: point-particle action in the Lifshitz black hole background}
    \begin{align}
        S_{LO} &= -\frac{m r_h^2}{2}\int \left[ \lb \dot x^\theta \rb^2 + \sin^2 \lb x^\theta\rb\lb \dot x^\phi \rb^2 \right]\,d\lambda\, , \label{eq: LO point-particle action in the Lifshitz black hole background} \\
        S_{NLO} &= -\frac{m}{2} \int \Bigg[ -\A^2 \lb x^\rho \rb^2 \lb \dot x^t \rb^2 + \lb \dot x^\rho \rb^2 + \B^2 \lb x^\rho \rb^2 \left\{ \lb \dot x^\theta \rb^2 + \sin^2 \lb x^\theta\rb\lb \dot x^\phi \rb^2 \right\} \nn \\
        & \quad + r_h^2 y^\theta \sin \lb 2 x^\theta \rb \lb \dot x^\phi \rb^2 + 2r_h^2 \left\{ \dot x^\theta \dot y^\theta + \dot x^\phi \dot y^\phi \right\} \Bigg] \, d\lambda \,, \label{eq: NLO point-particle action in the Lifshitz black hole background}
    \end{align}
\end{subequations}

where,
\begin{equation}
    \A^2 = \frac{4\lambda(z;\ell)^2}{\ell^{2(z-1)}r_h^{2-2z}}\,, \qquad \B^2 = 2r_h\lambda(z\,;\ell)\,.
\end{equation}

Varying \eqref{eq: point-particle action in the Lifshitz black hole background} with respect to $x^\theta$ and $x^\phi$ leads us to back to equations \eqref{eq:LO point-particle geodesics near Schwarzschild BH}, which are the usual great circle equations on $S^2$. We get the equations of motion for $x^t$ and $x^\rho$ after varying \eqref{eq: NLO point-particle action in the Lifshitz black hole background}. Its variation with respect to $x^\theta$, $x^\phi$, $x^t$ and $x^\rho$ yields us
\begin{subequations}\label{eq: NLO point-particle geodesic near Lifshitz BH}
    \begin{align}
        \frac{d}{d\lambda} \lb 2 \B^2 \lb x^\rho \rb^2 \dot x^\theta + 2r_h^2 \dot y^\theta \rb - 2r_h^2 \sin \lb 2 x^\theta \rb \dot x^\phi \dot y^\phi & \nn\\
        -\lb \B^2 \lb x^\rho \rb^2 \sin \lb 2 x^\theta \rb + 2r_h^2 y^\theta \cos \lb 2 x^\theta \rb \rb \lb \dot x^\phi \rb^2 &= 0 \,, \label{eq: NLO point-particle geodesic 1 near Lifshitz BH} \\
        2\B^2 \lb x^\rho \rb^2 \sin^2 \lb x^\theta \rb \dot x^\phi + 2r_h^2 \lb y^\theta \sin \lb 2 x^\theta \rb \dot x^\phi + \sin^2 \lb x^\theta \rb \dot y^\phi \rb &= \ell_\phi ^{(1)} \,, \label{eq: NLO point-particle geodesic 2 near Lifshitz BH} \\
        \lb x^\rho \rb^2 \dot x^t &= E^{(1)} \,, \label{eq: NLO point-particle geodesic 3 near Lifshitz BH} \\
        \ddot x^\rho + \A^2 x^\rho \lb \dot x^t \rb^2 - \B^2 x^\rho \lb \lb \dot x^\theta \rb^2 + \sin^2 \lb x^\theta\rb\lb \dot x^\phi \rb^2 \rb & = 0 \,. \label{eq: NLO point-particle geodesic 4 near Lifshitz BH}
    \end{align}
\end{subequations}

Solving \eqref{eq:LO point-particle geodesics near Schwarzschild BH}, \eqref{eq: NLO point-particle geodesic 3 near Lifshitz BH} and \eqref{eq: NLO point-particle geodesic 4 near Lifshitz BH} gives us
\begin{subequations}
    \begin{align}
        x^\rho(\lambda) &= \sqrt{ x^\rho_0 \pm \zeta(z)\sinh f(\lambda) } \,,  \\
        x^t(\lambda) &= x^t_0 \pm \frac{1}{\A}\ln \left[ \frac{ \lb \pm x^\rho_0 + \frac{\A E^{(1)}}{\B \ell_{\theta}^{(0)} } \rb e^{ f(\lambda)} - \zeta(z\,;\ell) }{ \lb \mp x^\rho_0 + \frac{\A E^{(1)}}{\B \ell_{\theta}^{(0)} } \rb e^{ f(\lambda)}+ \zeta(z\,;\ell) } \right] \,,
    \end{align}
\end{subequations}

where $f(\lambda)=\B \ell_\theta^{(0)} \lb \lambda - \lambda_0 \rb$, $\zeta(z\,;\ell) := \sqrt{ \lb \frac{\A E^{(1)}}{\B \ell_{\theta}^{(0)} } \rb^2 - \lb x^\rho_0 \rb^2 } $. These solutions are similar to near-horizon Schwarzschild geodesics.

\paragraph{NH limit of Lifshitz geodesics.} The geodesic equations in the full geometry of the 4D Lifshitz black hole are
\begin{subequations}\label{Geodesic equation of Lifshitz black hole}
    \begin{align}
    \dot{X}^t\left(X^r\right)^{2z}f(X^r)&=E\,\label{eq:Lifshitz Geodesic equation xt}\\
    \frac{\ell^2}{\left(X^r\right)^2f(X^r)}\ddot{X}^r-X^r\left(\dot{X}^\theta\right)^2-X^r\sin^2(X^\theta)\left(\dot{X}^\phi\right)^2\nonumber\\-\frac{\ell^2(2f(X^r)+X^r\d_rf(X^r))}{2\left(X^r\right)^3f(X^r)^2}\left(\dot{X}^r\right)^2+\frac{(2zf(X^r)+X^r\d_rf(X^r))}{2\ell^{2z}\left(X^r\right)^{1-2z}}\left(\dot{X}^t\right)^2&=0\,\label{eq:Lifshitz Geodesic equation xr}\\
    \left(X^r\right)^2\ddot{X}^\theta+2X^r\dot{X}^r\dot{X}^\theta-\left({X}^r\right)^2\cos(X^\theta)\sin(X^\theta)\left(\dot{X}^\phi\right)^2&=0\,,\label{eq:Lifshitz Geodesic equation xth}\\
        \left({X}^r\right)^2\sin^2(X^\theta)\dot{X}^\phi &=L_\phi\,\label{eq:Lifshitz Geodesic equation xphi},
    \end{align}
\end{subequations}
where $f(X^r)$ is defined in \eqref{eq:Lifshitz metric2b}, $\d_rf(X^r)$ implies derivative with respect to the $X^r$ coordinate, and $E$, $L_\phi$ are integration constants. The coordinates will be expanded as follows 
\begin{subequations}\label{eq:expansion of Lifshitz Coordinates}
\begin{align}
&X^\theta=x^\theta+\epsilon y^\theta+\cdots, ~~X^\phi=x^\phi+\epsilon y^{\phi}+\cdots,\\&
X^t=x^t+\epsilon y^t+\cdots,~~X^r=r_h+\lambda(z,\ell)\epsilon\lb x^\rho \rb^2+\cdots.
\end{align}
\end{subequations}
Plugging in the expansions of the coordinates \eqref{eq:expansion of Lifshitz Coordinates} in the geodesics \eqref{Geodesic equation of Lifshitz black hole}, we get, \eqref{eq:LO point-particle geodesics near Schwarzschild BH} and \eqref{eq: NLO point-particle geodesic near Lifshitz BH}. 
\subsection{Scalar field in the background of Lifshitz BHs}

\paragraph{Using the SC map.}\label{sec:Scalar field using the map (AdS Lifshitz Black hole)}
Using the map between the near-horizon Lifshitz metric and the SC metric \eqref{eq:Lifshitz Map inverse}, we explicitly write the action of a minimally coupled scalar field on the near-horizon region of a Lifshitz black hole
\begin{equation}\label{eq:LO Scalar Field action near LBH}
    \mathcal{S}_{LO}=-\frac{1}{2}\int d^4x\sqrt{-g}\Big(\left(\d_\rho\phi^{(0)}\right)^2-\frac{1}{\A^2\rho^2}\left(\d_t\phi^{(0)}\right)^2\Big)\,.
\end{equation}
The equation of motion for \eqref{eq:LO Scalar Field action near LBH} is
\begin{equation}\label{eq:LO Scalar Field EOM near LBH}
    \partial_t^2\phi^{(0)}-\A^2\rho\partial_\rho\phi^{(0)}-\A^2\rho^2\partial_\rho^2\phi^{(0)}=0\,.
\end{equation}
Next, we analyse the subleading action $\mathcal{S}_{NLO}$, which is given as 
\begin{multline}\label{eq:NLO Scalar Field action near LBH}
    \mathcal{S}_{NLO}=-\frac{1}{2}\int d^4x\:e\Big(2v^{\mu\nu}\partial_\mu\phi^{(0)}\partial_\mu\phi^{(1)}+\bar{h}^{\mu\nu}\partial_\mu\phi^{(0)}\partial_\mu\phi^{(0)}\\+m^2\left(\phi^{(0)}\right)^2+\left(\frac{(1+z)\lambda(z;\ell)}{r_h}\right)\rho^2v^{\mu\nu}\partial_\mu\phi^{(0)}\partial_\mu\phi^{(0)}\Big).
\end{multline}
Varying the $\phi^{(1)}$ field in the NLO action \eqref{eq:NLO Scalar Field action near LBH}, we obtain the LO equation of motion \eqref{eq:LO Scalar Field EOM near LBH}. The equation of motion of the $\phi^{(0)}$ field is 

\begin{multline}\label{eq:NLO Scalar Field EOM near LBH}
   \frac{3\left(\ell^2(1+z)+r_h^2z(2+z)(3+z)\right)\rho}{8\ell^2r_h^2z}\partial_\rho\phi^{(0)}+\frac{\left(\ell^2(1+z)+r_h^2z(2+z)(3+z)\right)\rho^2}{8\ell^2r_h^2z}\partial^2_\rho\phi^{(0)}\\+\frac{\ell^{2z-4}\left(\ell^2(-7+z)+r_h^2z(z^2-3z-10)\right)}{32r_h^{2z}z\lambda(z;\ell)^2}\partial_t^2\phi^{(0)}+ -\frac{1}{\A^2\rho^2}\d_t^2\phi^{(1)}\\+\frac{1}{\rho}\partial_\rho\phi^{(1)}+\partial_\rho^2\phi^{(1)}-\frac{1}{r_h^2}\hat{L}^2\phi^{(0)}-m^2\phi^{(0)} =0\,.
\end{multline}

\paragraph{Scalar field from NH limit.} The EOM of the scalar field in the Lifshitz black hole background is
\begin{equation}\label{eq:scalar field EOM LBH}
    \left(\frac{(3+z)rf(r)+r^2\d_rf(r)}{\ell^2}\right)\d_r\Phi-\frac{\ell^{2z}}{r^{2z}f(r)}\d^2_t\Phi+\frac{r^2f(r)}{\ell^2}\d_r^2\Phi-\frac{1}{r^2}\hat{L}^2\Phi-m^2\Phi=0\,,
\end{equation}
where $f(r)$ has been defined in \eqref{eq:Lifshitz metric2b}. Now, we will expand the scalar field as $\Phi\to\phi^{(0)}+\epsilon\phi^{(1)}+\cdots$. The near-horizon limit will be implemented by the following coordinate transformation of the radial coordinate $r\to r_h+\lambda(z\,;\ell)\epsilon\rho^2$. The EOM \eqref{eq:scalar field EOM LBH} can be arranged in powers of $\epsilon$ as follows
\begin{subequations}\label{eq:taking limits on scalar field EOM LBH}
    \begin{align}
        \mathcal{O}(\epsilon^{-1}):~&  \partial_t^2\phi^{(0)}-\A^2\rho\partial_\rho\phi^{(0)}-\A^2\rho^2\partial_\rho^2\phi^{(0)}=0\,,\label{eq:LO EOM from limiting analysis LBH}\\
        \mathcal{O}(\epsilon^{0}):~& \G\d^2_t\phi^{(0)}+\L\rho^2\d_\rho^2\phi^{(0)}-\left(\L\rho-\frac{\rho(z+2)}{\ell^2}\right)\d_\rho\phi^{(0)}\nn\\&-\frac{1}{r_h^2}\hat{L}^2\phi^{(0)}-m^2\phi^{(0)}
    -\frac{1}{\A^2\rho^2}\d_t^2\phi^{(1)}+\frac{1}{\rho}\d_\rho\phi^{(1)}+\d_\rho^2\phi^{(1)}=0\,.\label{eq:NLO EOM from limiting analysis LBH}
    \end{align}
\end{subequations}
The leading order EOM \cref{eq:LO Scalar Field EOM near LBH,eq:LO EOM from limiting analysis LBH} match exactly in both analyses. The NLO EOM \cref{eq:NLO Scalar Field EOM near LBH,eq:NLO EOM from limiting analysis LBH} match on adding the LO equations of motion
\begin{eqnarray}
    \eqref{eq:NLO Scalar Field EOM near LBH}+\frac{(z+1)\ell^{2(z-1)}}{4r_h^{2z-1}\lambda(z;\ell)}\eqref{eq:LO Scalar Field EOM near LBH}=\eqref{eq:NLO EOM from limiting analysis LBH}\,.
\end{eqnarray}


\bigskip \bigskip

\section{Closing onto the horizon}\label{sec:Closing onto the horizon}

In the previous sections, we have first laid the foundation of formulating the NH region of non-extremal black holes (and black branes) in terms of a fibre-bundled geometry with two-dimensional null fibres, which we have formalised as a SC geometry. We have then checked the validity of our formulation by considering various explicit examples of BHs and their NH geometries and probes on these geometries. 

\begin{figure}[hbt!]
    \centering
    \includegraphics[width=\linewidth]{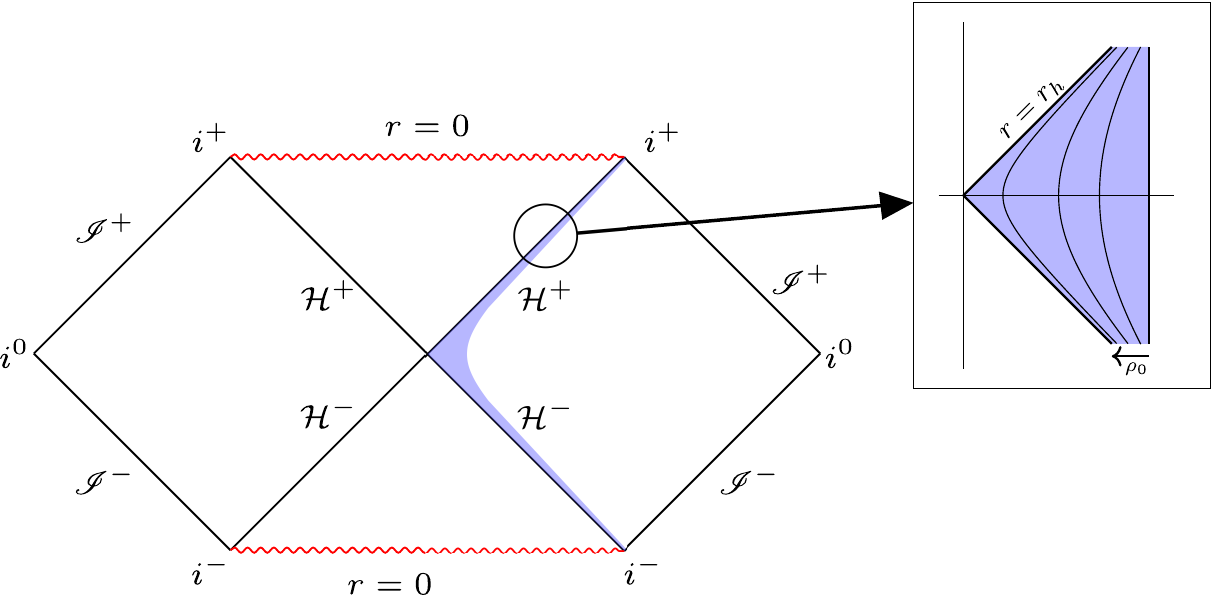}
    \vspace{-7mm}
    \caption{The NH region (blue) of the Schwarzschild BH in a Penrose diagram and the process of reaching the BH horizon from the NH in the inset.}
    \label{}
\end{figure}

\medskip

A natural question is how one can reach the horizon of the BH from the NH region.  We formulate the preliminary answer to this question in this section. 

\medskip

The technical question is the following. We know that the near horizon geometry is a SC geometry characterized by two null directions given by the 2d Rindler fibre. The horizon, on the other hand, is a Carroll manifold of one lower dimension and characterised by one null direction. How does one make the transition from a SC geometry to a Carroll geometry? 

\medskip

To do this, we will perform another Carroll expansion {\em inside} the SC geometry itself. Since we need a co-dimension one geometry emerging from the SC geometry, we will consider constant radial hypersurfaces of the 2d Rindler. On taking these hypersurfaces to the Rindler horizon, we will recover the horizon of the black hole. So, we will learn that the horizon of the black hole is encoded in the horizon of the 2d Rindler spacetime of the near-horizon region. While it is perhaps unsurprising that this will be the case for the field theories in NH backgrounds given the 2d Rindler dominates the inverse metric, the geodesics which are a priori controlled by the metric, also follow suit. This is because the LO geodesics are spatial and the dynamics again is controlled by the 2d Rindler. 

\medskip

Consider the SC expansion of a Lorentzian metric in the pre-Carrollian variables \eqref{SCmetric} and we put the spacelike longitudinal coordinate to a constant, say $x^1=x^1_0$. Since we have restricted ourselves to a codimension-one hypersurface of the spacetime manifold, we need to restrict the tangent space as well
\begin{subequations}
    \begin{align}
        g^{sc}_{\mu\nu}dx^\mu dx^\nu&\to\delta_{\bar{A}\bar{B}}e^{\bar{A}}_{\bar{\mu}} e^{\bar{B}}_{\bar{\nu}} dx^{\bar{\mu}} dx^{\bar{\nu}}+c^2\left(-\tau^0_{\bar{\mu}}\tau^0_{\bar{\nu}}+2\delta_{\bar A\bar B}e^{\bar{A}}_{(\bar{\mu}}\pi^{\bar{B}}_{\bar{\nu})}\right)dx^{\bar{\mu}} dx^{\bar{\nu}}+\cdots\,,\\
        g_{sc}^{\mu\nu}\partial_\mu \partial_\nu&\to -\frac{1}{c^2}v^{\bar{\mu}}_0v^{\bar{\nu}}_0\partial_{\bar{\mu}} \partial_{\bar{\nu}}+\delta^{\bar{A}\bar{B}}e_{\bar{A}}^{\bar{\mu}} e_{\bar{B}}^{\bar{\nu}} \partial_{\bar{\mu}} \partial_{\bar{\nu}}-2v^{(\bar{\mu}}_0m^{\bar{\nu})}_0+\cdots\,.
    \end{align}
\end{subequations}
The indices $\bar{\mu}\,,\bar{\nu}$ run over the d+1-dimensional hypersurface of the d+2-dimensional spacetime. Furthermore, the vielbeine are a function of the constant $x^1_0$, which can be scaled appropriately depending on the problem at hand. 

\medskip

We will now show how to explicitly get to the horizon from the near horizon region. We elucidate this for the 4D Schwarzschild black hole. The procedure remains similar for all black objects.

\subsection*{From NH to the horizon: Schwarzschild BH}
The SC expansion of the Schwarzschild black hole was given earlier in \cref{eq:Schwarzschild Map,eq:Schwarzschild Map inverse}. Consider a constant radial $\rho=\rho_0$ hypersurface, the near-horizon Schwarzschild metric takes the form,
\begin{subequations}\label{eq:Schwarzschild Map Particle-Carroll1}
\begin{align}
    g&=r_h^2d\Omega^2+\epsilon\left(-\frac{\rho_0^2}{4r_h^2}dt^2+\frac{\rho_0^2}{2}d\Omega^2\right)+\mathcal{O}(\epsilon^2)\,,\label{eq:SBH Particle-Carroll1}\\
    g^{-1}&=\frac{1}{\epsilon}\left(-\frac{4r_h^2}{\rho_0^2}\partial_t^2\right)+r_h^{-2}\partial^2_{\Omega}-\partial_t^2\,\label{eq:SBH inverse Particle-Carroll1}.
\end{align}
\end{subequations}
Now, we will scale the constant radial hypersurface with a dimensionless parameter $\lambda$, such that, $\rho=\lambda\rho_0$. Limiting $\lambda$ to be small, pushes the constant radial hypersurface towards the horizon. Implementing the scale in \eqref{eq:Schwarzschild Map Particle-Carroll1}, we get,
\begin{subequations}\label{eq:Schwarzschild Map Particle-Carroll2}
\begin{align}
    g&=r_h^2d\Omega^2+\tilde\epsilon\left(-\frac{\rho_0^2}{4r_h^2}dt^2+\frac{\rho_0^2}{2}d\Omega^2\right)+\mathcal{O}(\tilde\epsilon^2)\,,\label{eq:SBH Particle-Carroll}\\
    g^{-1}&=\frac{1}{\tilde\epsilon}\left(-\frac{4r_h^2}{\rho_0^2}\partial_t^2\right)+r_h^{-2}\partial^2_{\Omega}-\partial_t^2+\mathcal{O}(\tilde\epsilon)\,\label{eq:SBH inverse Particle-Carroll2},
\end{align}
\end{subequations}
where $\tilde\epsilon=\lambda^2\epsilon$. The horizon expansion of the Schwarzschild black hole \eqref{eq:Schwarzschild Map Particle-Carroll2} can be mapped to the particle-Carroll expansion \eqref{eq:particle Carroll expansion} on identifying $c^2=\tilde\epsilon$ as follows\footnote{In this section the Greek indices will run over the timelike and $x^{\bar{A}}$ coordinates.} 
\begin{subequations}\label{eq: particle carroll schwarschild data}
\begin{gather}
    h_{\mu\nu}dx^\mu dx^\nu=r_h^2d\Omega^2\,,~~\tau_\mu dx^\mu=\frac{\rho_0}{2r_h}dt\,,~~\Phi_{\mu\nu}dx^\mu dx^\nu=\frac{\rho_0^2}{2}d\Omega^2\,,\\
    v^\mu\partial_\mu=-\frac{2r_h}{\rho_0}\partial_t\,,~~\bar{h}^{\mu\nu}\partial_\mu\partial_\nu=r_h^{-2}\partial^2_{\Omega}-\partial_t^2\,.
    \end{gather}
\end{subequations}
The LO terms in \eqref{eq:Schwarzschild Map Particle-Carroll2} are  
\begin{eqnarray}
    g=0\cdot dt^2+r_h^2d\Omega^2~~,~~g^{-1}=-\frac{4r_h^2}{\rho_0^2}\partial_t^2\,,
\end{eqnarray}
where, $g$ is the metric on the horizon of the black hole \eqref{eq:degenerate metric}. The metric $g$ and the vector $v$ (clock-form) \cite{Duval:2014lpa} are the defining ingredients of the Carrollian geometry (non-Lorentzian) on the horizon of the black hole. 

\medskip

To analyse scalar fields on this background, we will start with the LO scalar field theory \eqref{eq:LO Scalar Field action near Sch BH} and further expand the scalar fields as $\phi^{(0)}\to\phi^{(0,0)}+\lambda\rho_0\phi^{(0,1)}+\cdots$. The determinant takes the form $e=\sqrt{-\tau_\mu\tau_\nu+h_{\mu\nu}}$. The scalar field theory on this background is given as 
\begin{eqnarray}\label{eq:SBH Scalar field Particle-Carroll}
    S_{LO}=\frac{1}{2}\int e~d^3x\frac{4r_h^2}{\rho_0^2}\left(\partial_t\phi^{(0,0)}\right)^2\,.
\end{eqnarray}
The equation of motion for the scalar field $\phi^{(0,0)}$ is $\partial_t^2\phi^{(0,0)}=0$.

\medskip

On the other hand, if one wants to analyse particle geodesics on this background, one needs to impose the geometric data \cref{eq: particle carroll schwarschild data} on \cref{eq:string-Carroll point particle action}. In this setting, we obtain the LO and NLO actions as follows
\begin{subequations}
    \begin{align}
        S_{Q,LO} &= -\frac{mr_h^2}{2} \int \left( \left( \dot x^\theta \right)^2 + \sin^2x^\theta \left( \dot x^\phi \right)^2 \right)\, d\lambda\,, \\
        S_{Q,NLO} &= -\frac{m}{2} \int \bigg( -\left( \frac{ \rho_0}{2 r_h} \right)^2 \left( \dot x^t \right)^2 + \frac{ \rho_0 ^2 }{2} \left( \left( \dot x^\theta \right)^2 + \sin^2 x^\theta \left( \dot x^\phi \right)^2 \right)\nonumber\\
    &\quad + r_h^2 y^\theta \sin\left( 2 x^\theta \right) \left( \dot x^\phi \right)^2 + 2r_h^2 \dot x^\theta\dot y^\theta + 2r_h^2\sin^2 \left(  x^\theta \right)\dot x^\phi\dot y^\phi \bigg) \, d\lambda \,.
    \end{align}
\end{subequations}

Variation of $S_{Q,LO}$ yields us the equations of motion for $x^\theta$ and $x^\phi$, given analogously to \cref{eq:LO point-particle geodesics near Schwarzschild BH}
\begin{equation}\label{eq: lo eom particle carroll schwarzschild}
\ddot x^\theta - \sin x^\theta\cos x^\theta\left( \dot x^\phi \right)^2 = 0\,,\quad
\dot x^\phi \,\sin^2 x^\theta = \ell_\phi^{(0)}\,.
\end{equation}

These are the usual geodesics on $S^2$ and are spatial geodesics. For the NLO theory, geodesic equations for $x^t$, $x^\theta$ and $x^\phi$ are
\begin{subequations}
    \begin{align}
        \ddot x^t &= 0\,, \\
        \frac{d}{d\lambda}\left[ \rho_0^2\dot x^\theta + 2r_h^2\dot y^\theta \right]-  \frac{ \rho_0^2 }{2} \sin\left( 2 x^\theta \right)\left( \dot x^\phi \right)^2  & \nn \\
        - 2r_h^2 \sin\left(2 x^\theta\right) \dot x^\phi\dot y^\phi - 2r_h^2  y^\theta \cos \left( 2 x^\theta \right) \left( \dot x^\phi \right)^2 &= 0 \,, \\
        \rho_0^2\sin^2 x^\theta \dot x^\phi + 2r_h^2 y^\theta\sin \left( 2  x^\theta \right) \dot x^\phi + 2 r_h^2\sin^2 \left(  x^\theta \right) \dot y^\phi  &=\ell^{(1)}_\phi \,.
    \end{align}
\end{subequations}

Thus, the solution of $x^t$ is
\begin{equation}\label{eq: xt particle carroll schwarschild}
    x^t (\lambda) = x^t_0 + \omega\lambda \,.
\end{equation}

From this relationship, one can write the LO geodesics as follows
\begin{equation}
\frac{d}{d x^t}\lb \frac{d x^\theta}{d x^t} \rb - \sin x^\theta\cos x^\theta\left( \frac{d x^\phi}{d x^t} \right)^2 = 0\,,\quad
\lb \frac{d x^\phi}{d x^t} \rb\,\sin^2 x^\theta  = \ell_\phi^{(0)}\,.
\end{equation}

From this, we can infer two facets. First, $x^t$ is an affine parameter for the particle geodesics, which are constrained to the constant radius hypersurface on the event horizon of the Schwarzschild black hole. Secondly, with asymptotic time, this particular particle dynamics does not necessarily freeze with coordinate time, which bears a stark difference from the results we have seen in the previous sections.


\section{Conclusions: Towards the horizon with Carroll} \label{Conclusions}

\subsection{Summary}

Black holes are the theoretical laboratory for high energy physicists. In this paper, we have revisited an age-old problem, that of understanding what happens when one nears the horizon of generic black holes. This question has been addressed very carefully for the case of extremal black holes where an exponentially long throat appears and the dynamics of the near horizon decouples from the system. Studies of symmetries and indeed the initial construction of holography in terms of the AdS/CFT correspondence was based on this realisation. A similar de-coupling mechanism is absent for non-extremal black holes which are also the realistic black holes we now observe in the sky. An important question is whether we can find a symmetry based approach to the near horizon of such realistic black holes.  

\medskip

We have answered this question in the affirmative in this paper, and have laid the foundation for a symmetry based formulation for the near horizon of generic non-extremal black holes and black branes. Our main claim is that the near horizon becomes a SC geometry, -- a fibre bundle with a spatial base and 2d Rindler as null fibres. The spacetime is doubly degenerate, with a zero appearing in front of the 2d Rindler. This is a generalization of the usual Carrollian structure which had only one null direction, but the ideas and methods applied to Carrollian physics, now used widely in theoretical physics, generalize in a straightforward manner. 

\medskip

We used these intuitions and earlier lessons learnt from Carrollian symmetries to study carefully the NH region of non-extremal black holes. The first part of the paper dealt with providing a general symmetry based analysis of the structure of the near horizon and probe point particles and scalar fields in the SC geometry. Following this, we looked at an almost exhaustive collection of examples where the same results were reproduced in independent ways providing thorough cross-checks of our general symmetry based analyses, firmly establishing the central role doubly null SC geometries would play in black hole physics going forward. We then formulated how to go from the near horizon to the horizon of the black hole geometries in terms of Carrollian expansions.

\medskip

It is good to keep in mind a caveat in our analysis. As we stated in Sec~\ref{sec:Probes in String Carroll}, all Carrollian expansions, including the SC expansions that we have performed in our analysis in this paper, are contingent to the assumption that all fields are expandable in powers of the (effective) speed of light. There may be (and perhaps will be) situations where this assumption breaks down. Various non-perturbative effects e.g. instanton-like $e^{-1/c^2}$ terms may spoil our ansatz. While these have not shown up in our analysis in the current paper, a systematic investigation as to when one can perform Carroll expansions is an important question going forward \footnote{We thank an anonymous referee for stressing this point.}. 

\subsection{Universality near the horizon}
Near horizon geometries of extremal black holes in $d$-dimensions are $AdS_2 \times \mathbb{S}^{d-2}$. This has been fundamental to the formulation of the AdS/CFT correspondence and other areas of holography. In our paper, we have built on \cite{Bagchi:2023cfp, Bagchi:2024rje, Banerjee:2025bkg} and shown that the near-horizon geometries of non-extremal black holes can be understood in terms of String Carroll geometries, Carrollian structures with two null directions.  

\paragraph{NH geometry and geodesics.} For all non-extremal black objects, the near horizon SC geometry involves a leading Euclidean doubly degenerate metric $h_{\mu\nu}$ reflecting the spatial topology of the object. The subleading metric is Lorentzian and contains a 2d Rindler along with pieces from the spatial metric. For black objects in AdS (dS), this spatial metric components at the subleading order will contain the AdS (dS) radius.

\medskip

We derived geodesics of the near horizon region and showed how the probe point particles in the SC geometry give rise to spatial geodesics to LO and how the 2d Rindler plays an important role in the NLO. This was because the probe action involved the metric and not its inverse. The horizon of the black hole was encoded in the horizon of this 2d Rindler spacetime. 

\paragraph{NH Field theories.} Unlike LO SC geodesics which become spatial, the dynamics of the LO SC scalar fields in all the cases is governed by the 2d Rindler, since now in the case of field theories in non-trivial backgrounds, it is the inverse metric appears in the action and at LO the 2d Rindler metric $v^{\mu\nu}$ contributes to the inverse metric. Some of the interesting features were the following. 

\medskip

{\em{Massless fields in NH}}: A massive scalar field in the full black hole geometry becomes massless to LO in the NH limit indicating that the LO field propagates along the null directions. This ties in nicely with the observation that the Rindler directions are null with respect to the transverse metric, remembering the orthogonality condition $h_{\mu\rho}v^{\rho\nu}=0$. 

\medskip

{\em Correlators and Rindler ultra-locality}: A very direct consequence of the LO 2d Rindler metric is the form of the scalar field propagator. One can simply invert the kinetic piece of the LO actions in the SC framework, e.g. any of the actions given in the main text, like \eqref{eq:LO Scalar Field action near Sch BH} to get 
\begin{align}
G^{(2)}(x^a, x^i) = \braket{\phi^{(0)}(x)\phi^{(0)}(x')}\sim f(|x^a-x'^a|)\delta^{(d-2)}(x^i-x'^i),   
\end{align}
where $x^i$ are the coordinates along the transverse directions and $x^a$ are the coordinates along the longitudinal directions. The most important thing to keep in mind here is the simple fact that the fields in the NH region of a $d$-dimensional black object are also $d$-dimensional. The spatial delta-function indicates that the $d$-dimensional fields localise on the 2d Rindler submanifold. This is a tell-tale sign of Carrollian physics. 

\medskip

The propagators of usual Carrollian field theories \cite{Bagchi:2022emh,carrollstories} (in the electric branch) take the form 
\begin{align}
    G^{(2)}(u, x^i) \sim f(u) \delta^{(d-1)}(x^i-x'^i). 
\end{align}
where $u$ is the null direction and $x^i$ are the $(d-1)$ non-null directions in a $d$-dimensional Carrollian manifold. 

\medskip

The horizon of the black hole is a Carrollian manifold \cite{Penna:2018gfx,Donnay:2019jiz} and appearance of delta-function propagators on the horizon would not be surprising. But at first glance, appearance of these structure in the near horizon geometry may have not been expected. These delta-function propagators in the SC manifold points to ultra-locality of the fields not only on the horizon but also in the near-horizon regime. 

\medskip

This is indicative of another important lesson. At first glance, $d$-dimensional Carrollian correlators may look one-dimensional (dependent on only the null time direction) but are actually representative of $d$-dimensional fields. We know that the ultra-local correlation functions do not mean that d-dimensional Carrollian theories are effective 1d theories. In an analogous way, the SC correlators look 2d, but the spatial directions would play a crucial role in the understanding of the field theories and {\emph{should not be neglected}}. 

\medskip

The ultralocality of the correlators in the NH region suggests that one should be able to smoothly transition from the SC structures in the near horizon region to the Carrollian structures on the horizon of the black object. We have made preliminary remarks on how to achieve this in \cref{sec:Closing onto the horizon}. 

\subsection{Comments and future directions}

We now end with a list of immediate questions and projects that we would like to address. 

\begin{itemize}
    \item{\em Spinning fields near BHs}: In our analysis of classical fields near BHs, we have only considered scalar fields so far. The natural generalisation is to include fields with spin e.g. fermions, and gauge fields. Carroll fermions come hand in hand with modified Clifford algebras and a lot of interesting structure \cite{Bagchi:2022eui}. The modifications of these structures for SC geometries would perhaps be even more intriguing. When considering gauge fields, the natural setting would be to build this formulation for charged black holes. We will consider these generalisations in upcoming work. 
    
    \item {\em Extremal BHs and Carroll}: As has been stressed, the NH regions of extremal BHs have been studied in a lot of detail in the existing literature and of late there also has been a lot of work in trying to understand near extremal BHs and their NH symmetries. One of the immediate questions for our framework is figuring out how the structures we have encountered here, viz. the SC geometry, naturally evolves to a Lorentzian geometry on take an extremal limit. 
    
    \item {\em Quantum fields near BHs}: Perhaps the most important question in the context of black hole physics is the understanding of the quantization of fields near BHs, now armed with these new structures. We would like to revisit the famous works of Unruh and Hawking, now with the insight of SC structures. Just as Carrollian quantum field theories are very different from quantum mechanics, even though there may be superficial structural similarities, it is expected that 4d QFTs in a SC background would have many features very different from a relativistic QFT in 2d Rindler spacetime while retaining some similarities. We will begin our investigations for scalar field theories before generalising to QFTs with spinning fields. 
    
    \item {\em  Isometries of NH}: As we emphasised at the beginning of the paper, Carrollian manifolds come equipped with an infinite dimensional isometry algebra because of the null structure. It is expected that such infinite dimensional isometries are generic features of all higher Carroll structures and in particular would also show up in SC manifolds and hence BH NH geometries. 
    
    \item {\em Asymptotic symmetries and connections with Kerr/CFT}: For near horizon extreme Kerr (NHEK), the asymptotic symmetries were investigated and a dual description of the extreme Kerr BH was proposed in terms of a 2d relativistic CFT, where a Cardy counting of the states reproduced the entropy of the Kerr BH. This holographic correspondence is known as the Kerr/CFT correspondence \cite{PhysRevD.80.124008,Bredberg:2009pv,Compere:2012jk}. In an attempt to build a similar holographic correspondence, now for a non-extremal BH, we will investigate the asymptotic symmetries of the NH region. This would be the asymptotic symmetries of a SC geometry which itself would have infinite dimensional isometries. The structures we encounter and the asymptotic analysis itself would be very rich and rewarding. We should mention here that there is already quite a large body of literature on understanding the nature of near horizon symmetries of black holes following \cite{Donnay:2015abr, Hawking:2016msc, Donnay:2016ejv, Grumiller:2019fmp}. It would be good to compare our SC approach with this existing literature.  

     \item {\em Hyperscaling violating Lifshitz singularities}:  As we have seen, non-extremal black holes with Lifshitz/hvLif asymptotics continue to exhibit SC expansions, similar to other cases: this is consistent with other physics aspects that are known to be reasonable at finite temperature \cite{Taylor:2015glc,Hartnoll:2016apf}.
     A fascinating question pertains to the deep IR singularities that arise in 
     these theories at zero temperature. In the recasting (\ref{hvL-2d}), we see that the transverse space scales nontrivially towards the horizon for $z>1$: this might suggest novel structures beyond the finite temperature Carroll expansion we have seen. It would be very interesting to understand this zero temperature limit here and possible degenerate Carroll-like structures which underlie Lifshitz singularities. Note that this is likely to be somewhat different from the zero temperature limit for 
     {\em charged} hvLif theories which give rise to $AdS_2$ factors and JT gravity \cite{Kolekar:2018sba}.
\end{itemize}
We hope to report on some of these intriguing points in the near future.

\subsection*{Acknowledgements}
We thank Priyadarshini Pandit for initial collaboration, Pushkar Soni for his assistance with the expansion of the determinants and Deepanshu Bisht for his help with the figures.  We would also like to thank Aritra Banerjee, Stephane Detournay, Daniel Grumiller, Jelle Hartong, Emil Have, Kedar Kolekar, Alok Laddha, Mangesh Mandlik, Joan Simon, Abhinava Bhattacharjee and Ansh Mishra for discussions on related issues. 

\medskip

KN thanks the warm hospitality of the IIT Kanpur String Group for a fun visit and numerous discussions in the early stages of this collaboration.
AB and SRI thank Chennai Mathematical Institute and its members for warm hospitality and for fruitful discussions during a visit in the middle of the collaboration. 

\medskip

AB is partially supported by a Swarnajayanti Fellowship from the Science and Engineering Research Board (SERB) under grant SB/SJF/2019-20/08 and also by an ANRF grant CRG/2022/006165. Arkachur and SRI are supported by the IIT Kanpur Institute Assistantship. The work of KN is partially supported by a grant to CMI from the Infosys Foundation.

\bigskip \bigskip

\appendix

\section*{APPENDICES}

\section{Carroll symmetries}\label{Carroll-sym}
Historically, Carrollian symmetries were obtained in the vanishing speed of light $c\to0$ limit of relativistic Poincare symmetries \cite{Leblond65, SenGupta:1966qer}. The limit, in terms of group theory, is an In{\"o}n{\"u}-Wigner contraction. The generators of the underlying symmetry algebra can be written as 
\begin{subequations}\label{Carr-gen}
\begin{align}
&\text{Null ``time'' translation:} \quad H =\partial_u, \\ 
&\text{Carroll boosts:} \quad C_i=x_i\partial_u \\
&\text{Spatial rotations:} \quad J_{ij}=-x_i\partial_j+x_j\partial_i \\
&\text{Spatial translation:} \quad T_i=\partial_i. 
\end{align} 
\end{subequations}
These generators close to form the Carroll algebra: 
\begin{align}\label{Car-al}
[J_{ij},J_{kl}]=\delta_{ik}J_{jl}-\delta_{jk}J_{il}+\delta_{jl}J_{ik}-\delta_{il}J_{jk}\,,\quad[T_i,C_j]=\delta_{ij}H\,,\nonumber\\
[J_{ij},T_k]=\delta_{ki}T_{j}-\delta_{kj}T_{i}\,,\quad[J_{ij},C_k]=\delta_{ki}C_{j}-\delta_{kj}C_{i}\,.
\end{align}
We now remind the reader of the definition of a Carroll manifold. We start with the Minkowski spacetime and send $c\to0$. As mentioned in the intrduction, the metric and its inverse degenerate in the limit and give us \eqref{carr-met}. It is clear from here that $h_{\mu\nu} \theta^\nu = 0$. A flat Carroll manifold is thus defined with both $(h_{\mu\nu}, \theta^\nu)$ make up the structure. We now define a general Carroll structure based on these ideas. 

\medskip

As defined in the main body of our paper, a $d$-dimensional Carrollian manifold is a smooth manifold with a degenerate metric $h_{\mu\nu}$ and a nowhere-vanishing vector field $\theta^\mu$ that generates the kernel of the metric. For flat Carroll manifolds, the isometries of this structure \eqref{Carr-Kill} generates the Carroll algebra. The vector field is explicitly given by \eqref{xi} where $\omega^i_{\, j}$ denotes spatial rotations, $a^i$ spatial translations and $a^0$ temporal translations. $f(x)$ is a function that is a priori unfixed hence generically there is an infinite dimensional symmetry for Carrollian manifolds in all spacetime dimensions. We recover the finite Carrollian algebra \eqref{Car-al} when we restrict the function $f(x)$ to be linear in $x$. 

\medskip

A flat $(d+1)$-dimensional Carrollian manifold is a fibre-bundle $\mathbb{R}_u \times \mathbb{R}^d$, where $\mathbb{R}^d$ forms a d-dimensional spatial base and the null direction $\mathbb{R}_u$ forms ``temporal'' fibres. On top of the finite symmetry transformations  mentioned above in \eqref{Carr-gen}, when $f(x)$ in \eqref{xi} is unrestricted, we will have infinite dimensional ``supertranslations'' which are translations of the null directions dependent on the spatial coordinates. The generator of these supertranslations is 
\begin{align}
    M(f) = f(x^i) \partial_u 
\end{align}

The supertranslations form an abelian subalgebra:
\begin{align}
    [M(f), M(g)]=0. 
\end{align}
The geometric structure that we have encountered here is a fibre-bundle with a $d$-dimensional spatial base and one dimensional null ``temporal'' fibres. 

\medskip

Let us contrast Carrollian manifolds with the perhaps more familiar Galilean set-up. In the non-relativistic world, where the speed of light goes to infinity instead of zero, as mentioned in the introduction, it is the upper version of the metric which survives c.f. \eqref{1.1}. 
A $(d+1)$ dimensional Galilean or a Newton-Cartan manifold is thus given by $(\h^{\mu\nu}, \tau_\nu)$ subject to \begin{align}\label{htnc}
\h^{\mu\nu} \tau_\nu = 0.
\end{align}
This is a fibre-bundle with 1d base which is the time direction and $d$-dimensional spatial fibres. The well-known Galilean algebra arises as the isometry algebra of flat Newton-Cartan manifolds. As we can now see, Carrollian spacetimes are the base$\leftrightarrow$fibre flipped version of these more familiar Galilean (or Newton-Cartan) structures. Potential dualities between Galilean and Carrollian theories are currently being explored \cite{Duval:2014uoa, Figueroa-OFarrill:2022pus}.

\section{Diffeomorphisms and string-Carroll Transformations}\label{app: string-Carroll expansion}
Expanding the background metric in powers of $c^2$, alters the local structures. In this section, we will derive the transformation properties of the decomposed vielbeine under local tangent space transformations and diffeomorphisms upto the first order.   

\medskip

The vielbeine $E^\m_{A'}$ and $E_\m^{A'}$ transform under diffeomorphisms $\zeta$ and local Lorentz transformations $L^{A'}_{~~~B'}$ as follows 
\begin{eqnarray}\label{eq:transformation of vielbeine}
    \delta E^{A'}_\m=\pounds_\zeta E^{A'}_\m+L^{A'}_{~~~B'}E^{B'}_\m\,,~~~~\delta E_{A'}^\m=\pounds_\zeta E_{A'}^\m+L^{~~~B'}_{A'}E_{B'}^\m\,.
\end{eqnarray}
The local Lorentz transformations can be decomposed into longitudinal and transverse components in the following way 
\begin{eqnarray}
    \begin{aligned}
L^{A'}_{~~~B'}&=\delta^{A'}_{\bar{A}}\delta^{\bar{B}}_{B'}L^{\bar{A}}_{~~~\bar{B}}+\delta^{A'}_{A}\delta^{B}_{B'}L^{A}_{~~~B}+\delta^{A'}_{\bar{A}}\delta^{B}_{B'}L^{\bar{A}}_{~~~B}+\delta^{A'}_{A}\delta^{\bar{B}}_{B'}L^{A}_{~~~\bar{B}}\,.
    \end{aligned}
\end{eqnarray}
Next, we introduce factors of $c$ such that SC symmetries emerge when $c$ is dialled to 0
\begin{eqnarray}
L^{A'}_{~~B'}&=\delta^{A'}_{\bar{A}}\delta^{\bar{B}}_{B'}\lambda^{\bar{A}}_{~~\bar{B}}+\delta^{A'}_{A}\delta^{B}_{B'}L\epsilon^{A}_{~~B}+c\delta^{A'}_{\bar{A}}\delta^{B}_{B'}\lambda^{\bar{A}}_{~~B}+c\delta^{A'}_{A}\delta^{\bar{B}}_{B'}\lambda^{A}_{~~\bar{B}}\,. 
\end{eqnarray}
Note that we wrote $L^A{_B} = L\epsilon^A{_B}$ for the parameters of the local Lorentz transformation in the longitudinal sector, where $\epsilon^A{_B} = \eta^{AC}\epsilon_{CB}$ is the two-dimensional Levi-Civita symbol. We also note the relation
\begin{equation}
    \lambda^A{_{B'}} = \eta^{AB} \delta_{A'B'} \lambda_B{^{A'}}\,.
\end{equation}
The decomposed vielbeine are related to the vielbeine $E^{A'}_\m$ and $E_{A'}^\m$ as,
\begin{eqnarray}
E^{A'}_\m=cT^A_\m\delta^{A'}_A+E^{\bar{A}}_\mu\delta^{A'}_{\bar{A}}\,,~~E_{A'}^\m=\frac{1}{c}V_A^\m\delta_{A'}^A+E_{\bar{A}}^\mu\delta_{A'}^{\bar{A}}\,.
\end{eqnarray}
The transformation of the decomposed vielbeine can be found using the transformation \eqref{eq:transformation of vielbeine} as follows
\begin{eqnarray}\label{eq:transformation of the preCarrollian variables}
    \begin{aligned}
        \delta T_\mu^A&=\pounds_\zeta T_\mu^A+L\epsilon^A_{~~B}T^B_\mu+\lambda^A_{~~\bar{B}}E_\mu^{\bar{B}}\,,\\
        \delta E_{\mu}^{\bar{A}}&=\pounds_\zeta E_\mu^{\bar{A}}+\lambda^{\bar{A}}_{~~\bar{B}}E^{\bar{B}}_\m+c^2\lambda^{\bar{A}}_{~~B}T^B_\m\,,\\
        \delta V^\mu_A&=\pounds_\zeta V^\m_A+L\epsilon_A^{~~B}V^\m_B+c^2\lambda_{A}^{~~\bar{B}}E_{\bar{B}}^\m\,,\\
        \delta E^{\m}_{\bar{A}}&=\pounds_\zeta E^{\m}_{\bar{A}}+\lambda_{\bar{A}}^{~~\bar{B}}E_{\bar{B}}^\mu+\lambda_{\bar{A}}^{~~B}V_B^\m\,.
    \end{aligned}
\end{eqnarray}

The expansion of the diffeomorphisms and local Lorentz transformations can be expressed as a systematic expansion in the parameter $c$ as 
\begin{eqnarray}\label{eq:expansions of the transformations}
\begin{aligned}
    \zeta^\mu&=\zeta^\mu_{(0)}+c^2\zeta^\mu_{(2)}+\mathcal{O}(c^4)\,,\\
    L&=L_{(0)}+c^2L_{(2)}+\mathcal{O}(c^4)\,,\\
    \lambda^{\bar{A}}_{~~\bar{B}}&=\lambda^{\bar{A}}_{~(0)\bar{B}}+c^2\lambda^{\bar{A}}_{~(2)\bar{B}}+\mathcal{O}(c^4)\,,\\
    \lambda^{\bar{A}}_{~~B}&=\lambda^{\bar{A}}_{~(0)B}+c^2\lambda^{\bar{A}}_{~(2)B}+\mathcal{O}(c^4)\,.
\end{aligned}
\end{eqnarray}
Here, $\zeta^\mu_{(0)}$ generate the LO diffeomorphisms and $\zeta^\mu_{(2)}$ generate the NLO diffeomorphisms. $L_{(0)}$ is the LO local Lorentz transformation (LO SC boost) along the longitudinal direction while $L_{(2)}$ is the next-to-leading order. Likewise, $\lambda^{\bar{A}}_{~(n)B}$ are the Carroll boosts and $\lambda^{\bar{A}}_{~(n)\bar{B}}$ are the $SO(d)$ rotations at the nth order. The transformation of the decomposed vielbeine at every order can be derived by substituting \eqref{eq:precarrollian variables expansion stringCarroll geometry} and \eqref{eq:expansions of the transformations} into \eqref{eq:transformation of the preCarrollian variables}
\begin{eqnarray}\label{eq:transformations}
    \begin{aligned}
        \delta \tau_\mu^A&=\pounds_{\zeta_{(0)}}\tau_\mu^A+L_{(0)}\epsilon^A_{~~B}\tau^B_\mu+\lambda^A_{~(0)\bar{B}}e_\mu^{\bar{B}}\,,\\
        \delta e_{\mu}^{\bar{A}}&=\pounds_{\zeta_{(0)}}e_\mu^{\bar{A}}+\lambda^{\bar{A}}_{~(0)\bar{B}}e^{\bar{B}}_\m\,,\\
        \delta \pi_{\mu}^{\bar{A}}&=\pounds_{\zeta_{(0)}}\pi_\mu^{\bar{A}}+\pounds_{\zeta_{(2)}}e_\mu^{\bar{A}}+\lambda^{\bar{A}}_{~(0)\bar{B}}\pi^{\bar{B}}_\m+\lambda^{\bar{A}}_{~(2)\bar{B}}e^{\bar{B}}_\m+\lambda^{\bar{A}}_{~(0)B}\tau^B_\m\,,\\
        \delta v^\mu_A&=\pounds_{\zeta_{(0)}} v^\m_A+L_{(0)}\epsilon_A^{~~B}v^\m_B\,,\\
        \delta m^\mu_A&=\pounds_{\zeta_{(0)}}m^\m_A+\pounds_{\zeta_{(2)}} v^\m_A+L_{(0)}\epsilon_A^{~~B}m^\m_B+\lambda_{(0)A}^{~~~~~\bar{B}}e_{\bar{B}}^\m+L_{(2)}\epsilon_A^{~~B}v^\mu_B\,,\\
         \delta e^{\m}_{\bar{A}}&=\pounds_{\zeta_{0}} e^{\m}_{\bar{A}}+\lambda_{(0)\bar{A}}^{~~~~~\bar{B}}e_{\bar{B}}^\mu+\lambda_{(0)\bar{A}}^{~~~~~B}v_B^\m\,.
    \end{aligned}
\end{eqnarray}
From these transformations, it can be deduced that $h_{\mu\nu}$ and $v^{\mu\nu}$ are invariant under tangent space SC transformations. The particle-Carroll expansion and the transformation of the corresponding vielbeine can be deduced by suppressing one of the directions along the longitudinal space.

\section{Particle Carroll Expansion}\label{app:Particle Carroll Expansion}
The particle Carroll expansion is an expansion of a Lorentzian metric in orders of $c$ where only a time-like direction is singled out. Therefore, we partition a d+1 dimensional spacetime into a d-dimensional transverse one-dimensional longitudinal subspaces. As shown in \cite{Hansen:2021fxi}, any d+1-dimensional Lorentzian metric $g_{\mu\nu}$ and its inverse can be decomposed in terms of pre-Carrollian variables as
\begin{eqnarray}\label{eq:meric decomposition}
    g_{\m\n}=-c^2T_{\mu}T_{\nu}+\Pi_{\m\n}\,,~~g^{\m\n}=-\frac{1}{c^2}V^\mu V^\nu+\Pi^{\m\n}\,,
\end{eqnarray}
such that, 
\begin{eqnarray}
    \Pi^{\m\n}T_{\m}=0\,,~~~\Pi_{\m\n}V^\m=0\,.
\end{eqnarray}
Here, $\Pi_{\m\n}$ is a non-degenerate metric in the transverse space. In the full spacetime $\Pi_{\mu\nu}$ is a degenerate metric. The metric $\Pi$ can be decomposed into vielbeine as follows 
\begin{eqnarray}
    \Pi_{\m\n}=\delta_{AB}E^A_\m E^B_\n\,,~~~~\Pi^{\m\n}=\delta^{AB}E_{A}^\m E^\n_{B}\,.
\end{eqnarray}
Here the indices $A,B$ are along the transverse directions. Now, each pre-Carrollian variable can be expanded in orders of small $c^2$ as 
\begin{eqnarray}\label{eq:expansion of pre-Carrollian variables in PCE}
\begin{aligned}
    T_\m &=\tau_\mu+\mathcal{O}(c^2)\,,\\
    E^A_\m &=e^A_\mu+c^2\pi^A_\mu+\mathcal{O}(c^4)\,,\\
    V^\m &=v^\m+c^2m^\m+\mathcal{O}(c^2)\,,\\
    E^\m_A &=e^\m_A+\mathcal{O}(c^2)\,.
\end{aligned}
\end{eqnarray}
Rewriting the metric $g_{\m\n}$ and its inverse in terms of these expanded pre-Carrollian variables we get a $c^2$ expansion of the $g_{\m\n}$ as follows
\begin{eqnarray}\label{eq:particle Carroll expansion}
g_{\m\n}=h_{\m\n}-c^2\tau_\m\tau_\n+c^2\Phi_{\m\n}+\cdots\,,~~~g^{\m\n}=-\frac{1}{c^2}v^\m v^\n+\bar{h}^{\m\n}+\cdots\,.
\end{eqnarray}
The metric components appearing in the above expansion are defined as 
\begin{eqnarray}
        &h_{\m\n}=\delta_{AB}e^A_\m e^B_\n\,,~~\Phi_{\m\n}=2\delta_{AB}e_{(\m}^A\pi^B_{\n)}\,,~~\bar{h}^{\m\n}=\delta^{AB}e^\m_A e^\n_B-2v^{(\m}m^{\n)}\,.
\end{eqnarray}

\section{Scalar fields on flat Carroll background}\label{App:scalar field on Carroll background}
In this appendix, we will derive the scalar field Lagrangians on a Carroll background, following \cite{deBoer:2021jej,carrollstories}, and compare it to the scalar field Lagrangians on the SC background derived in section \ref{sec:Scalar Field}. A scalar field action on flat Minkowski background is given as 
\begin{eqnarray}\label{eq:flat scalar field}
    S=-\frac{1}{2}\int d^4x\left(\eta^{\mu\nu}\partial_\mu\Phi\partial_\nu\Phi+m^2\Phi^2\right)\,.
\end{eqnarray}
To obtain the leading and the next-to-leading order actions, we will scale the time coordinate $t\to\epsilon t$ and expand the scalar field as 
\begin{eqnarray}
\Phi=\phi^{(0)}+\epsilon^2\phi^{(1)}+\cdots\,.
\end{eqnarray}
Substituting the scalar field expansion and the scaled time coordinate into the action \eqref{eq:flat scalar field}, we get the following leading and next-to-leading order actions 
\begin{subequations}
    \begin{align}
        S_{LO}&=\frac{1}{2}\int d^4x\partial_t\phi^{(0)}\partial_t\phi^{(0)}\,,\label{eq:LO flat particle Carroll}\\
        S_{NLO}&=\frac{1}{2}\int d^4x\left(2\partial_t\phi^{(0)}\partial_t\phi^{(1)}-\partial_i\phi^{(0)}\partial_i\phi^{(0)}-m^2\left(\phi^{(0)}\right)^2\right)\,\label{eq:NLO flat particle Carroll}.
    \end{align}
\end{subequations}
Note that the LO action \eqref{eq:LO flat particle Carroll} is Carroll invariant, while the NLO action \eqref{eq:NLO flat particle Carroll} is not Carroll invariant. However, the NLO action can be made Carroll invariant by introducing a Lagrange multiplier and assigning it the right transformation \cite{deBoer:2021jej,carrollstories}.

\medskip

The LO SC action \eqref{leading} involves derivatives of the scalar field along the longitudinal directions, whereas the LO particle-Carroll action \eqref{eq:LO flat particle Carroll} singles out only the time coordinate. This distinction persists at the level of the NLO actions as well. The only additional term in \eqref{nextleading} arises from the NLO contribution to the determinant. A similar term also appears in the NLO scalar field action on a curved Carroll background. The NLO scalar field $\phi^{(1)}$ in both cases, does not transform as a scalar. In the Carroll case it transforms under Carroll boosts as 
\begin{eqnarray}
\delta\phi^{(1)}=\vec{c}\cdot\vec{x}\partial_t\phi^{(1)}+t\vec{c}\cdot\vec{\partial}\phi^{(0)}\,,
\end{eqnarray}
where, $\vec{c}$ is the boost parameter. In the SC case, the NLO scalar field transforms under diffeomorphisms as
\begin{eqnarray}
\delta_{\zeta}\phi^{(1)}=\zeta_{(0)}^\mu\partial_{\mu}\phi^{(1)}+\zeta_{(2)}^\mu\partial_{\mu}\phi^{(0)}\,.
\end{eqnarray}

\section{Rotating BTZ Black hole}\label{D}
We now move on beyond asymptotically flat spacetimes. To consider a very different example, let us take the Ba\~nados-Teitelboim-Zanelli (BTZ) black holes in three-dimensional Anti de Sitter (AdS$_3$) spacetimes. Gravity in three dimensions is of course known to have no propagating degrees of freedom and indeed the presence of non-trivial black hole solutions was a big surprise when these were initially discovered. It was soon understood that these black holes were orbifolds of AdS$_3$. We will see below that our formulation of SC geometries encompasses the near horizon of such black holes as well. For making the analysis even more non-trivial, we will consider rotating BTZ black holes.

\subsection{Explicit map}\label{sec:Explicit map (Rotating BTZ Black hole)}

As before, we begin with the metric. The metric of a 2+1 dimensional rotating BTZ black hole with mass $M>0$ and angular momentum $J \neq 0$ in the Boyer-Lindquist-like coordinate system is given as
\begin{equation}\label{eq: rotating BTZ}
	ds^2 = -N^2(r)\, dt^2 + \frac{dr^2}{N^2(r)} + r^2 \left(N^\phi(r) dt + d\phi \right)^2\,,
\end{equation}
where $N^2(r)$ and $N^\phi(r)$ are the lapse and the angular drag functions, respectively, and are given as follows
\begin{equation}
    N^2(r) = -M + \frac{r^2}{\ell^2} + \frac{J^2}{4r^2}\,,\quad N^\phi (r)= -\frac{J}{2r^2}\,.
\end{equation}
The rotating BTZ black hole has two event horizons at $r_\pm$. 

\medskip

Throughout the following analysis, we will work with a non-extremal rotating BTZ black hole. In order to write the near-horizon expansion of this metric in a diagonal coordinate system up to $\mathcal{O}(\epsilon)$, we do the following coordinate transformations
\begin{equation}\label{eq: non extremal coordinate transform}
t\to T=\left(1+\frac{J^2\alpha_+}{2r_+^3}  \right)t - \left(\frac{J \alpha_+}{r_+}\right) \phi,~~~\phi \to \varphi= \phi - \frac{J}{2r_+^2} t,
\end{equation}
where
\begin{equation} \label{alpha plus}
	    \alpha_+ = \lb \frac{2r_+}{\ell^2} - \frac{J^2}{2r_+^3} \rb^{-1} = \frac{\ell^2 r_+}{2 \left(r_+^2-r_-^2\right)} \, .
\end{equation}
In this coordinate system, the metric diagonalises as follows
\begin{equation}
    ds^2 = -\frac{ \left(r^2-r_+^2\right) \left(r_+^2-r_-^2\right)}{\ell^2 r_+^2}dT^2 + \frac{ \ell^2 r^2}{\left(r^2-r_-^2\right) \left(r^2-r_+^2\right)} dr^2 + \frac{ r_+^2 \left(r^2-r_-^2\right) }{r_+^2-r_-^2} d\varphi^2 \,.
\end{equation}
A few remarks are in order regarding the new coordinate system \cite{Krishnan:2009kj}. Firstly, note that we are transitioning to a co-rotating frame, which rotates along with the outer horizon. The new coordinates do not suffer from any causal pathologies outside the outer horizon. In the usual interpretation of the AdS space as a cylinder, the time coordinate chalks out a helix. We can diagonalise the rotating BTZ black hole metric as a consequence of the fact that the spacetime is locally AdS.       

Next, we do the near-horizon probe
\begin{equation}
    r = r_+ + \lb\frac{\epsilon}{4 \alpha_+}\rb\rho^2\,.
\end{equation}
The resulting metric becomes
\begin{equation}\label{eq: btz non extremal expansion}
	ds^2 = r_+^2 d\varphi^2 + \epsilon 
		\left(
		-\left(\frac{\rho^2}{4\alpha_+^2}\right) dT^2 
		+  d\rho^2 
		+ \left( \frac{r_+^2}{\ell^2} \right) \rho^2 d\varphi^2
		\right)
		+ \mathcal{O}(\epsilon^2)\,.
\end{equation}
The near-horizon metric can be mapped to the SC metric \eqref{SCmetric} by identifying $c^2=\epsilon$ as follows
\begin{eqnarray}\label{eq:rotating BTZ BH Map}
    h_{\mu\nu}dx^\mu dx^\nu = r_+^2 d\varphi^2\,,~~
            \Phi_{\mu\nu} dx^\mu dx^\nu = \frac{r_+^2\rho^2}{\ell^2}d\varphi^2,~~\tau_{\mu\nu}dx^\mu dx^\nu=-\frac{\rho^2}{4\alpha_+^2}dT^2 
		+  d\rho^2 \,.
\end{eqnarray}
The inverse near-horizon metric can be mapped likewise
\begin{eqnarray}\label{eq:rotating BTZ BH Map inverse}
    v^{\mu\nu}\partial_\mu\partial_\nu=-\frac{4\alpha_+^2}{\rho^2}\partial_T^2+\partial_\rho^2\,,~~~\bar{h}^{\mu\nu}\partial_\mu\partial_\nu=\frac{1}{r_+^2}\partial_\varphi^2+\frac{\alpha_+}{2r_+}\partial_T^2+\frac{\rho^2(3r_-^2+r_+^2)}{4\ell^2r_+^2}\partial_\rho^2\,.
\end{eqnarray}

\subsection{Geodesics of BTZ: using the map}

The LO and NLO point particle actions in the near-horizon limit (about outer horizon) of a non-extremal BTZ black hole in the co-rotating frame $(T,r,\varphi)$ are given as
\begin{subequations}\label{eq: point-particle action in NE BTZ}
    \begin{align}
        S_{LO} &= -\frac{m r_+^2}{2}\int \lb \dot x^\varphi \rb^2\,d\lambda\, , \label{eq: LO point-particle action in NE BTZ} \\
        S_{NLO} &= -\frac{m}{2} \int \left[  -\lb \frac{\xr}{2\alpha_+}\rb^2 \lb \dot x^T \rb^2 + \lb \frac{r_+\xr}{\ell} \rb^2 \lb \dot x^\varphi \rb^2 + 2r_+^2 \dot x^\varphi \dot y^\varphi  \right] \, d\lambda \,. \label{eq: NLO point-particle action in NE BTZ}
    \end{align}
\end{subequations}

Variation with respect to $\xff$ gives us
\begin{equation}\label{eq:LO point-particle geodesics near NE BTZ}
    \ddot x^\varphi = 0 \, ,
\end{equation}

which is solved by
\begin{equation}\label{eq:LO point-particle geodesics solutions near NE BTZ}
    \xff(\lambda) = \xff_0 + \ell_\varphi^{(0)} \lb \lambda - \lambda_0 \rb \,.
\end{equation}

Observe that $\ell_\varphi^{(0)}$ is an integral of motion for our system. Now, we vary \eqref{eq: NLO point-particle action in NE BTZ} with respect to $\xff$, $\xtt$ and $\xr$ to get the NLO equations of motion
\begin{subequations}\label{eq: NLO point-particle geodesics near NE BTZ}
    \begin{align}
        \frac{d}{d\lambda} \lb \dot y^\varphi + \lb \frac{\xr}{\ell}\rb^2 \dot x^\varphi \rb &= 0\,, \label{eq: NLO point-particle geodesics 1 near NE BTZ} \\
        \lb \xr \rb^2 \dot x^T &= E^{(1)}\, , \label{eq: NLO point-particle geodesics 2 near NE BTZ}\\
        \ddot x^\rho + \frac{\xr}{4\alpha_+^2}\lb \dot x^T \rb^2 - \frac{r_+^2 \xr}{\ell^2} \lb \dot x^\varphi \rb^2 &= 0 \,. \label{eq: NLO point-particle geodesics 3 near NE BTZ}
    \end{align}
\end{subequations}

\eqref{eq: NLO point-particle geodesics 2 near NE BTZ} and \eqref{eq: NLO point-particle geodesics 3 near NE BTZ} are solved by
\begin{subequations}
    \begin{align}
        \xr(\lambda) & = \sqrt{\xr_0 \pm A \sinh f(\lambda)  }\,, \label{eq: NLO point-particle geodesics solutions 1 near NE BTZ}  \\
        \xtt(\lambda) &= \xtt_0 \pm \alpha_+ \ln \left[ \frac{\lb \pm \xr_0 + \frac{\ell E^{(1)}}{2\alpha_+r_+\ell_\varphi^{(0)}} \rb e^{f(\lambda)} - A }{ \lb \mp \xr_0 + \frac{\ell E^{(1)}}{2\alpha_+r_+\ell_\varphi^{(0)}} \rb e^{f(\lambda)} + A } \right] , \label{eq: NLO point-particle geodesics solutions 2 near NE BTZ}
    \end{align}
\end{subequations}

where $f(\lambda)=\frac{2 r_+ \,\ell_\varphi^{(0)}}{\ell}\lb \lambda - \lambda_0 \rb$ and $A = \sqrt{ \lb \frac{\ell E^{(1)}}{2\alpha_+r_+\ell_\varphi^{(0)}} \rb^2 - \lb \xr_0 \rb^2 }$. Solutions \eqref{eq: NLO point-particle geodesics solutions 1 near NE BTZ}, \eqref{eq: NLO point-particle geodesics solutions 2 near NE BTZ} and \eqref{eq:LO point-particle geodesics solutions near NE BTZ} can be written in the usual non-corotating frame of BTZ as follows
\begin{subequations}
    \begin{align}
        \xt(\lambda) &= x^t_0 + \frac{J\ell\alpha_+}{2r_+^2}f(\lambda) \pm \alpha_+ \ln \left[ \frac{\lb \pm \xr_0 + \frac{\ell E^{(1)}}{2\alpha_+r_+\ell_\varphi^{(0)}} \rb e^{f(\lambda)} - A }{ \lb \mp \xr_0 + \frac{\ell E^{(1)}}{2\alpha_+r_+\ell_\varphi^{(0)}} \rb e^{f(\lambda)} + A } \right] \,,  \\
        \xr(\lambda) & = \sqrt{  \xr_0 \pm A \sinh f(\lambda)  }\,, \\
        \xf (\lambda) &= \xf_0 + \ell_\varphi^{(0)} \lb \lambda - \lambda_0 \rb + \frac{J}{2r_+^2} x^t(\lambda) \,,
    \end{align}
\end{subequations}

where $\xt_0 = \xtt_0 + \frac{J\alpha_+}{r_+}\xff_0$  and  $\xf_0 = \xff_0 + \frac{J}{2r_+^2}\xt_0$.

\subsubsection*{Intrinsic geodesic equations in non-corotating frame}

In the non-corotating frame of a rotating non-extremal BTZ black hole \eqref{eq: rotating BTZ}, the near-horizon metric looks as
\begin{equation}
    ds^2 = r_+^2 \lb d\phi - \frac{J}{2r_+^2} dt \rb^2 + \epsilon \left[ -\lb \frac{r_+ \rho^2}{2\alpha_+} \rb dt^2 + d\rho^2 + \lb \frac{r_+ \rho^2}{2\alpha_+} \rb d\phi^2  \right] + \mathcal O (\epsilon^2)\,.
\end{equation}

In this coordinate system the LO and NLO quadratic point particle actions looks as
\begin{subequations}\label{eq: point-particle action in NE BTZ NCR}
    \begin{align}
        S_{LO} &= - \frac{mr_+^2}{2} \int \lb \dot x^\phi - \frac{J}{2r_+^2} \dot x^t\rb^2 d\lambda \,, \label{eq: LO point-particle action in NE BTZ NCR} \\
        S_{NLO} &= -\frac{m}{2} \int \Bigg[ -\lb \frac{r_+  (\xr)^2 }{2\alpha_+} \rb \lb \dot x^t \rb^2 + \lb \dot x^\rho \rb^2 + \lb \frac{r_+ (\xr)^2}{2\alpha_+} \rb \lb \dot x^\phi \rb^2 \nn \\
        &\quad + 2r_+^2 \lb \dot x^\phi - \frac{J}{2r_+^2} \dot x^t \rb \lb \dot y^\phi - \frac{J}{2r_+^2} \dot y^t \rb \Bigg] \, d\lambda \,. \label{eq: NLO point-particle action in NE BTZ NCR}
    \end{align}
\end{subequations}

The LO equations of motion for \eqref{eq: LO point-particle action in NE BTZ NCR} for $x^t$ and $x^\phi$ are the same. It is given below
\begin{equation}\label{eq: LO point-particle geodesics near NE BTZ NCR}
    \frac{d}{d\lambda} \lb \dot x^\phi - \frac{J}{2 r_+^2} \dot x^t \rb = 0 \,.
\end{equation}

This is equivalent to LO equation of motion \eqref{eq:LO point-particle geodesics near NE BTZ} in corotating frame, once we make the proper identification
\begin{equation}\label{eq: NCR BTZ relation 1}
    \xff(\lambda) = \xf(\lambda) - \frac{J}{2r_+^2} \xt(\lambda)\,.
\end{equation}

The equations of motion for $\xf$, $\xt$ and $\xr$ for the NLO action \eqref{eq: NLO point-particle action in NE BTZ NCR} are 
\begin{subequations}\label{eq: NLO point-particle geodesics near NE BTZ NCR}
    \begin{align}
        \frac{d}{d\lambda} \left[ \lb \frac{r_+ \lb \xr\rb^2}{\alpha_+} \rb \dot x^\phi + 2r_+^2 \lb \dot y^\phi - \frac{J}{2r_+^2} \dot y^t \rb  \right] & = 0 \,, \label{eq: NLO point-particle geodesics 1 near NE BTZ NCR} \\
        \frac{d}{d\lambda} \left[ \lb \frac{r_+\lb \xr\rb^2}{\ell^2 \alpha_+} \rb \dot x^t + J \lb \dot y^\phi - \frac{J}{2r_+^2} \dot y^t \rb  \right] &= 0 \,. \label{eq: NLO point-particle geodesics 2 near NE BTZ NCR} \\
        \ddot x^\rho + \lb \frac{r_+\lb \xr\rb^2}{\ell^2 \alpha_+} \rb \lb\dot x^t \rb^2 -  \lb \frac{r_+ \xr}{2\alpha_+} \rb \lb \dot x^\phi \rb^2 &= 0 \,. \label{eq: NLO point-particle geodesics 3 near NE BTZ NCR}
    \end{align}
\end{subequations}

Equations \eqref{eq: NLO point-particle geodesics 1 near NE BTZ NCR} and \eqref{eq: NLO point-particle geodesics 1 near NE BTZ NCR} can be linearly combined to obtain (and using \eqref{eq: LO point-particle geodesics near NE BTZ NCR} and \eqref{eq: NCR BTZ relation 1})
\begin{subequations}\label{eq: NLO BTZ geodesic eqn}
    \begin{align}
        \frac{d}{d\lambda} \left[ \lb \frac{\xr}{\ell} \rb^2 \lb \dot x^\phi - \frac{J}{2r_+^2} \dot x^t \rb + \lb y^\phi - \frac{J}{2r_+^2} \dot y^t \rb \right] &= 0 \,, \label{eq: NLO BTZ geodesic eqn 1}\\
        \frac{d}{d\lambda} \left[ \lb \xr \rb^2 \lb \dot x^t - \frac{J\alpha_+}{r_+}\dot x^\varphi \rb  \right] &= 0\,. \label{eq: NLO BTZ geodesic eqn 2}
    \end{align}
\end{subequations}

These equations are equivalent to \eqref{eq: NLO point-particle geodesics 1 near NE BTZ} and \eqref{eq: NLO point-particle geodesics 2 near NE BTZ} after the identifications
\begin{equation}\label{eq: NCR BTZ relation 2}
    \xtt (\lambda) = \xt(\lambda) - \frac{J\alpha_+}{r_+}\xff(\lambda) \,, \qquad \yff(\lambda) = \yf(\lambda) - \frac{J}{2r_+^2}y^t(\lambda) \,.
\end{equation}

Plugging \eqref{eq: NCR BTZ relation 1} and \eqref{eq: NCR BTZ relation 2} in equation \eqref{eq: NLO point-particle geodesics 3 near NE BTZ NCR}, we get
\begin{equation}
    \ddot x^\rho + \frac{\xr}{4\alpha_+^2}\lb \dot x^T \rb^2 - \lb \frac{r_+^2 \xr}{\ell^2} \rb \lb \dot x^\varphi \rb^2 = 0 \,.
\end{equation}

This exercise concludes the fact that the analysis in corotating frame and non corotating frame of the rotating BTZ gives us identical results.

\subsection{Geodesics from NH limit of Rotating BTZ geodesics}\label{Geodesics from near horizon limit of Rotating BTZ geodesics}

In this section, we analyse the geodesic equations in BTZ black hole in non-corotating frame $(t,r,\phi)$. The geodesic equations are
\begin{subequations}
    \begin{align}
        \ddot X^\phi - \frac{2\, r_+ r_- X^r}{\ell\, \Xi(X^r) }\dot X^t\dot X^r + \frac{2X^r((X^r)^2-r_-^2-r_+^2)}{\Xi(X^r)} \dot X^r \dot X^\phi &= 0 \,,  \\
        \ddot X^t-\frac{2 \ell r_- r_+ X^r}{\Xi(X^r)}\dot X^\phi \dot X^r   +\frac{2  (X^r)^3}{\Xi(X^r)} \dot X^t \dot X^r &= 0 \,,  \\
        \ddot X^r   + \frac{ \left(r_-^2 r_+^2-(X^r)^4\right)}{\Xi(X^r) X^r}(\dot X^r)^2 
          -\frac{ \Xi(X^r)}{\ell^2 X^r} (\dot X^\phi)^2 + \frac{ \Xi(X^r)}{\ell^4 X^r}(\dot X^t)^2 & = 0 \,,
    \end{align}
\end{subequations}
where
\begin{equation}
    \Xi(X^r) := \left( (X^r)^2-r_-^2\right) \left( (X^r)^2-r_+^2\right) \,.
\end{equation}

In the near-horizon setting, we do the following transformations on the worldline coordinates
\begin{equation}\label{eq:expansion of Schwarzschild Coordinates for RBTZ}
\begin{aligned}
& X^t=x^t+\epsilon y^t+\cdots,~~~~X^r=r_h+\frac{\epsilon}{4\alpha_+} \lb x^\rho \rb^2+\cdots \,,\\&
 ~~~~~~~~~~~~~~~~~~~~~~~~X^\phi=x^\phi+\epsilon y^{\phi}+\cdots .
\end{aligned}
\end{equation}

The geodesic equations expand as (upto $\mathcal{O}(\epsilon) $)
\begin{subequations}\label{eq: geodesic expansion BTZ NCR}
    \begin{align}
       \ddot x^\phi +  \lb \frac{2 r_+ r_- }{ \ell\xr \Delta_r}\rb \dot x^t\dot x^\rho  -  \lb \frac{2 r_-^2 }{\xr\Delta_r}\rb \dot x^\phi \dot x^\rho  &= 0 \, , \label{eq: geodesic expansion 1 BTZ NCR}  \\
        \ddot y^\phi + \lb \frac{\left(r_-^4-r_+^2 r_-^2+4 r_+^4\right) \xr }{2 \ell^2 r_+^2  \Delta_r }\rb  \dot x^\phi \dot x^\rho -\lb \frac{\left(r_-^3+3 r_+^2 r_-\right) \xr }{2 \ell^3 r_+\Delta_r} \rb \dot x^\rho \dot x^t & \nn\\+ \lb \frac{2 r_- r_+ }{ \ell  \Delta_r  \xr}\rb \dot x^\rho \dot y^t   -  \lb \frac{2 r_-^2 }{ \Delta_r  \xr }\rb \dot x^\rho \dot y^\phi &= 0\, ,  \label{eq: geodesic expansion 2 BTZ NCR} \\
        \ddot x^t - \lb \frac{2 \ell r_- r_+ }{\xr\Delta_r }\rb  \dot x^\phi \dot x^\rho + \lb \frac{2 r_+^2 }{\xr\Delta_r}\rb \dot x^\rho \dot x^t  &= 0 \,, \label{eq: geodesic expansion 3 BTZ NCR} \\
        \ddot y^t + \lb \frac{\left(r_-^3+3 r_+^2 r_-\right) \xr }{2 \ell r_+  \Delta_r }\rb \dot x^\rho \dot x^\phi - \lb \frac{\left(5 r_-^2-r_+^2\right) \xr }{2 \ell^2 \Delta_r}\rb  \dot x^t \dot x^\rho & \nn \\
          + \lb \frac{2 r_+^2 }{ \Delta_r  \xr}\rb  \dot x^\rho \dot y^t - \lb \frac{2 \ell r_- r_+ }{ \Delta_r  \xr }\rb \dot x^\rho \dot y^\phi  &= 0 \,, \label{eq: geodesic expansion 4 BTZ NCR}  \\
        \ddot x^\rho + \lb \frac{r_+\lb \xr\rb^2}{2 \alpha _+ \ell^2} \rb \lb \dot x^t \rb^2 - \lb \frac{r_+ \xr }{2 \alpha_+}\rb  \lb \dot x^\phi \rb^2 &= 0 \,, \label{eq: geodesic expansion 5 BTZ NCR}
    \end{align}
\end{subequations}
where $\Delta_r=r_+^2-r_-^2$. Equation \eqref{eq: geodesic expansion 5 BTZ NCR} is equivalent to \eqref{eq: NLO point-particle geodesics 3 near NE BTZ NCR}. Equations \eqref{eq: geodesic expansion 1 BTZ NCR} and \eqref{eq: geodesic expansion 3 BTZ NCR} can be linearly combined to obtain
\begin{align}
     & \ddot x^\phi - \frac{J}{2 r_+^2} \ddot x^t = 0 \,, \label{eq: geodesic expansion 6 BTZ NCR}  \\
     &\lb  \frac{\alpha _+ J^2}{2 r_+^3}+1  \rb \ddot x^t - \frac{\alpha _+ J}{r_+}\ddot x^\phi 
     + \frac{2\dot x^\rho}{\xr} \left[ \lb  \frac{\alpha _+ J^2}{2 r_+^3}+1  \rb \dot x^t - \frac{\alpha _+ J}{r_+}\dot x^\phi \right]= 0\label{eq: geodesic expansion 7 BTZ NCR}\,.
\end{align}

\eqref{eq: geodesic expansion 6 BTZ NCR} is equal to \eqref{eq: LO point-particle geodesics near NE BTZ NCR}. Moreover, using \eqref{eq: NCR BTZ relation 2}, one can easily see that \eqref{eq: geodesic expansion 7 BTZ NCR} is equivalent to \eqref{eq: NLO BTZ geodesic eqn 2}. Similarly \eqref{eq: geodesic expansion 2 BTZ NCR} and \eqref{eq: geodesic expansion 4 BTZ NCR} can be linearly combined to get
\begin{equation} \label{eq: geodesic expansion 8 BTZ NCR}
    \lb \ddot y^\phi -\frac{J}{2r_+^2} \ddot y^t \rb +\frac{2 \xr\dot x^\rho }{\ell^2} \lb \dot x^\phi - \frac{J}{2 r_+^2} \dot x^t \rb = 0 \, .
\end{equation}

\eqref{eq: geodesic expansion 8 BTZ NCR} is actually \eqref{eq: NLO BTZ geodesic eqn 1} in disguise after using \eqref{eq: geodesic expansion 6 BTZ NCR} (or equivalently \eqref{eq: LO point-particle geodesics near NE BTZ NCR}). Thus we establish that our near-horizon expansions of geodesic equations for a rotating non-extremal BTZ black hole gives matches with our intrinsic analysis of point particle in the near-horizon limit. This calculation explicitly establishes this correspondence in the usual Boyer-Lindquist-like (BL) coordinate system. However, we have already shown that the analysis in BL coordinate system and co-rotating coordinate system are the same. Hence we can also extend this argument in that coordinate system.


\subsection{Scalar fields in NH BTZ}

\subsection*{Using the map}
Using the map between the near-horizon rotating BTZ metric and the SC metric \eqref{eq:rotating BTZ BH Map inverse}, we explicitly write the action of a minimally coupled scalar field on the near-horizon region of a rotating BTZ black hole\footnote{In this and the following sections, we will analyse the rotating BTZ black hole in the new coordinate system. We will relabel $T$ to $t$ instead of using $T$ as the time coordinate in the new coordinate system.}
\begin{equation}\label{eq:LO Scalar Field action near RBTZ BH}
    \mathcal{S}_{LO}=-\frac{1}{2}\int d^4x~e\Big(-\frac{4\alpha_+^2}{\rho^2}\left(\partial_t\phi^{(0)}\right)^2+\left(\partial_\rho\phi^{(0)}\right)^2\Big),
\end{equation}
The equation of motion for \eqref{eq:LO Scalar Field action near RBTZ BH} is
\begin{equation}\label{eq:LO Scalar Field EOM near RBTZ BH}
    \partial_t^2\phi^{(0)}-\frac{\rho}{4\alpha_+^2}\partial_\rho\phi^{(0)}-\frac{\rho^2}{4\alpha_+^2}\partial_\rho^2\phi^{(0)}=0\,.
\end{equation}

Next, we analyse the subleading action $\mathcal{S}_{NLO}$, which is given as 
\begin{multline}\label{eq:NLO Scalar Field action near RBTZ BH}
    \mathcal{S}_{NLO}=-\frac{1}{2}\int d^4x\:e\Big(2v^{\mu\nu}\partial_\mu\phi^{(0)}\partial_\mu\phi^{(1)}+\bar{h}^{\mu\nu}\partial_\mu\phi^{(0)}\partial_\mu\phi^{(0)}\\
    +m^2\left(\phi^{(0)}\right)^2+\frac{\rho^2}{4r_+^2\alpha_+}v^{\mu\nu}\partial_\mu\phi^{(0)}\partial_\mu\phi^{(0)}\Big).
\end{multline}
Varying the $\phi^{(1)}$ field in the NLO action \eqref{eq:NLO Scalar Field action near RBTZ BH}, we obtain the LO equation of motion \eqref{eq:LO Scalar Field EOM near RBTZ BH}. The equation of motion of the $\phi^{(0)}$ field is 
\begin{multline}\label{eq:NLO Scalar Field EOM near RBTZ BH}
\frac{1}{\rho}\partial_\rho\phi^{(1)}+\left(\frac{9r_+^2+3r_-^2}{4\ell^2r_+^2}\right)\rho\partial_\rho\phi^{(0)}-m^2{\phi^{(0)}}+\frac{1}{r_+^2}\partial_\varphi^2\phi^{(0)}-\frac{4\alpha^2_+}{\rho^2}\partial_t^2\phi^{(1)}\\-\frac{\alpha_+}{2r_+}\partial_t^2\phi^{(0)}+\partial_\rho^2\phi^{(1)}+\left(\frac{3r_+^2+r_-^2}{4\ell^2r_+^2}\right)\rho^2\partial_\rho^2\phi^{(0)}=0\,.
\end{multline}

\subsection*{From NH limit}\label{sec:Scalar field from near horizon limit (Rotating BTZ Black hole)}
In this section, we will verify that, if we take near-horizon limits on the scalar field equations in the background of a rotating BTZ black hole, we can retrieve the LO equations of motion \eqref{eq:LO Scalar Field EOM near RBTZ BH}. The equation of motion of the scalar field in the BTZ black hole background is
\begin{multline}\label{eq:scalar field EOM RBTZ}
    \frac{(r^2-r_+^2)(r^2-r_-^2)+2(r^4-r_+^2r_-^2)}{\ell^2r^3}\d_r\Phi-\frac{\ell^2 r_+^2}{\left(r^2-r_+^2\right) \left(r_+^2-r_-^2\right)}\d^2_t\Phi\\+\frac{(r^2-r_+^2)(r^2-r_-^2)}{\ell^2r^2}\d_r^2\Phi+\frac{1}{r_+^2}\left(\frac{r_+^2-r_-^2}{r^2-r_-^2}\right)\d_\varphi^2\Phi-m^2\Phi=0\,.
\end{multline}
Now, we will expand the scalar field as $\Phi\to\phi^{(0)}+\epsilon\phi^{(1)}+\cdots$. The near-horizon limit will be implemented by the following coordinate transformation of the radial coordinate $r\to r_+ + \frac{\epsilon}{4 \alpha_+}\rho^2$. The equation of motion \eqref{eq:scalar field EOM RBTZ} can be arranged in powers of $\epsilon$ as follows
\begin{subequations}\label{eq:taking limits on scalar field EOM RBTZ}
    \begin{align}
        \mathcal{O}(\epsilon^{-1}):~& \partial_t^2\phi^{(0)}-\frac{\rho}{4\alpha_+^2}\partial_\rho\phi^{(0)}-\frac{\rho^2}{4\alpha_+^2}\partial_\rho^2\phi^{(0)}=0\,,\label{eq:LO EOM from limiting analysis RBTZ}\\
        \mathcal{O}(\epsilon^{0}):~& \frac{1}{\rho}\partial_\rho\phi^{(1)}+\left(\frac{7r_+^2+5r_-^2}{4\ell^2r_+^2}\right)\rho\partial_\rho\phi^{(0)}-m^2{\phi^{(0)}}+\frac{1}{r_+^2}\partial_\varphi^2\phi^{(0)}-\frac{4\alpha^2_+}{\rho^2}\partial_t^2\phi^{(1)}\nonumber\\&+\frac{\alpha_+}{2r_+}\partial_t^2\phi^{(0)}+\partial_\rho^2\phi^{(1)}+\left(\frac{r_+^2+3r_-^2}{4\ell^2r_+^2}\right)\rho^2\partial_\rho^2\phi^{(0)}=0\,\label{eq:NLO EOM from limiting analysis RBTZ}.
    \end{align}
\end{subequations}
The leading order equations of motion \cref{eq:LO Scalar Field EOM near RBTZ BH,eq:LO EOM from limiting analysis RBTZ} match exactly in both analyses. Whereas the NLO equations of motion \cref{eq:NLO Scalar Field EOM near RBTZ BH,eq:NLO EOM from limiting analysis RBTZ} match on adding the LO equations of motion
\begin{eqnarray}
    \eqref{eq:NLO Scalar Field EOM near RBTZ BH}+\frac{\alpha_+}{r_+}\eqref{eq:LO Scalar Field EOM near RBTZ BH}=\eqref{eq:NLO EOM from limiting analysis RBTZ}\,.
\end{eqnarray}

\section{String Carroll Maps}\label{String Carroll Maps}
In this section, we will list the SC metric data for various black bodies apart from those covered in the previous sections.\\

\underline{\textit{(A)dS Schwarzschild Black hole}}:\\
The metric for the Schwarzschild black hole in asymptotically AdS and dS spacetimes is given as 
\begin{eqnarray}
    ds^2=-\left(1-\frac{2M}{r}+(-1)^s\frac{r^2}{\ell^2}\right)dt^2+\frac{1}{\left(1-\frac{2M}{r}+(-1)^s\frac{r^2}{\ell^2}\right)}dr^2+r^2d\Omega^2\,,
\end{eqnarray}
where $s=0$ corresponds to the Schwarzschild black hole in asymptotically AdS spacetime, whereas $s=1$ corresponds to the Schwarzschild black hole in asymptotically dS spacetime. The event horizon for these black holes will be denoted by $r_h$. The SC metric data for this black hole are given below with $\alpha=\frac{\ell^2r_h}{\ell^2+3(-1)^sr_h^2}$
\begin{equation}
        h_{\mu\nu}dx^\mu dx^\nu=r_h^2d\Omega_{S^2}^2\,,~~\tau_{\mu\nu}dx^\mu dx^\nu=-\frac{\rho^2}{4\alpha^2}dt^2+d\rho^2\,,~~\Phi_{\mu\nu}dx^\mu dx^\nu=\frac{r_h}{2\alpha}\rho^2d\Omega_{S^2}^2\,.
\end{equation}
The Schwarzschild $dS$ black hole \cite{Gibbons:1977mu} ($s=1$ above) has in addition the cosmological horizon at the second physical zero of the blackening factor: the focus here is solely on the (outer) black hole horizon $r_h$.

\medskip

\underline{\textit{AdS Kerr Black hole}}:\\
The metric for the Kerr black hole in asymptotically AdS spacetimes is given as 
\begin{multline}
ds^2 = -{\frac{\Delta_r}{\Sigma}}
\left[dt-\frac{a\sin^2\theta}{\Xi}\ d\phi\right]^2
+\frac{\Sigma}{\Delta_r}\ dr^2+\frac{\Sigma}{\Delta_\theta}\ d\theta^2
\\+\frac{\Delta_\theta\sin^2\theta}{\Sigma}
\left[a\ dt-\frac{r^2+a^2}{\Xi}\ d\phi\right]^2\,,
\end{multline}
where,
\begin{equation}
\Sigma=r^2+a^2\cos^2\theta\,,~~\Xi=1-\frac{a^2}{l^2}\,,~~\Delta_r=(r^2+a^2)\left( 1+\frac{r^2}{l^2}\right)-2mr\,,~~\Delta_\theta=1-\frac{a^2}{l^2}\cos^2\theta\:.    
\end{equation}
Here $a$ denotes the rotational parameter. The black hole horizons are located at $\Delta_r=0$, and the outer event horizon for this black hole will be denoted by $r_h$. The SC metric data for this black hole are given below with $\lambda=\frac{(a^2+\ell^2)r_h^2+3r_h^4-a^2\ell^2}{4\ell^2r_h\Sigma(r_h)}$
\begin{equation}
\begin{split}
    h_{\mu\nu}dx^\mu dx^\nu&=\frac{\Sigma(r_h)}{\Delta_\theta}d\theta^2+\frac{\Delta_\theta \sin^2\theta}{\Sigma(r_h)}\left(adt-\frac{(a^2+r_h^2)}{\Xi}d\phi\right)^2\,,\\
    \tau_{\mu\nu}dx^\mu dx^\nu&=d\rho^2-4\lambda^2\rho^2\left(dt-\frac{a\sin^2\theta}{\Xi}d\phi\right)^2\,,\\
    \Phi_{\mu\nu}dx^\mu dx^\nu&=-\frac{2r_h\Delta_\theta\lambda\rho^2a^2\sin^2\theta}{\Sigma(r_h)^2}\Bigg(dt^2-\frac{2a\sin^2\theta}{\Xi}dtd\phi\\&-\frac{\left(a^2+r_h^2\right)\left(a^2 \cos 2\theta+r_h^2\right)}{a^2\Xi^2}d\phi^2\Bigg)+\frac{2r_h\lambda\rho^2}{\Delta_\theta}d\theta^2\,.
\end{split}
\end{equation}

\medskip

\underline{\textit{HSV Lifshitz Black hole}}:\\
The metric for the HSV Lifshitz black hole is given as 
\begin{eqnarray}\label{eq:HSV_lifshitz_BH}
    ds^2=\left(\frac{r_F}{r}\right)^\Theta\left(\frac{l^2}{r^2f(r)}dr^2-\frac{r^{2z}}{l^{2z}}f(r)dt^2+r^2d\Omega^2\right),
\end{eqnarray}
where,
\begin{eqnarray}
    f(r)=1-\frac{m}{r^{2+z-\Theta}}+\frac{l^2}{r^2(z-\Theta)^2}.
\end{eqnarray}
The SC metric data for this black hole are given below
\begin{equation}
\begin{split}
    h_{\mu\nu}dx^\mu dx^\nu&=r_h^2 \left(\frac{r_F}{r_h}\right)^{\Theta }d\Omega_{S^2}^2\,,\\
    \tau_{\mu\nu}dx^\mu dx^\nu&=-4\lambda^2 \rho ^2\left(\frac{r_F}{r_h}\right)^{2\Theta}\left(\frac{\ell}{r_h}\right)^{2-2 z}dt^2+d\rho^2\,,\\
    \Phi_{\mu\nu}dx^\mu dx^\nu&=(2-\Theta) \lambda  \rho ^2 r_h\left(\frac{r_F}{r_h}\right)^{\Theta}d\Omega_{S^2}^2\,,
\end{split}
\end{equation}
where $$\lambda=\frac{\left(\ell^2+r_h^2 (z-\Theta) (-\Theta +z+2)\right)}{4 \ell^2 r_h (z-\Theta)\left(\frac{r_F}{r_h}\right)^{\Theta }}\,.$$

\medskip

\underline{\textit{Plebanski-Demianski Black holes}}:\\
The Plebanski-Demianski solutions \cite{Plebanski:1976ab} is a seven parameter family of stationary and axisymmetric black holes of which Schwarzschild, Kerr and Reissner Nordstr\"om are special cases. The seven parameters are mass ($M$), angular momentum ($J=aM$) electric and magnetic charges ($Q_e$ and $Q_m$), cosmological constant ($\Lambda$), Taub-NUT charge ($l$) and acceleration ($\alpha$). If the parameters are set appropriately, the outermost horizon would be the event horizon. In this section, we will show that this family of black holes admits a string-Carroll expansion in the vicinity of their event horizon assuming that it is the outermost horizon. The metric of the Plebanski-Demianski family in the Boyer-Lindquist coordinates is given as 
\begin{equation}\label{eq:PD Metric}
    g=\frac{1}{\Omega^2}\Bigg(-\frac{\Delta}{\Sigma}\Big(dt-\Gamma~d\phi\Big)^2+\frac{\Sigma}{\Delta}dr^2+\frac{\Sigma}{P}d\theta^2+\frac{P\sin^2\theta}{\Sigma}\Big(a~dt-(r^2+a^2+l^2)~d\phi\Big)^2\Bigg)\,.
\end{equation}
The metric functions and constants are given as 
\begin{subequations}
    \begin{align}
    P =&~1+ak_1\cos\theta+a^2k_2\cos^2\theta\,,\\
    \Omega =&~1-\alpha(l+a\cos\theta)~r\,,\\
    \Sigma =&~r^2+(l+a\cos\theta)^2\,,\\
    \Delta =&~ \left(\left(k+Q_e^2+Q_m^2\right)\left(2 \alpha  l r+1\right)-2 M r+\frac{kr^2}{a^2-l^2}\right)\Big(1-\alpha(a+l)r\Big) \Big(1+\alpha(a-l)r\Big)\nonumber\\
    &~-\frac{\Lambda}{3}r^2 \Big(2 \alpha  l r \left(a^2-l^2\right)+a^2+3 l^2+r^2\Big)\,,\\
    \Gamma=&~a\sin^2\theta-2l\cos\theta\,,\\
    k=&~(a^2-l^2)\frac{1+\alpha l(2M-3\alpha(Q_e^2+Q_m^2))-\Lambda l^2}{1+3\alpha^2l^2(a^2-l^2)}\,,\\
    k_1=&~4\alpha^2l\left(k+Q_e^2+Q_m^2\right)-2\alpha M+4l\frac{\Lambda}{3}\,,\\
    k_2=&~\alpha^2\left(k+Q_e^2+Q_m^2\right)+\frac{\Lambda}{3}\,.
    \end{align}
\end{subequations}
The horizons of the black hole are the roots of the metric function $\Delta$. The metric function $\Delta$ is a quartic polynomial in the $r$-coordinate, and the event horizon ($r_h$) is one of the roots. In the following analysis, we will subtitute the mass parameter in terms of the horizon radius and the remaining parameters using the relation $\Delta(r_h)=0$. The metric \eqref{eq:PD Metric} can be split into two doubly degenerate metrics as follows 
\begin{subequations}\label{eq:PD metric split}
\begin{align}
    g=~&H+T\,,\\
    H=~&\frac{1}{\Omega^2}\Bigg(\frac{\Sigma}{P}d\theta^2+\frac{P\sin^2\theta}{\Sigma}\Big(a~dt-(r^2+a^2+l^2)~d\phi\Big)^2\Bigg)\,,\\
    T=~&\frac{1}{\Omega^2}\Bigg(-\frac{\Delta}{\Sigma}\Big(dt-\Gamma~d\phi\Big)^2+\frac{\Sigma}{\Delta}dr^2\Bigg)\,.
\end{align}    
\end{subequations}

To probe the near-horizon region of the black holes, a new radial coordinate $\rho$ is introduced, which is related to $r$ by the coordinate transformation $r=r_h+\epsilon\rho^2$, where $\epsilon$ captures the closeness to the horizon. The near-horizon expansion of the metric \eqref{eq:PD Metric} is obtained by expanding the doubly-degenerate metrics $H$ and $T$ in $\epsilon$
\begin{eqnarray}\label{eq:PD expansion}
H=H_1+\epsilon H_2+\cdots\,,\quad\quad T=\epsilon~T_1+\epsilon^2~T_2+\cdots\,.
\end{eqnarray}
Similarly, the inverse metric $g^{-1}$ can be split into doubly degenerate structures $g^{-1}=V+h$ such that 
\begin{subequations}
    \begin{align}
        V=~&\Omega^2\Bigg(-\frac{\Sigma}{\left(a^2+l^2+r^2-a\Gamma\right)^2\Delta}\Big((a^2+l^2+r^2)dt+a~d\phi\Big)^2+\frac{\Delta}{\Sigma}dr^2\Bigg)\,,\\
        h=~&\Omega^2\Bigg(\frac{P}{\Sigma}d\theta^2+\frac{\Sigma\csc^2\theta}{\left(a^2+l^2+r^2-a\Gamma\right)^2P}\Big(\Gamma~dt+d\phi\Big)^2\Bigg)\,.
    \end{align}
\end{subequations}
Following the same procedure and expanding the structures $V$ and $h$ in small $\epsilon$, we obtain
\begin{eqnarray}
    V=\frac{1}{\epsilon}V_1+V_2+\cdots\,,\quad\quad h=h_1+\epsilon h_2+\cdots\,.
\end{eqnarray}
These expansions can be mapped to the string-Carroll expansion given in \eqref{SCmetric} after identifying $\epsilon=c^2$. The string-Carroll metric data for \eqref{eq:PD metric split} is found to be
\begin{subequations}
\begin{align}
h_{\mu\nu}dx^\mu dx^\nu=~&H_1=\frac{1}{\Omega(r_h)^2}\Bigg(\frac{\Sigma(r_h)}{P}d\theta^2+\frac{P\sin^2\theta}{\Sigma(r_h)}\Big(a~dt-(r_h^2+a^2+l^2)~d\phi\Big)^2\Bigg)\,,\\
\tau_{\mu\nu}dx^\mu dx^\nu=~&T_1=\frac{1}{\Omega(r_h)^2}\Bigg(-\frac{\rho ^2 \partial_r \Delta\Big|_{r=r_h}}{\Sigma \left(r_h\right)}\Big(dt-\Gamma d\phi\Big)^2+ \frac{4 \Sigma \left(r_h\right)}{\partial_r \Delta\Big|_{r=r_h} } d\rho^2\Bigg)\,.
\end{align}
\end{subequations}

\section{Determinant of the string- Carroll Metric}\label{App:Determinant of the Metric}
In this section, we will discuss the expansion of the determinant of the SC metric. This section has been worked out in collaboration with Pushkar Soni. The determinant of a $d\times d$ metric is given as 
\begin{eqnarray}
    g=\varepsilon^{\mu_1\cdots\mu_d}g_{1\mu_1}\cdots g_{d\mu_d},
\end{eqnarray}
where $\varepsilon^{\mu_1\cdots\mu_d}$ is a Levi-Civita symbol in d-dimensions. Substituting the metric in terms of the decomposed vielbeine \eqref{eq:SC metric decomposition}, we get 
\begin{eqnarray}
    g=\varepsilon^{\mu_1\cdots\mu_d}(c^2 T_{1\mu_1}+\Pi^\perp_{1\mu_1})\cdots (c^2 T_{d\mu_d}+\Pi^\perp_{d\mu_d})\,,
\end{eqnarray}
such that $T_{\mu\nu}$ and $\Pi_{\mu\nu}$ are doubly degenerate. Arranging the metric components in orders of $c$ and considering four dimensional spacetime we get,
\begin{multline}
g=\underbrace{\varepsilon^{\mu_1\cdots\mu_4}\prod_{i=1}^4\Pi^\perp_{i\mu_i}}_{\det(\Pi^\perp)}+c^2\varepsilon^{\mu_1\cdots\mu_4}\sum_{n=1}^4T_{n\mu_n}\prod_{\substack{i=1\\i\neq n}}^4\Pi^\perp_{i\mu_i}\\+c^4\varepsilon^{\mu_1\cdots\mu_4}\sum_{n=1}^4\sum_{m>n}^4T_{n\mu_n}T_{m\mu_m}\prod_{\substack{i=1\\i\neq n\neq m}}^4\Pi^\perp_{i\mu_i}+\mathcal{O}(c^6).
\end{multline}
The first term is 0 because it is the determinant of a degenerate metric. The terms of $\mathcal{O}(c^2)$, $\mathcal{O}(c^6)$ and above are also zero because these terms contain an unequal number of $T_{\mu\nu}$ 's and $\Pi^{\perp}_{\mu\nu}$ 's. Therefore, we are left with only the $\mathcal{O}(c^4)$ term. Expanding $\Pi^\perp_{\mu\nu}$ as $\Pi^\perp_{\mu\nu}=h_{\mu\nu}+c^2\Phi_{\mu\nu}$ and $c^2T_{\m\n}=c^2\tau_{\m\n}+c^4m_{\m\n}$ in the $\mathcal{O}(c^4)$ term we get,
\begin{equation}
    g=c^4\varepsilon^{\mu_1\cdots\mu_4}\sum_{n=1}^4\sum_{m>n}^4(\tau_{n\mu_n}+c^2m_{n\mu_n})(\tau_{m\mu_m}+c^2m_{m\mu_m})\prod_{\substack{i=1\\i\neq n\neq m}}^4(h_{i\mu_i}+c^2\Phi_{i\mu_i}).
\end{equation}
Rearranging this expansion, we get
\begin{multline}
    g=c^4\varepsilon^{\mu_1\cdots\mu_4}\sum_{n=1}^4\sum_{m>n}^4\tau_{n\mu_n}\tau_{m\mu_m}\prod_{\substack{i=1\\i\neq n\neq m}}^4h_{i\mu_i}\\+c^6\varepsilon^{\mu_1\cdots\mu_4}\sum_{n=1}^4\sum_{m>n}^4(\tau_{n\mu_n}m_{m\mu_m}+\tau_{m\mu_m}m_{n\mu_n})\prod_{\substack{i=1\\i\neq n\neq m}}^4h_{i\mu_i}\\+c^6\varepsilon^{\mu_1\cdots\mu_4}\sum_{n=1}^4\sum_{m>n}^4(h_{n\mu_n}\Phi_{m\mu_m}+h_{m\mu_m}\Phi_{n\mu_n})\prod^4_{\substack{i=1\\i\neq n\neq m}}\tau_{i\mu_i}.
\end{multline}
The $\mathcal{O}(c^4)$ term can be written as 
\begin{equation}
    g=c^4\varepsilon^{\mu_1\cdots\mu_d}\sum_{n=1}^d\sum_{m>n}^d\tau_{n\mu_n}\tau_{m\mu_m}\prod_{\substack{i=1\\i\neq n\neq m}}^dh_{i\mu_i}=\det(c\tau_{\mu\nu}+ch_{\mu\nu})\,.
\end{equation}
The terms up to $\mathcal{O}(c^6)$ can be written as 
\begin{eqnarray}\label{eq:determinant of g upto order six}
g=\det\Big[c(\tau_{\mu\nu}+h_{\mu\nu})+c^3(m_{\mu\nu}+\Phi_{\mu\nu})\Big]=c^4\det\Big[\tau_{\mu\nu}+h_{\mu\nu}+c^2(m_{\mu\nu}+\Phi_{\mu\nu})\Big]\,.
\end{eqnarray}
For ease of notation, we will define $A_{\mu\nu}=\tau_{\m\n}+h_{\m\n}$ such that its inverse is $A^{\m\n}=v^{\m\n}+h^{\m\n}$. The equation \eqref{eq:determinant of g upto order six} can be written as 
\begin{eqnarray}\label{eq:rearranging the determinant}
g=c^4\det[A_{\mu\rho}]\det\Big[\delta^\rho_\n+c^2A^{\rho\sigma}(m_{\sigma\nu}+\Phi_{\sigma\nu})\Big]\,.   
\end{eqnarray}
For the next step, we will require the following identity for the determinant of a matrix $\mathbb{1}+\epsilon A$, where $\mathbb{1}$ is the identity matrix and $\epsilon$ is small
\begin{eqnarray}\label{eq:identity}
    \det(\mathbb{1}+\epsilon A)=\exp(\Tr(\log(\mathbb{1}+\epsilon A)))=1+\epsilon\Tr(A)+\cdots
\end{eqnarray}
Using the identity \eqref{eq:identity} in \eqref{eq:rearranging the determinant} for small $c^2$ we get,
\begin{eqnarray}
g=c^4\det[\tau_{\mu\n}+h_{\m\n}]\Big[1+c^2(v^{\m\n}+h^{\m\n})(m_{\m\nu}+\Phi_{\m\nu})\Big]\,.
\end{eqnarray}
Finally, we can write down the following expansion for the square root of the determinant of the metric
\begin{eqnarray}\label{e:det-Expansion}
    \sqrt{-\det g_{\m\n}}=c^2\sqrt{-\det(\tau_{\mu\n}+h_{\m\n})}\left(1+\frac{c^2}{2}(v^{\m\n}+h^{\m\n})(m_{\m\nu}+\Phi_{\m\nu})+\cdots\right)\,.
\end{eqnarray}
The subleading factor ($\mathcal{O}(c^2)$) is the one that appears in the subleading SC action.
\newpage
\bibliographystyle{JHEP.bst}
\bibliography{ccft}

\end{document}